\documentclass[usenatbib]{mnras}
\usepackage{amsmath}
\usepackage{amssymb}
\usepackage{graphicx}
\usepackage[utf8]{inputenc}
\usepackage{listings}
\usepackage{geometry}
\usepackage{xcolor}

\hypersetup{draft} %to ignore some warnings for now

\newcommand{\equ}[1]{eq.~(\ref{eq:#1})}
\newcommand{\equs}[1]{eqs.~(\ref{eq:#1})}
\newcommand{\equm}[1]{(\ref{eq:#1})}

\newcommand{\Equs}[1]{Eqs.~(\ref{eq:#1})}
\newcommand{\equnp}[1]{eq.~\ref{eq:#1}}
\newcommand{\equsnp}[1]{eqs.~\ref{eq:#1}}
\newcommand{\equmnp}[1]{\ref{eq:#1}}
\newcommand{\se}[1]{\S\ref{sec:#1}}
\newcommand{\fig}[1]{Fig.~\ref{fig:#1}}
\newcommand{\figs}[1]{Figs.~\ref{fig:#1}}
\newcommand{\figss}[1]{\ref{fig:#1}}
\newcommand{\Fig}[1]{Figure~\ref{fig:#1}}

\newcommand{\tab}[1]{Table~\ref{tab:#1}}
\newcommand{\be}{\begin{equation}}
\newcommand{\ee}{\end{equation}}
\newcommand{\bea}{\begin{eqnarray}}
\newcommand{\eea}{\end{eqnarray}}

\newcommand{\no}{\noindent}

\newcommand{\msun}{{\rm M}_\odot}

\newcommand{\ifm}[1]{\relax\ifmmode#1\else$\mathsurround=0pt #1$\fi}
\newcommand{\kms}{\ifmmode\,{\rm km}\,{\rm s}^{-1}\else km$\,$s$^{-1}$\fi}

\newcommand{\Mpc}{\,{\rm Mpc}}
\newcommand{\kpc}{\,{\rm kpc}}
\newcommand{\pc}{\,{\rm pc}}

\newcommand{\K}{\,{\rm K}}

\newcommand{\ltsima}{$\; \buildrel < \over \sim \;$}
\newcommand{\lsim}{\lower.5ex\hbox{\ltsima}}
\newcommand{\gtsima}{$\; \buildrel > \over \sim \;$}
\newcommand{\gsim}{\lower.5ex\hbox{\gtsima}}

\def\sy{\,M_\odot\, {\rm yr}^{-1}}

\def\M*{M_{\rm *}}

\def\Rv{R_{\rm v}}

\def\tff{t_{\rm ff}}

\def\Rs{R_{\rm s}}
\def\Pc{P_{\rm c}}

\def\deltac{\delta_{\rm c}}
\def\mucr{\mu_{\rm cr}}
\def\rhoc{\rho_{\rm c}}
\def\rhob{\rho_{\rm b}}
\def\rhos{\rho_{\rm s}}

\def\cb{c_{\rm b}}
\def\cs{c_{\rm s}}
\def\csz{c_{\rm s,0}^2}

\def\Vs{V_{\rm s}}

\def\tsc{t_{\rm sc}}
\def\Mb{M_{\rm b}}

\def\hb{h_{\rm b}}
\def\hs{h_{\rm s}}

\def\Pi{\varpi_{_{\rm I}}}
\def\rmd{{\rm d}}

\usepackage{color}
\begin{document} 

\large
\title[KHI in Self-Gravitating Cylinders]
{Kelvin-Helmholtz Instability in Self-Gravitating Streams}
%Nonlinear Evolution of Kelvin-Helmholtz Instabilities in Self-Gravitating Cylindrical Streams

\author[Aung et al.] 
{\parbox[t]{\textwidth} 
{
Han Aung$^1$\thanks{E-mail: han.aung@yale.edu },
Nir Mandelker$^{2,3}$\thanks{E-mail: nir.mandelker@yale.edu },
Daisuke Nagai$^{1,2}$,
Avishai Dekel$^4$,
Yuval Birnboim$^4$
} 
\\ \\ 
$^1$Department of Physics, Yale University, New Haven, CT 06520, USA;\\
$^2$Department of Astronomy, Yale University, PO Box 208101, New Haven, CT, USA;\\
$^3$Heidelberger Institut f{\"u}r Theoretische Studien, Schloss-Wolfsbrunnenweg 35, 69118 Heidelberg, Germany;\\
$^4$Centre for Astrophysics and Planetary Science, Racah Institute of Physics, The Hebrew University, Jerusalem 91904, Israel}
\date{} 

\pagerange{\pageref{firstpage}--\pageref{lastpage}} \pubyear{0000} 

\maketitle 

\label{firstpage} 

\begin{abstract}
Self-gravitating gaseous filaments exist on many astrophysical scales, from sub-pc filaments in the interstellar medium to Mpc scale streams feeding galaxies from the cosmic web. These filaments are often subject to Kelvin-Helmholtz Instability (KHI) due to shearing against a confining background medium. We study the nonlinear evolution of KHI in pressure-confined self-gravitating gas streams initially in hydrostatic equilibrium, using analytic models and hydrodynamic simulations, not including radiative cooling. We derive a critical line-mass, or mass per unit length, as a function of the stream Mach number and density contrast with respect to the background, $\mucr(\Mb,\deltac)\le 1$, where $\mu=1$ is normalized to the maximal line mass for which initial hydrostatic equilibrium is possible. For $\mu<\mucr$, KHI dominates the stream evolution. A turbulent shear layer expands into the background and leads to stream deceleration at a similar rate to the non-gravitating case. However, with gravity, penetration of the shear layer into the stream is halted at roughly half the initial stream radius by stabilizing buoyancy forces, significantly delaying total stream disruption. % disintegration. 
Streams with $\mucr<\mu\le 1$ fragment and form round, long-lived clumps by gravitational instability (GI), with typical separations roughly 8 times the stream radius, similar to the case without KHI. When KHI is still somewhat effective, these clumps are below the spherical Jeans mass and are partially confined by external pressure, but they approach the Jeans mass as $\mu\rightarrow 1$ and GI dominates. We discuss potential applications of our results to streams feeding galaxies at high redshift, filaments in the ISM, and streams resulting from tidal disruption of stars near the centres of massive galaxies.
\end{abstract}
%add TDE


%%%%%%%%%%%%%%%%%%%%%%%%%%%%%%%%%%%%%%%%%%%%%%%%%%%%%%%%%%
\section{Introduction}
\label{sec:intro}

\smallskip
Filamentary structures are present across many astrophysical scales, from $\Mpc$ to sub-$\pc$. On the largest scales, structure formation occurs in the ``cosmic web", a network of sheets and filaments that connect dark matter haloes \citep{Zeldovich70,Bond96,Springel05}, and is also evident in the distributions of galaxies \citep[e.g.][]{Colless03,Tegmark04,Huchra05}. 
Intergalactic gas cools and condenses towards the centres of the dark matter filaments, forming a network of baryon-dominated intergalactic gas streams \citep{db06,Birnboim16}. There have been several recent attempts to model such streams self-gravitating Mpc-scale gaseous cylinders, which seems consistent with cosmological simulations \citep{Harford08,Harford11,Freundlich14,M18a}. 

\smallskip
At the nodes of the cosmic web, the most massive haloes reside at the intersection of several filaments and are penetrated by the gas streams residing at their centres. These streams constitute the main mode of gas accretion onto the central galaxies \citep{Keres05,Dekel09,Danovich12,Zinger16}. At redshifts $z\gsim 2$, simulations suggest that streams feeding galactic haloes remain dense and cold, with temperatures of $\sim 10^4\K$, as they travel through the hot circumgalactic medium (CGM) towards the central galaxy (\citealp{Keres05,db06,Ocvirk08,Dekel09,CDB,FG11,vdv11}, though see also \citealp{Nelson13,Nelson16}). The filamentary structure in such systems can thus be maintained down to scales of tens of $\kpc$ around galaxies (though see below), where it has been suggested that they may fragment due to gravitational instability \citep[hereafter GI;][]{DSC,Genel12,M18a}. While these cold circumgalactic streams are difficult to directly detect, recent observations have revealed massive extended cold components in the CGM of high-redshift galaxies, whose spatial and kinematic properties are consistent with predictions for cold streams \citep{Bouche13,Bouche16,Prochaska14,Cantalupo14,Martin14a,Martin14b,Borisova16,Fumagalli17,Leclercq17,Arrigoni18}. 

\smallskip
Within galactic discs, spiral arms have been modeled as one dimensional filaments whose gravitational fragmentation leads to the formation of giant molecular clouds (GMCs) or star-forming clumps \citep{Inoue18}. 
Within individual GMCs, \textit{Herschel} observations of star forming regions reveal a multi-scale network of filamentary structures and dense cores aligned with them like beads on a string \citep{Andre10,Jackson10,Arzoumanian11,Kirk13,Palmeirim13}.
%observations show filamentary structures in the interstellar medium (ISM) of galaxies 
This has led to the suggestion that turbulence-driven formation of filaments in the interstellar medium (ISM) is the first step towards core and star-formation \citep{Molinari10,Andre10,Andre14}, a connection which had been speculated for some time \citep[e.g.][]{Schneider79,Larson85}. In this scenario, the densest filaments with widths of order $\sim 0.1\pc$ \citep{Arzoumanian11,Hennebelle13} collapse due to GI and lead to the formation of dense cores where star-formation occurs. Simulations of molecular clouds in the ISM reveal similar multi-scale filamentary structures, arising from a variety of mechanisms such as turbulence, gravitational collapse of larger structures, thermal instabilities, or colliding flows \citep[e.g.][]{Padoan01,Banerjee09,Gomez14,Moeckel15,Smith16}.

\smallskip
Studies of the structure and stability of self-gravitating filaments have a long history, mostly in the context of star-formation in ISM filaments. Early analytic work investigated the stability of an infinite incompressible cylinder with and without an axial magnetic field \citep{Chandrasekhar53}, a compressible yet still homogeneous infinite cylinder \citep{Ostriker64b}, a homogeneous stream of finite radius \citep{Mikhailovskii72,Fridman84}, and a uniformly rotating isothermal cylinder \citep{hansen76}. Hydrostatic equilibrium of a self-gravitating isothermal cylinder is only possible if its mass per unit length (hereafter line-mass) is less than a critical value which depends only on its temperature (\citealp{Ostriker64}; see \equ{isothermal_line_mass} below). For non-isothermal filaments, the critical line-mass is similar (\se{prof}). Filaments with line-mass larger than the critical value must collapse radially. For line-masses smaller than the critical value, a hydrostatic solution exists, but is unstable to long wavelength axisymmetric perturbations. The fastest growing wavelength is roughly eight times radius of the stream, $\lambda\sim 8\Rs$ \citep[][hereafter N87]{Nagasawa}, resulting in stream fragmentation as described in more detail below. A collapsing filament with a line mass slightly exceeding the critical value, as may eventually be the case for a filament growing via radial accretion, is also unstable to axisymmetric perturbations and will fragment at a similar wavelength to the hydrostatic case \citep{Inutsuka92}. Both cases eventually lead to the formation of bound clumps with masses of order the local Jeans mass \citep{Clarke16,Clarke17}. However, if the line-mass greatly exceeds the critical value the filament collapses towards its axis without fragmenting \citep{Inutsuka92}. 
On scales smaller than the filament radius, the local stability criterion reduces to the classical Jeans criterion, even in the presence of rotation \citep{Freundlich14}. This implies that such local collapse is only possible if the filament is larger than its Jeans length.

\smallskip
N87 studied the stability of a self-gravitating isothermal cylinder with line-mass below the critical value, pressure confined by a low density external medium. He found that the system is always unstable to long-wavelength axisymmetric perturbations even at low values of the line-mass. Similarly, \citet[][hereafter H98]{Hunter98} found that a self-gravitating cylinder which is pressure confined by an external medium, with a density discontinuity at the boundary, is always unstable to long-wavelength axisymmetric perturbations. These results are contrary to the spherical case, where a hydrostatic sphere with mass below the critical Bonner-Ebert mass \citep{Ebert55,Bonnor56} is stable against gravitational collapse. We elaborate further on these two studies in \se{sgi}.

\smallskip
In addition to GI, cylindrical streams or jets are susceptible to Kelvin-Helmholtz Instability (KHI) whenever there is a shearing motion between the stream and its surroundings. Numerous authors have studied KHI in cylinders, typically focusing on light or equidense jets meant to represent protostellar or AGN jets \citep[e.g.][]{Birkinshaw84,Payne_Cohn85,Hardee95,Bassett95,Bodo98,Bogey11}. Several authors have also addressed the effects of magnetic fields and/or radiative cooling on KHI in cylindrical jets \citep{Ferrari81,Massaglia92,Micono00,Xu00}. However, none of the aforementioned studies accounted for the self-gravity of the gas, as this is expected to be negligible for the systems being considered, namely jets from young stars or AGN. It has also been noted that tidally disrupted streams, resulting from stars tidally destroyed by black holes, may also experience KHI \citep{Bonnerot2016}. Recently, in a series of several papers, \citet{M16,P18}; and \citet{M18b} (hereafter M16, P18 and M19, respectively) presented a detailed study of KHI, without self-gravity or radiative cooling, in a dense supersonic cylinder representing the cold circumgalactic streams feeding high redshift galaxies. These can be up to 100 times denser than their surroundings. They found that KHI can be important in the evolution of such streams, leading to significant deceleration and energy dissipation, and in certain cases to total stream disruption in the CGM. We elaborate further on these studies in \se{khi}.

\smallskip
Clearly, extensive work has been done studying separately the effects of GI and of KHI in filaments and streams. While the evolution of KHI in a self-gravitating fluid has been studied in planar \citep[][hereafter H97]{Hunter97} and spherical geometry \citep[][hereafter M93]{Murray93}, we are unaware of any such work in cylindrical geometry. 
Since the evolution of KHI in cylindrical geometry is qualitatively different than in planar geometry (M19, and references therein), while GI in cylinders 
is qualitatively different than in spheres 
%manifests in a qualitatively different way than in spherical systems 
(e.g. N87; H98), it is worth explicitly studying the combined effects of KHI and self-gravity in cylindrical systems, which is the focus of this paper. 

\smallskip
This has important astrophysical implications as well, as there are several filamentary systems where both effects are likely to be important. For instance, it has been shown that the cold circumgalactic streams are likely gravitationally unstable in the inner haloes of massive galaxies at high redshift, potentially resulting in star formation and even globular cluster formation along the streams in the CGM \citep{M18a}. This may explain recent \textit{ALMA} observations of dense star-forming gas at distances of tens of $\kpc$ away from a massive galaxy at $z\sim 3.5$, which does not appear to be associated with the galaxy or any of its satellites \citep{Ginolfi17}. Additionally, filaments in GMCs in the ISM occasionally exhibit shearing flows with respect to their background \citep{HilyBlant09,Federrath16,Kruijssen19}, suggesting that KHI may be important in their evolution.

\smallskip
The rest of this paper is organised as follows.
In \se{theory}, we review the current theoretical understanding of GI and KHI in pressure-confined cylinders, and present predictions for how the two may behave in unison. In \se{sim}, we describe a suite of numerical simulations used to study GI and KHI in cylinders. In \se{res} we present the results of our numerical analysis and compare these to our analytical predictions. In \se{disc} we discuss our results and their astrophysical applications, present caveats to our analysis and outline future work. Finally, we summarise our main conclusions in \se{conc}.
%%%%%%%%%%%%%%%%%%%%%%%%%%%%%%%%%%%%%%%%%%%%%%%%%%%%%%%%%%%%%%

\section{Theory of instabilities} 
\label{sec:theory}
In this section we briefly review the existing theory of GI (\se{sgi}) and KHI (\se{khi}) in pressure confined cylinders. We then make new predictions for how the two effects may be combined in cylindrical systems (\se{combined}, to be tested using numerical simulations in \se{res}), and compare these to previous results of a combined analysis in spherical systems (\se{sphere}). 
%!!!!!!!!!!!!!!!!!!!!!!!!!!!!!!!!!!!!!!!!!!

\subsection{Gravitational instability}
\label{sec:sgi}
We focus here on the results of N87 and H98, as these are the most relevant for our current analysis. These studies both focus on the stability of a self-gravitating cylinder with finite radius and line-mass below the critical value for hydrostatic equilibrium, pressure confined by a uniform external medium.

\smallskip
N87 consider an isothermal cylinder initially in hydrostatic equilibrium, with the density profile 
\be 
\label{eq:isothermal_density}
\rho(r)=\rho_{\rm c}\left[1+\frac{1}{8}\left(\frac{r}{H}\right)^2\right]^{-2}, \quad H=\frac{c_s}{\sqrt{4\pi G \rho_{\rm c}}},
\ee
{\no}\citep{Ostriker64}. $\rho_{\rm c}$ is the central density of the cylinder, $H$ is its scale height, $\cs$ is the isothermal sound speed, and $G$ is the gravitation constant. The line-mass of such a cylinder out to radius $\Rs$ is 
\be 
\label{eq:line_mass_def}
\Lambda=\int_0^{\Rs} 2\pi r\rho (r)~{\rm dr}.
\ee 
{\no}For $\Rs=\infty$, this yields the critical line-mass for hydrostatic equilibrium \citep{Ostriker64}, 
\be 
\label{eq:isothermal_line_mass}
\Lambda_{\rm cr,\,iso}=2\cs^2/G.
\ee 
{\no}An equilibrium initial condition is only possible for $\Lambda\le \Lambda_{\rm cr,iso}$. For a cylinder truncated at a finite radius $\Rs$, the density and line mass profiles at $r<\Rs$ are still given by \equs{isothermal_density} and \equm{line_mass_def}. Thus, the ratio of the cylinder's line-mass to the critical line-mass is related to the ratio of the cylinder's radius to its scale height,
\be 
\label{eq:iso_line_mass_radius}
\frac{\Lambda}{\Lambda_{\rm cr,\,iso}}=\left[1+8\left(\frac{H}{\Rs}\right)^2\right]^{-1}.
\ee 
{\no}Increasing the central density, $\rhoc$, or decreasing the temperature and thus the sound speed, $\cs$, reduces the scale height, $H$. For a fixed stream radius, $\Rs$, this results in an increase of the ratio $\Lambda/\Lambda_{\rm cr,\,iso}$. 

\smallskip
In terms of the external pressure confining the truncated cylinder, pressure equilibrium at the boundary dictates that 
\be 
P_{\rm ext}=P(\Rs)=\cs^2\rho(\Rs)=\cs^2\rhoc \left[1+\frac{\Rs^2}{8H^2}\right]^{-2}.
\ee 
{\no}Inserting this into \equ{iso_line_mass_radius} yields 
\be 
\label{eq:line_mass_pressure}
\frac{\Lambda}{\Lambda_{\rm cr,\,iso}}=1-\frac{P_{\rm ext}}{\rhoc \cs^2}.
\ee
{\no}This shows that for a given temperature and external pressure, a cylinder can have any line mass from $0$ to $\Lambda_{\rm cr,\,iso}$, by decreasing the central density from $\rhoc=P_{\rm ext}/\cs^2$ to $0$. The critical line-mass therefore does not depend on the external pressure. This is fundamentally different from the spherical case where the maximal mass for which a hydrostatic equilibrium solution exists depends on the external pressure. This is the Bonnor-Ebert mass, 
\be 
\label{eq:BE_mass}
M_{\rm BE}=1.18\frac{\cs^4}{P_{\rm ext}^{1/2}G^{3/2}}
\ee
\citep{Ebert55,Bonnor56}.
{\no}For further comparison of the structure and properties of self-gravitating cylinders and spheres confined by external pressure, see \citet{Fischera12}. For the remainder of our analysis we will use the scale-height, $H$, and the stream radius, $\Rs$, rather than the external pressure.

\smallskip
N87 analyzed perturbations about hydrostatic equilibrium in a cylinder with radius $\Rs$, pressure confined by an external medium with constant pressure and effectively zero density, $\rho_{\rm ext}<<\rho(\Rs)$. The dispersion relation was numerically evaluated for several values of $\Lambda/\Lambda_{\rm cr,\,iso}$. All cases were found to be stable to non-axisymmetric modes. For axisymmetric modes, the system was found to be unstable at long wavelengths, with longitudinal wavenumber $k<k_{\rm cr}$. The system attains a maximal growth rate, $\omega_{\rm max}$, at a finite wavenumber, $k_{\rm max}$, hereafter the fastest growing mode, and then stabilises again at infinite wavelengths, $\omega\rightarrow 0$ as $k\rightarrow 0$. This is unlike the spherical Jeans instability where the growth rate diverges as $k\rightarrow 0$. There is no closed analytic expression for $k_{\rm cr}$, $k_{\rm max}$ or $\omega_{\rm max}$ for the general case, but it is useful to consider two limiting cases.

\smallskip
In the limit $\Lambda\rightarrow\Lambda_{\rm cr,\,iso}$, equivalent to $\Rs>>H$ (\equnp{iso_line_mass_radius}), the solution converges to that of an infinite cylinder. In this case, one obtains $k_{cr} \simeq 0.56 H^{-1}$, $k_{\rm max}\simeq  0.28 H^{-1}$, and $\omega_{\rm max}\simeq 0.60\left(4 G\rho_{\rm c}\right)^{1/2}$. For comparison, the free-fall time of a cylinder with average density $<\rho>=\Lambda/(\pi\Rs^2)$ is
\be 
\label{eq:tff}
t_{\rm ff}=(4G<\rho>)^{-1/2}.
\ee
{\no}For an isothermal cylinder with radius $\Rs$, 
\be 
\label{eq:rho_bar}
<\rho>=\rho_{\rm c}\left(1-\frac{\Lambda}{\Lambda_{\rm cr,\,iso}}\right)=\rho_{\rm c}\left(1+\frac{\Rs^2}{8H^2}\right)^{-1}.
\ee
{\no}For $\Lambda = 0.90\Lambda_{\rm cr,\,iso}$, we thus have $\Rs\simeq 8.5H$ and $\omega_{\rm max}/t_{\rm ff}^{-1}\simeq 1.9$. For larger values of $\Lambda$ the ratio $\omega_{\rm max}/t_{\rm ff}^{-1}$ increases. 

\smallskip
In the opposite limit, when $\Lambda<<\Lambda_{\rm cr,\,iso}$ or $\Rs<<H$, the density is roughly constant within $\Rs$ and the solution converges to that of an incompressible cylinder, first studied by \citet{Chandrasekhar53}. The dispersion relation for an incompressible cylinder is given by\footnote{Note that there is a minus sign missing from the corresponding equation (4.10) in N87.}
\be 
\label{eq:nagasawa}
\frac{\omega^2}{4\pi G \rho} = -\frac{xI_1}{I_0}\left[K_0I_0-\frac{1}{2}\right] ,
\end{equation}
{\no}where $I_{\nu}(x)$ and $K_{\nu}(x)$ are modified Bessel functions of the first and second kind of order $\nu$, evaluated at the argument $x=k\Rs$. This yields $k_{cr} \simeq 1.1 \Rs^{-1}$, $k_{max} \simeq 0.6 \Rs^{-1}$, and $\omega_{\rm max}\simeq 0.4t_{\rm ff}^{-1}$. 

\smallskip
To summarise, the shortest unstable wavelength is $\lambda_{\rm cr}=2\pi/k_{\rm cr}\sim 4\pi H$ and $2\pi\Rs$ in the limits $\Lambda \rightarrow \Lambda_{\rm cr,\,iso}$ and $\Lambda<<\Lambda_{\rm cr,\,iso}$ respectively. In all cases, the most unstable mode occurs at $\lambda_{\rm max}\sim 2\lambda_{\rm cr}$, while $\omega_{\rm max}/t_{\rm ff}^{-1}$ is within a factor $\sim 2$ of unity. 
Note that since in the latter limit $\Rs<<H$, we arrive at the somewhat counterintuitive result that for smaller values of the line-mass the shortest and most unstable wavelengths are much shorter. As noted by N87, the instability manifests itself in different ways in these two limits. For large values of the line-mass the system is unstable to body-modes which are maximal near the stream axis and are similar to the classic Jeans instability. On the other hand, for small values of the line-mass the instability is dominated by surface modes, which are maximal near the stream interface and lead to its deformation. In the non-linear regime, these two modes of instability lead to different shapes and orientations of collapsed clumps within the stream \citep{Heigl18}.

\smallskip
H98 generalised this analysis by allowing for a finite background density, $\rhob$, confining the stream. However, they assumed a constant stream density, $\rhos$, rather than an isothermal profile. Their scenario is thus analogous to the limit $\Lambda<<\Lambda_{\rm cr,\,iso}$ from N87. H98 derive the following dispersion relation\footnote{This is equivalent to equation (68) from H98 using the identity $I_0(x)K_1(x) + I_1(x)K_0(x) = 1/x$.} 
\be 
\label{eq:hunter}
\begin{array}{c}
\dfrac{\omega^2}{4\pi G \bar{\rho}}= -\left[\dfrac{x(\rhos-\rhob)^2 I_0K_0}{\bar{\rho}^2}-\dfrac{x\rhos(\rhos-\rhob)}{2\bar{\rho}^2}\right]\\
\times\left[\dfrac{\rhos I_0}{\bar{\rho}I_1}+\dfrac{\rhob K_0}{\bar{\rho}K_1}\right]^{-1},
\end{array}
\ee 
{\no}where $\bar{\rho}=0.5(\rhos+\rhob)$, and $I_{\nu}(x)$ and $K_{\nu}(x)$ are again modified Bessel functions with $x=kR_s$. This converges to \equ{nagasawa} in the limit $\rhob\rightarrow 0$. 

\smallskip
From \equ{hunter}, the condition for instability is 
\be 
\label{eq:hunter_unstable}
I_0K_0>\frac{1}{2\left(1-\delta^{-1}\right)},
\ee
{\no}where $\delta=\rhos/\rhob$ is the density contrast between the stream and the background. If $\delta<1$, such that the background is denser than the stream, the system is unstable at all wavelengths due to Rayleigh-Taylor instability (RTI). If $\delta>1$, such that the stream is denser than the background, the system is unstable at long wavelengths, i.e. small values of the argument of the Bessel functions on the left-hand side of \equ{hunter_unstable}, $x=k\Rs$. Furthermore, H98 find that the instability always manifests itself as a surface mode, leading to the deformation of the stream-background interface, similar to the conclusion of N87 for the low line-mass case. For $\delta\rightarrow\infty$, corresponding to $\rhob\rightarrow 0$, the system is unstable for $k<k_{\rm cr}\simeq 1.07\Rs^{-1}$, as for \equ{nagasawa}. For $\delta=1$ such that there is no density discontinuity at the interface, $k_{\rm cr}=0$ and the system is stable for all finite wavelengths. This highlights the fact that this is an interface instability, caused by a density discontinuity between the stream and the background. For $\delta=4,~10,~100$, we have $k_{\rm cr}\Rs\simeq 0.79,~0.96,~1.06$. The maximal growth rate for these cases is $\omega_{\rm max}/t_{\rm ff}^{-1}\simeq 0.26,~0.36,~0.43$. 

\smallskip
H98 also note that in the case of a dense sphere pressure confined by a lower density background, the analogous surface mode is always stable, in agreement with the known fact that spheres less massive than the Bonner-Ebert mass are stable. However, in planar geometry, such as a dense slab pressure confined by a lower density background, a similar surface instability exists above a critical wavelength (H97).

\smallskip
The GI surface modes can be thought of as RTI analogues, induced by the self-gravity of the fluid rather than by an external gravitational field. An intuitive explanation was offered by H97 for the planar case, and can be adapted to cylindrical geometry as follows. Consider a dense cylinder with constant density $\rhos$  
pressure confined by a background medium with constant density $\rhob<\rhos$. Such a system is stable to classical RTI. Now consider an axisymmetric perturbation to the interface of the cylinder with longitudinal wavelength $\lambda$. In some region, say $0<z<\lambda/2$, there is an outward distortion of the interface, $\xi(z)$, which results in a mass excess just outside the original interface, proportional to $(\rhos-\rhob)\xi(z)$. Through Poisson's equation, this leads to a more negative gravitational potential in this region, resulting in a perturbation $\Phi_1<0$ to the initial potential. As the fluid is incompressible and at rest, Bernoulli's equation tells us that $P+\rho\Phi={\rm const}$ along any streamline in either fluid, where $P$ is the pressure. 
%and $\Phi$ the gravitational potential. 
In the incompressible limit, where $\rho={\rm const}$ in each fluid, this implies that $P_{\rm 1,s}=-\rhos\Phi_{\rm 1}$ and $P_{\rm 1,b}=-\rhob\Phi_{\rm 1}$, where $P_{\rm 1,s}$ and $P_{\rm 1,b}$ are the perturbations to the pressure in the stream and the background respectively, on either side of the interface. Since $\rhos>\rhob$ and $\Phi_1<0$, we have that $P_{\rm 1,s}>P_{\rm 1,b}$, so the pressure in the stream just inside the interface is larger than the pressure in the background just outside the interface, causing the perturbation to continue growing. 

\smallskip
This~instability~only~manifests~at~long wavelengths, when the mass excess leading to the perturbation of the potential is large enough to overcome the stabilizing effect of RT modes induced by the unperturbed potential. As noted above, the shortest unstable wavelength for cylinders is $\sim 2\pi\Rs\sim 6.3\Rs$ (\equnp{hunter_unstable}). By contrast, the longest available wavelength on the surface of a sphere corresponds to the $l=2$ spherical harmonic, since the $l=0$ mode represents global expansion or contraction of the sphere while the $l=1$ mode represents a rigid displacement. The wavenumber associated with the $l=2$ mode is $k=[l(l+1)]^{1/2}/\Rs\sim 2.5/\Rs$, corresponding to a wavelength of $\lambda\sim 2.6\Rs$. This is too short for GI surface modes to overcome RT stabilization, which is why there are no GI surface modes for spherical systems (H98).

%!!!!!!!!!!!!!!!!!!!!!!!!!!!!!!!!!!!!!!!!!!!
\subsection{KH Instability}
\label{sec:khi}
KHI arises from shearing motion between the interfaces of two fluids, leading to efficient mixing and smoothing out the initial contact discontinuity. We focus here on the recent results of M19, who analysed the non-linear evolution of KHI in a dense 3d cylinder streaming through a static background, expanding on earlier work by M16 and P18. 
%\citet{M16} and \citet{P18}.
The system is characterised by two dimensionless parameters, the Mach number of the stream velocity with respect to the background sound speed, $\Mb=\Vs/\cb$, and the density contrast of the stream and the background, $\delta=\rhos/\rhob$. M19 analytically derived timescales for the non-linear mixing of the two fluids and eventual disruption of the stream, as well as for stream deceleration and the loss of bulk kinetic energy, as a function of these two parameters. 

\smallskip
We begin by noting that, similar to the dichotomy between surface modes and body modes in GI (N87), there are two modes of KHI. The nature of the instability depends primarily on the ratio of the stream velocity to the sum of the two sound speeds,
\be 
\label{eq:Mtot}
M_{\rm tot}=\frac{\Vs}{\cs+\cb}.
\ee

\smallskip
If $M_{\rm tot}<1$, the instability is dominated by surface modes. These are concentrated at the interface between the fluids, and lead to the growth of a shear layer which expands into both fluids. Within the expanding shear layer a highly turbulent medium develops, efficiently mixing the two fluids. Surface modes can have any longitudinal wavenumber\footnote{So long as the wavelength, $\lambda=2\pi/k$, is larger than the width of the transition region between the two fluids.}, $k$, and any azimuthal wavenumber, $m$, representing the number of azimuthal nodes along the stream-background interface. $m=0$ corresponds to axisymmetric perturbations, $m=1$ to helical perturbations, and $m\ge 2$ to more complicated fluting modes. Low order $m$ modes with wavelengths of order $\Rs$ dominate the early non-linear evolution of the instability, as their eddies reach the largest amplitudes before they break, but the shear layer between the fluids quickly develops into a highly turbulent mixing zone with no discernible symmetry.

\smallskip
The shear layer separating the fluids expands self-similarly through vortex mergers. Independent of the initial perturbation spectrum, the width of the shear layer, $h$, evolves as 
\be 
\label{eq:shear_growth}
h=\alpha \Vs t
\ee
{\no}where $\alpha$ is a dimensionless growth rate that depends primarily on $M_{\rm tot}$, and is typically in the range $\alpha\sim 0.05-0.25$ (P18; M19). 

\smallskip
The shear layer penetrates asymmetrically into the stream and background due to their different densities. The penetration depth of the shear layer in either medium can be derived from conservation of mass and momentum in the shear layer, and are given by (P18; M19): %\citep[][M19]{P18}
\be 
\label{eq:hs_growth}
\hs = \frac{\alpha \Vs t}{1+\sqrt{\delta}},\quad \hb = \frac{\sqrt{\delta}\alpha \Vs t}{1+\sqrt{\delta}}.
\ee
{\no}Stream disruption occurs when the shear layer encompasses the entire stream, namely when $\hs=\Rs$. This occurs at time
\be 
\label{eq:tau_diss}
t_{\rm dis} = \frac{\left(1+\sqrt{\delta}\right)\Rs}{\alpha\Vs}.
\ee
{\no}The contact discontinuity effectively disappears before the stream is completely disrupted, once the full width of the shear layer is of order the stream radius, namely $h=\Rs$. This occurs at time 
\be 
\label{eq:tau_shear}
t_{\rm shear} = \frac{\Rs}{\alpha\Vs}.
\ee

\smallskip
As the shear layer expands into the background, it entrains background mass. This causes the stream to decelerate as its initial momentum is distributed over more mass. As shown by M19, the stream velocity as a function of time is well fit by 
\be 
\label{eq:stream_deceleration}
V_{\rm s}(t) = \frac{V_{\rm s,0}}{1+t/t_{\rm dec}},
\ee
{\no}where $V_{\rm s,0}$ is the initial velocity of the stream, and
\be 
\label{eq:tau_dec}
t_{\rm dec} = \dfrac{\left(1+\sqrt{\delta}\right)\left(\sqrt{1+\delta}-1\right)}{\alpha\sqrt{\delta}}\frac{\Rs}{V_{\rm s,0}},
\ee
{\no}is the time when the background mass entrained in the shear layer equals the initial stream mass, such that momentum conservation implies the velocity is half its initial value.

\smallskip
An empirical expression for the dimensionless shear layer growth rate, $\alpha$, was proposed by \citet{Dimotakis}, 
\be 
\label{eq:alpha_fit}
\alpha \simeq 0.21\times \left[0.8{\rm exp}\left(-3 M_{\rm tot}^2\right)+0.2\right].
\ee
{\no}M19 found \equ{alpha_fit} to be a good fit to shear layer growth in simulations of 2d slabs, regardless of whether one measures $h$, $\hs$, or $\hb$. However, they found that $\hs$ expanded more rapidly in 3d cylinders due to an enhanced eddy interaction rate near the stream axis. This yielded $\alpha$ values $\sim 50\%$ larger than \equ{alpha_fit} when measuring $\hs$ and using \equ{hs_growth}. On the other hand, $\hb$ was found to expand at a similar rate in 2d and 3d so long as $\hb\lsim 2\Rs$. 
Since the shear layer width is dominated  by $\hb$ for $\delta>1$, we use \equ{alpha_fit} together with \equ{tau_shear} to evaluate the time when the contact discontinuity is destroyed. 

\smallskip
Once $\hb\gsim 2\Rs$, its growth rate is reduced by roughly half, due to a turbulent cascade to small scales which removes energy from the largest eddies driving the expansion. For $\delta>8$, this occurs before the stream reaches half its initial velocity (\equsnp{stream_deceleration}-\equmnp{alpha_fit}). M19 found that in these cases, a good fit to the velocity evolution of streams can be obtained simply by using $0.5\alpha$ in \equ{tau_dec} with $\alpha$ taken from \equ{alpha_fit}.

\smallskip
When $M_{\rm tot}>1$, surface modes of low azimuthal order (low values of $m$) stabilise\footnote{The formal condition for stabilization of $m=0,\,1$ surface modes is $\Mb>(1+\delta^{-1/3})^{3/2}$, similar to $M_{\rm tot}>1$.}. The nature of the instability then depends on the width of the initial transition region between the fluids (which is likely set by transport processes such as viscosity and thermal conduction).
If this is relatively narrow, the instability becomes dominated by high-$m$ surface modes, and the above description, summarised in \equs{shear_growth}-\equm{tau_dec}, remains valid, with $\alpha\sim 0.05$ according to \equ{alpha_fit}. However, if the initial transition region is wide, of order $\gsim 0.25\Rs$ or larger, high-$m$ surface modes are also stable and the instability becomes dominated by body modes. These do not result in shear layer growth but rather in the global deformation of the stream into a helical, $m=1$, shape with a characteristic wavelength of $\sim 10\Rs$ and an amplitude of $\gsim \Rs$. The timescale for this to occur depends on the initial perturbation amplitude and spectrum, though it is almost always longer than the timescale for stream disruption by surface modes when these are unstable. Following the formation of the sinusoid, small scale turbulence develops near its peaks and leads to stream disruption within roughly one stream sound crossing time. Interestingly M19 find that \equs{stream_deceleration}-\equm{tau_dec} are a good description of stream deceleration due to body modes as well, despite the different processes involved. 

\smallskip
We will hereafter ignore KHI body modes, and assume that KHI is dominated by surface modes of some order $m$ for all Mach numbers. If KHI surface modes are suppressed by a large initial transition region, then GI surface modes will also likely be suppressed, based on the analysis of H97 and H98.

%%%%%%%%%%%%%%%%%%%%%%%%%%%%%%%%%%%%%%%%%%%%%%%%%%%%%%%%%%%%%%

\subsection{Combined treatment}
\label{sec:combined}
We now wish to combine the above two processes, and discuss the evolution of a pressure-confined self-gravitating cylinder undergoing KHI. In addition to $\Mb$ and $\delta$, a third parameter is required to describe such a system, namely the line-mass of the cylinder in units of the critical line-mass for hydrostatic equilibrium, $\mu\equiv\Lambda/\Lambda_{\rm cr}$. 
We begin by making the assumption, to be justified below, that any coupling between GI and KHI in the linear regime is relatively small, such that the region of parameter space where each process results in instability is unchanged, and the linear growth rates are only mildly altered. 
Under this assumption, it is clear from \se{sgi} and \se{khi} that for all values of $(\Mb,\delta,\mu)$, the system is unstable over some wavelength range. We assume that the initial perturbation spectrum spans this range. 

\smallskip
GI enhances density contrasts and leads to the formation of long-lived collapsed clumps, while KHI smooths the interface between the fluids and dilutes the mean density of the stream. The question is which process will win. The timescale for GI is the inverse growth rate of the fastest growing mode discussed in \se{sgi}, $t_{\rm max}\equiv \omega_{\rm max}^{-1}$. At low values of $\mu$, GI is dominated by surface modes (N87), which require the presence of a contact discontinuity (H98). Thus, the timescale for KHI to prevent gravitational collapse is $t_{\rm shear}$ (\equnp{tau_shear}), the timescale for nonlinear KHI to destroy the contact discontinuity. On the other hand, for high values of $\mu$, GI is dominated by body modes which are unrelated to the contact discontinuity (N87). In this case, the relevant timescale for KHI to prevent collapse is $t_{\rm dis}$ (\equnp{tau_diss}), the timescale for nonlinear KHI to disrupt the stream itself. 

\smallskip
Since $t_{\rm shear}<t_{\rm dis}$ for all $\delta>1$, we distinguish between three regimes. If $t_{\rm max}<t_{\rm shear}<t_{\rm dis}$, we expect GI to win and the stream to fragment into long-lived clumps. If $t_{\rm shear}<t_{\rm dis}<t_{\rm max}$, we expect KHI to win and disrupt the stream by mixing it into the background. We hereafter refer to this process as ``shredding the stream". In the intermediate case where $t_{\rm shear}<t_{\rm max}<t_{\rm dis}$, the outcome may depend on the value of $\mu$. If $\mu$ is small, such that GI is dominated by surface modes, then we expect KHI to win and shred the stream since $t_{\rm shear}<t_{\rm max}$. On the other hand, if $\mu$ is large such that GI is dominated by body modes, GI may still win and lead to stream fragmentation and the formation of bound clumps, since $t_{\rm max}<t_{\rm dis}$. However, this is uncertain, since the shear layer will penetrate somewhat into the stream within $t_{\rm max}$, reducing the effective line-mass of the unperturbed (non-turbulent) region. If this is reduced below the threshold for GI body modes to be effective, KHI may still win and suppress clump formation.

\smallskip
Since $t_{\rm max}\propto \rho_{\rm c}^{-1/2}\propto \mu^{-1/2}$, as $\mu$ is increased at fixed $(\Mb,\delta)$, $t_{\rm max}$ decreases while $t_{\rm shear}$ (and $t_{\rm dis}$) remain constant. Thus, for each $(\Mb,\delta)$ there exists a critical value of $\mu$, $\mu_{\rm cr}\equiv \mu_{\rm cr}(\Mb,\delta)$, such that $t_{\rm max}<t_{\rm shear}$ for $\mu>\mu_{\rm cr}$ (see \fig{criticalmu} in \se{khivg} below).  Therefore, GI will win and lead to stream fragmentation and clump formation whenever $\mu>\mu_{\rm cr}$. If $\mu_{\rm cr}$ is small enough to be in the regime where GI is dominated by surface modes, then KHI will win and shred the stream for $\mu<\mu_{\rm cr}$. On the other hand, if $\mu_{\rm cr}$ is in the regime where GI is dominated by body modes, the fate of the stream at $\mu<\mu_{\rm cr}$ depends on the ratio of $t_{\rm max}$ to $t_{\rm dis}$.

\smallskip
At first glance, it may seem inconsistent to compare a linear timescale for GI, $t_{\rm max}=\omega_{\rm max}^{-1}$, to a nonlinear timescale for KHI, $t_{\rm shear}$ or $t_{\rm dis}$. While $t_{\rm max}$ is formally the timescale for the growth of linear perturbations, once density perturbations grow the free-fall times become ever shorter and the collapse accelerates. Full collapse is thus dominated by the linear growth time. On the other hand, KHI tends to saturate following the linear phase, because it is driven by the presence of a contact discontinuity which is destroyed by the instability. Continued growth in the nonlinear regime is dominated by the merger of eddies within the shear layer on timescales of $t_{\rm shear}$ and $t_{\rm dis}$, as described in \se{khi}. %(see discussion in P18 and M19). 

\smallskip
The above discussion notwithstanding, one may ask whether density fluctuations within the stream induced by KHI can trigger local gravitational collapse when $\mu<\mucr$. Note that this is different than the global fragmentation of the stream induced by GI. Such local collapse can occur in filaments on scales larger than the spherical Jeans length, $\lambda_{\rm J}=[\pi\cs^2/(G\rho)]^{1/2}$, but smaller than the stream radius, $\Rs$ \citep{Freundlich14}. This implies that this is only possible if $\lambda_{\rm J}<\Rs$. A lower limit to the Jeans length is obtained by inserting $\rho=\rhoc$, the density along the stream axis. This yields $\lambda_{\rm J}=2\pi H$, with $H$ given by \equ{isothermal_density}. The condition that $\lambda_{\rm J}<\Rs$ thus implies that $\Rs>>H$, so GI is dominated by body modes 
(N87). We conclude that KHI induced density fluctuations can only trigger local gravitational collapse if $\mu<\mucr$ but GI is still dominated by body modes.

\smallskip
We must now justify our initial ansatz that the linear coupling between GI and KHI does not fundamentally alter the instability region of parameter space. We rely here on the analysis of H97, who derived the dispersion relation of a self-gravitating system undergoing KHI in the vortex sheet limit, i.e. two semi-infinite fluids separated by a single, planar interface. In their derivation they made the simplifying assumption that the gravitational field in the unperturbed system was weak compared to the perturbed forces induced by both pressure and potential perturbations. This is equivalent to assuming that the wavelengths are much shorter than the gravitational scale-height of the unperturbed system, which itself is equivalent to assuming constant density and pressure in both fluids. The resulting dispersion relation contains terms associated with KHI, RTI, and surface mode GI.
%, and can be cast into a 24th order polynomial with complex coefficients. 
We refer the reader to H97 for the expression and its derivation. %The main take-away
Relevant to our discussion is the fact that the coupling between self-gravity and shearing motions does not modify the stability region of the system, only mildly affects the linear growth rates of KH modes at short to intermediate wavelengths, and does not suppress GI surface modes at long wavelengths. Deriving an analogous dispersion relation for cylinders is beyond the scope of this paper. Rather, we assume that the same conclusions hold for cylindrical systems, in particular because KHI in cylinders is even more unstable than for planar vortex sheets (M16; M19). %\citep[][M19]{M16}. 
The validity of this assumption and our subsequent analysis will be tested with numerical simulations in \se{res}.

%%%%%%%%%%%%%%%%%%%%%%%%%%%%%%%%%%%%%%%%%%%%%%%%%%%%%%%%%%%%%%

\subsection{Comparison to the Spherical Case}
\label{sec:sphere}
%\subsubsection{Comparison to Previous Work}
It is worth comparing our analysis to that of M93, who addressed the question of when self-gravity would prevent KHI from disrupting a cold, dense spherical cloud moving through a hot, dilute background. They assumed that the cloud was pressure confined by the background fluid, and that its mass was less than the Bonnor-Ebert mass, making it gravitationally stable and in hydrostatic equilibrium. In this case, unlike for self-gravitating cylinders, there is no GI, and the only effect of the self-gravity is to induce RT modes at the cloud surface. Since the cloud is denser than the background, these RT modes can counteract the KHI and stabilise the system, due to the restoring buoyancy force. They showed this by considering the combined dispersion relation of KHI and RTI in the incompressible limit, 
\be 
\label{eq:Murray}
\omega^2=-\frac{\rhos \rhob}{(\rhos+\rhob)^2}V^2 k^2 + \frac{\rhos-\rhob}{\rhos+\rhob} kg,
\ee
{\no}where $V$ is the velocity of the cloud in the static background and $g$ is the gravitational acceleration at its surface. This implies that KHI is stable for all wavelengths greater than 
\be 
\label{eq:Murray2}
\lambda_{\rm max}=\frac{2\pi \rhos\rhob V^2}{\left(\rhos^2-\rhob^2\right)g}.
\ee

\smallskip
M93 then assumed that KHI would only disrupt the cloud if $\lambda_{\rm max}>R_{\rm cl}$, the cloud radius. This was based on the assumption that KHI surface modes saturate at an amplitude comparable to their wavelength, thus neglecting the subsequent shear layer growth. This assumption together with $g=GM_{\rm cl}/R_{\rm cl}^2$ and $M_{\rm cl}=(4\pi/3)\rho_{\rm cl} R_{\rm cl}^3$ results in a minimum mass for self-gravity to stabilise the sphere against KHI. For velocities of order the background sound speed, the critical mass is of order the Bonnor-Ebert mass, $M_{\rm BE}$. Such a system is thus always unstable, either to KHI at $M_{\rm cl}<M_{\rm BE}$ or to global gravitational collapse at $M_{\rm cl}>M_{\rm BE}$. 

\smallskip
Our main prediction for the cylindrical case is qualitatively similar. We predict that a self-gravitating stream will always be unstable either to KHI at $\mu<\mu_{\rm cr}$ or to GI at $\mu>\mu_{\rm cr}$, depending on whether the timescale for GI, $t_{\rm max}$, is longer or shorter than the timescale for KHI to destroy the contact discontinuity, $t_{\rm shear}$, and/or the stream itself, $t_{\rm dis}$. However, unlike M93, we do not rely on a similar criterion of gravity stabilizing wavelengths longer than $\Rs$. First of all, unlike in spherical systems, self-gravity actually destabilises cylinders at long wavelengths (N87; H98; \se{sgi}). Furthermore, even if KHI is stable for wavelengths longer than $\Rs$ in the linear regime, it can still lead to stream disruption in the nonlinear regime by shear layer growth caused by initially shorter wavelength perturbations. 

%%%%%%%%%%%%%%%%%%%%%%%%%%%%%%%%%%%%%%%%%%%%%%%%%%%%%%%%%%%%%%
\section{Numerical Methods}
\label{sec:sim}
\smallskip
In this section we describe the details of our simulation code and setup, as well as our analysis method. We use the Eulerian AMR code \texttt{RAMSES} \citep{Teyssier02}, with a piecewise-linear 
reconstruction using the MonCen slope limiter \citep{vanLeer77}, an HLLC approximate Riemann solver \citep{Toro94}, and a multi-grid Poisson solver.

\subsection{Hydrostatic Cylinders}
\label{sec:prof}

\smallskip
Unlike the isothermal cylinder described in \se{sgi}, there is no closed analytic expression for the density profile of an isentropic 
cylinder in hydrostatic equilibrium, so this must be evaluated numerically. We briefly review here how this is done, beginning with the equilibrium solution of an isolated cylinder 
%with a polytropic EoS, 
following \citet{Ostriker64}. The equation of hydrostatic equilibrium, 
\be
\label{eq:hydrostatic_eq}
\vec{\nabla} P = -\rho\vec{\nabla} \Phi,
\ee
{\no}is solved together with Poisson's equation 
\be
\label{eq:poisson}
\nabla^2 \Phi = 4\pi G \rho,
\ee
{\no}and an isentropic
equation of state (EoS),
\be 
\label{eq:polytrope}
P = K \rho^\gamma,
\ee 
{\no}where we assumed $K$ to be constant and the adiabatic index of ideal monoatomic gas, $\gamma=5/3$, throughout. These equations can be combined to yield 
\be
\label{eq:gen_stream}
\frac{1}{r}\frac{\partial}{\partial r}\left[\frac{r}{\rho} \frac{\partial \left(K\rho^\gamma\right)}{\partial r}\right] = -4\pi G \rho,
\ee 
{\no}with the boundary conditions
\be
\label{eq:bc}
\rho(r=0) = \rhoc, \quad \frac{\partial \rho}{\partial r}\bigg|_{r=0}=0.
\ee

\smallskip
\Equs{gen_stream}-\equm{bc} can be cast into unitless form by defining $y=\rho/\rhoc$ and $x=r/H$, with %\citep{Ostriker64} 
\be
\label{eq:H_general}
H^2 = \frac{\csz}{(\gamma-1)4\pi G\rhoc},
\ee
{\no}the scale radius of the cylinder, where $\csz=\gamma \Pc/\rhoc=\gamma K \rhoc^{\gamma -1}$ is the sound speed along the filament axis, with $\Pc=P(r=0)$ the pressure along the filament axis. The resulting equation is 
\be
\label{eq:gen_stream_unitless}
\frac{1}{x}\frac{\partial}{\partial x}\left(x \frac{\partial y^{\gamma-1}}{\partial x}\right) = - y, \quad y(0) = 1, \quad \frac{\partial y}{\partial x}\bigg|_0=0.
\ee
{\no}Analytic solutions exist only for $\gamma=1$ (isothermal cylinder), $\gamma=2$, and $\gamma=\infty$ (incompressible cylinder) \citep{Ostriker64}. For other values of $\gamma$ \equ{gen_stream_unitless} must be solved numerically. 

\begin{figure}
\includegraphics[width=0.49\textwidth]{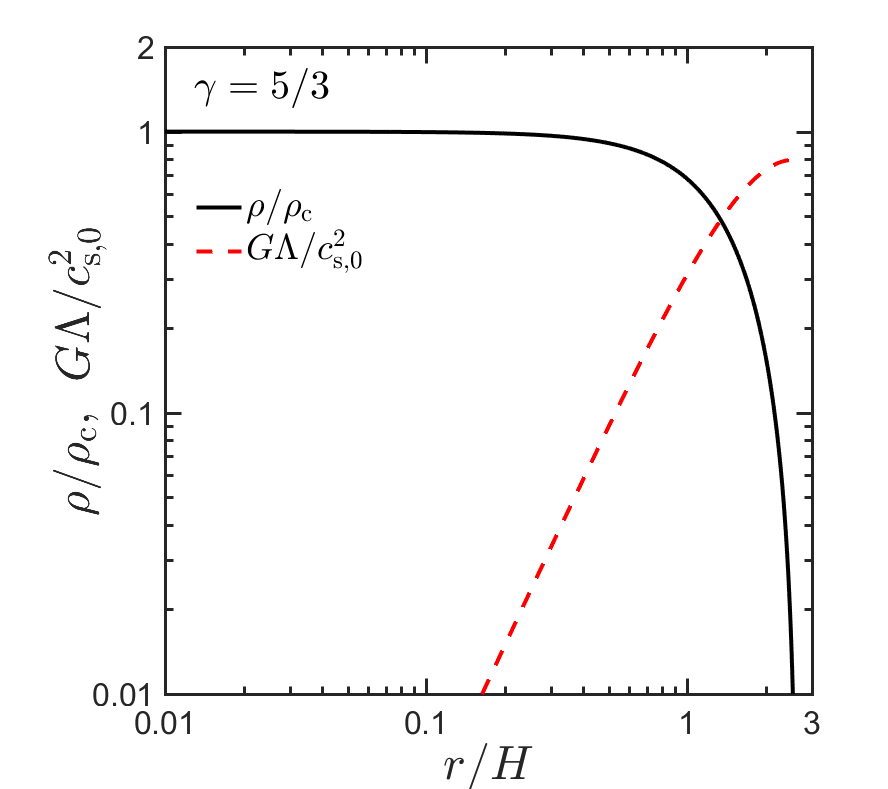}
\caption{Normalised density and line-mass profiles for a self-gravitating, isentropic 
cylinder with $\gamma=5/3$. The radial coordinate has been normalised by $H$ given in \equ{H_general}, the density (solid black line) has been normalised by its central value, and the line-mass (dashed red line) has been normalised by $c_{\rm s,0}^2/G$ following \equ{lambda_crit_gen}. The cylinder has a finite radius $R_{\rm equ}\simeq2.65H$, and a finite line-mass equal to $\Lambda_{\rm cr}\simeq 0.80c_{\rm s,0}^2/G$. The half-mass radius of the cylinder is $R_{\rm 1/2}\simeq 1.17H$.}
\label{fig:adiabatic_profile}
\end{figure}

\smallskip
While the isothermal cylinder discussed in \se{sgi} extends to $r=\infty$, all cases with $\gamma>1$ have a finite radius, $R_{\rm equ}$, defined as the radius where the density profile first reaches $\rho=0$ \citep{Ostriker64}. We can thus generalise the notion introduced in \se{sgi} of a critical line-mass above which hydrostatic equilibrium is not possible
\be
\label{eq:lambda_crit_gen}
\Lambda_{\rm cr} = \frac{\csz}{2(\gamma-1)G} \int_0^{R_{\rm equ}/H} y(x)\times x~{\rm d}x = a \frac{\csz}{G},
\ee
{\no}where $y(x)$ is the solution to \equ{gen_stream_unitless}. 
The factor $a$ on the right-hand-side of \equ{lambda_crit_gen} depends on the EoS. For $\gamma=5/3$, $R_{\rm equ}\simeq 2.648H$, the half-mass radius is $R_{\rm 1/2}\simeq 1.168H$, and $a\simeq 0.796$. For comparison, an isothermal cylinder has $R_{\rm 1/2}\simeq 2.828H$, with $H$ defined in \equ{isothermal_density}, and $a=2$ (\equnp{isothermal_line_mass}). 
In \fig{adiabatic_profile} we show the normalised equilibrium density and line-mass profiles of an isolated, isentropic, $\gamma=5/3$ cylinder. 

\begin{figure}
\includegraphics[width=0.49\textwidth]{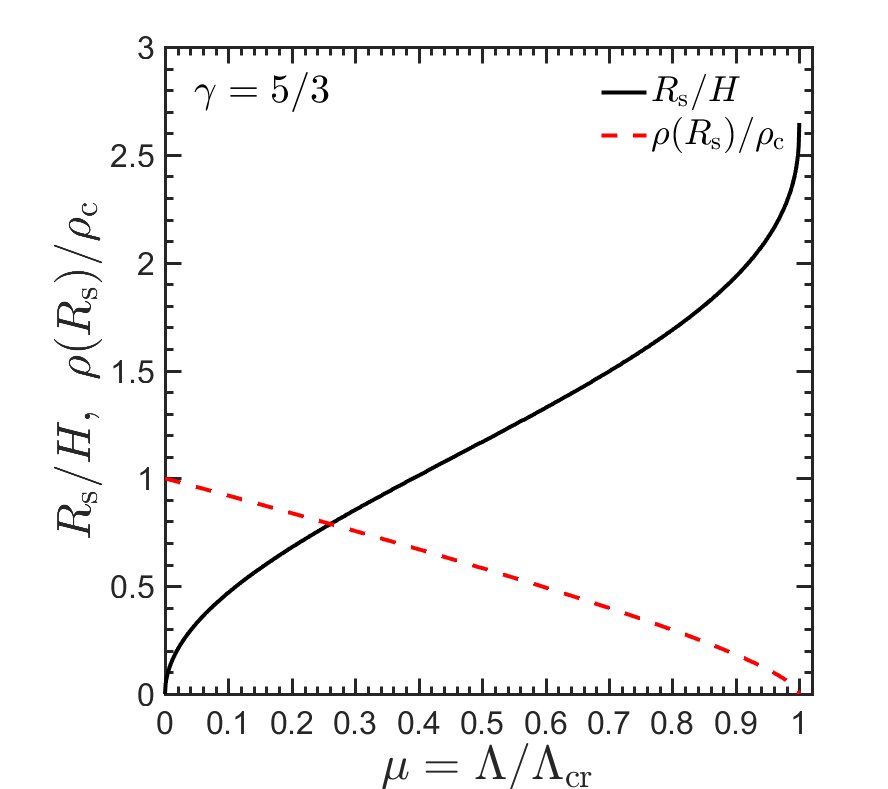}
\caption{Properties of a truncated $\gamma=5/3$ cylinder in hydrostatic equilibrium. The x-axis shows the line-mass divided by the critical line-mass, $\mu=\Lambda/\Lambda_{\rm cr}$. On the y-axis we show the stream radius, $\Rs$, divided by the scale radius, $H$ (\equnp{H_general}, black solid line), and the density at the stream radius divided by the central density, $\rho(\Rs^-)/\rho_{\rm c}$ (red dashed line).}
\label{fig:muh}
\end{figure}

\smallskip
Equilibrium profiles with $\Lambda<\Lambda_{\rm cr}$ can be constructed for cylinders pressure confined by an external medium and truncated at some radius $\Rs<R_{\rm equ}$. %We define $\mu=\Lambda/\Lambda_{\rm cr}$. 
In \fig{muh} we show the stream radius, $\Rs/H$, as a function of $\mu=\Lambda/\Lambda_{\rm cr}$. For $\mu=0,\,1$ we have $\Rs=0,\,R_{\rm equ}$ respectively. For $\mu=0.5$ we have $\Rs=R_{1/2}\simeq 1.17H$. We adopt model units where $G=\rho_{\rm c}=1$ and $\Rs=1/32$. For a given value of $\mu$, we can obtain $H$ in model units from \fig{muh} and then \equ{H_general} can be used to obtain $c_{\rm s,0}=(8\pi/3)^{1/2}H$ and $P_{\rm c}=K_{\rm s}=3c_{\rm s,0}^2/5$. Note that the stream and the background fluid have different entropy, and hence different values of $K$.

\smallskip
In addition to $\mu$, the system is defined by 
\be
\label{eq:deltac}
\deltac = \frac{\rhoc}{\rho(\Rs^+)},
\ee 
{\no}the ratio of the density along the stream axis to the background density just outside the stream. For a given $\mu$ and $\deltac$ we may evaluate the density contrast between the stream and background on either side of the interface, 
\be
\label{eq:delta}
\delta = \frac{\rho(\Rs^-)}{\rho(\Rs^+)} = \deltac \frac{\rho(\Rs^-)}{\rhoc}.
\ee 
{\no}We show the ratio $\rho(\Rs^-)/\rho_{\rm c}=\delta/\deltac$ as a function of $\mu$ in \fig{muh}. For $\mu=0.1,\,0.5,\,0.9$ we have $\delta/\deltac\simeq 0.92,\,0.58,\,0.18$ respectively. 

\smallskip
To construct equilibrium profiles for pressure confined cylinders with given values of $\mu$ and $\deltac$, we first evaluate $\Rs/H$ and $\delta$ from \fig{muh}. We then solve \equ{gen_stream_unitless} separately for $r<\Rs$ and $r>\Rs$. For $r<\Rs$, the boundary conditions are $y(0) = 1$, and ${\rm d}y/{\rm d}x|_0=0$, and the profile is unchanged from the isolated cylinder. For $r>\Rs$, the boundary conditions are given in terms of the pressure, rather than the density. Specifically, the pressure is continuous at the interface, $P(\Rs^-)=P(\Rs^+)$, while the pressure gradient is discontinuous, with 
\be
\label{eq:pressure_discontinuity}
\frac{{\rm d}P/{\rm d}R|_{\Rs^-}}{{\rm d}P/{\rm d}R|_{\Rs^+}} = \frac{\rho(\Rs^-)}{\rho(\Rs^+)} = \left(\frac{K(\Rs^+)}{K(\Rs^-)}\right)^{1/\gamma} = \delta,
\end{equation}
{\no}which follows from \equ{hydrostatic_eq}. %\hancomment{HA: The pressure continuity and density contrast also introduces entropy contrast at the boundary of the stream.} \nmr{Han, see the comment I added before equation 31. I think that is enough to address the point, so that we do not need this additional sentence. Let me know if you disagree.}
 
\begin{figure}
\includegraphics[trim={0cm 0.5cm 0cm 0cm},clip,width=0.49\textwidth]{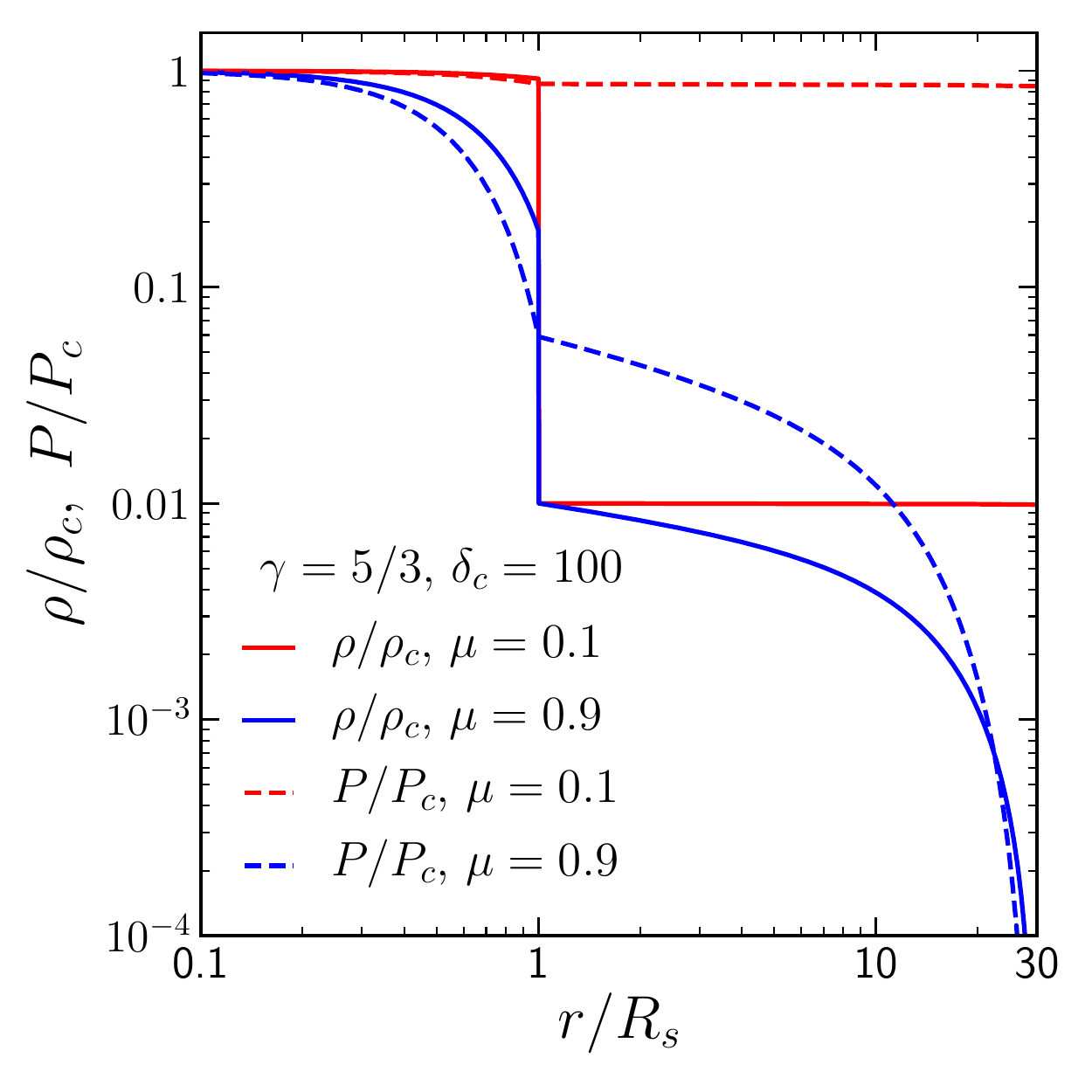}
\caption{Equilibrium density and pressure profiles of pressure confined cylinders with two different values of the stream line-mass, 
$\mu=0.9$ (in blue) and $\mu=0.1$ (in red).
The solid (dashed) lines show the density (pressure) profiles. All cases correspond to $\deltac=100$ and $\gamma=5/3$. For $\mu=0.1$, $\rho(\Rs^-)\simeq 0.92\rho_{\rm c}$ (\fig{muh}) and the density and pressure are nearly constant in both the stream and background. For $\mu=0.9$, $\rho(\Rs^-)\simeq 0.18\rho_{\rm c}$ (\fig{muh}), and there are strong density and pressure gradients within the stream. 
}
\label{fig:profiles}
\end{figure}

\fig{profiles} shows the resulting density and pressure profiles for $\mu=0.1$ and $0.9$. For $\mu=0.1$ the density and pressure are nearly constant in either medium, while for $\mu=0.9$ there are strong gradients 
%in both density and pressure 
within the stream.

%!!!!!!!!!!!!!!!!!!!!!!!!!!!!!!!!!!!!!!!!!!!
\subsection{Initial Conditions}
\label{sec:methods}

%!!!!!!!!!!!!!!!!!!!!!!!!!!!!!!!!!!!!!!!!!!!
\subsubsection{Simulation Domain \& Boundary Conditions}
\label{sec:bc}

The simulation domain is a cube of side $L=1$, extending from $-0.5$ to $0.5$ in all directions. We hereafter adopt the standard cylindrical coordinates, $(r,\varphi,z)$. The axis of our cylindrical stream is placed along the $z$ axis, at $r=0$, and we adopt a stream radius of $\Rs=1/32$. The stream fluid occupies the region $r<\Rs$ while the background fluid occupies the rest of the domain. 
The equation of state (EoS) of both fluids is that of an ideal monoatomic gas with adiabatic index $\gamma=5/3$. 

We use periodic boundary conditions at $z=\pm 0.5$, and zero force boundary conditions, often called outflow boundary conditions, at $x=\pm 0.5$ and $y=\pm 0.5$, such that gas crossing the boundary is lost from the simulation domain. At these boundaries, the gradients of density and velocity are set to 0, while the pressure gradient is taken from the hydrostatic profile computed following \se{prof}. The potential at the boundary is set to be that at the outer edge of an isolated and infinitely long cylinder with total mass $M$, equal to the total mass in the simulation domain, $\Phi(r) =2G(M/L){\rm ln} (r)$ with $r=(x^2+y^2)^{1/2}$ on the boundary. We note that this does not produce perfect equilibrium due to fitting a cylindrical profile in a cubic box. However, we find that our configuration is extremely stable in simulations with no initial perturbations and no shear flow, exhibiting $\lsim 3\%$ change in the density and pressure profiles after 4 stream free-fall times.

%------------------------------------------
\subsubsection{Smoothing the Discontinuity}

\smallskip
As noted by many previous studies of KHI, the presence of a sharp discontinuity at the interface of two fluids leads to numerical perturbations on the grid scale. These grow faster than the intended perturbations in the linear regime, and may dominate the instability at late times depending on their amplitude. Furthermore, since smaller scales grow more rapidly in the linear regime, these numerical perturbations become more severe as the resolution is increased, preventing convergence of the solution. To alleviate this issue, we smooth the velocity and density around the interfaces using the ramp function proposed by \citet{Robertson10}, also used by M16, P18, and M19.
%\citet{M16,P18}; and M19. 
Specifically, we normalise each quantity in the stream and the background by its value at $\Rs$, denoted $f_{\rm s}$ and $f_{\rm b}$ respectively. We then smooth between these values using 
\be 
\label{eq:ramp1}
f(r)=f_{\rm b} + \left(f_{\rm s}-f_{\rm b}\right)\times \mathcal{R}(r),
\ee 
\be 
\label{eq:ramp2}
\mathcal{R}(r)=\frac{1}{2}\left[1-{\rm tanh}\left(\frac{r-\Rs}{\sigma}\right)\right],
\ee
{\no}and multiply the normalised profiles in either medium by $f(r)$. The parameter $\sigma$ determines the width of the transition zone. The function $\mathcal{R}(r)$ transitions from $0.05$ to $0.95$ over a full width of $\sim 3\sigma$ in $(r-\Rs)$. We adopt $\sigma=\Rs/32$ for all of our simulations, which is sufficient to suppress artificial perturbations with small longitudinal wavelength, while still allowing azimuthal modes with $m\lsim 12$ to grow (M19).

%!!!!!!!!!!!!!!!!!!!!!!!!!!!!!!!!!!!!!!!!!!!
\subsubsection{Perturbations} 
\label{sec:methods-pert}

\smallskip
The stream is initialised with velocity $\vec{v_s} = \Mb\cb \hat{z}$, where $c_b=[\gamma P(\Rs^+)/\rho(\Rs^+)]^{1/2}$ is the sound speed %in the background 
at the outer boundary of the stream. 
The background gas is initialised at rest, with velocity to $\vec{v_b}=0$. 

\smallskip
We then perturb our simulations with a random realization of periodic perturbations in the radial component of the velocity, $v_{\rm r}=v_{\rm x}{\rm cos}(\varphi)+v_{\rm y}{\rm sin}(\varphi)$, as in M19. In practice, we perturb the Cartesian components of the velocity, 
\be 
\label{eq:pertx}
\begin{array}{c}
v_{\rm x}^{\rm pert}(r,\varphi ,z) = \sum_{j=1}^{N_{\rm pert}} v_{0,j}~{\rm cos}\left(k_{j}z+m_{j}\varphi + \phi_{j}\right)\\
\\
\times {\rm exp}\left[-\dfrac{(r-\Rs)^2}{2\sigma_{\rm pert}^2}\right]{\rm cos}\left(\varphi\right)
\end{array},
\ee
\be 
\label{eq:perty}
\begin{array}{c}
v_{\rm y}^{\rm pert}(r,\varphi ,z) = \sum_{j=1}^{N_{\rm pert}} v_{0,j}~{\rm cos}\left(k_{j}z+m_{j}\varphi + \phi_{j}\right)\\
\\
\times {\rm exp}\left[-\dfrac{(r-\Rs)^2}{2\sigma_{\rm pert}^2}\right]{\rm sin}\left(\varphi\right)
\end{array}.
\ee

\smallskip
The velocity perturbations are localised on the stream-background interface, with a penetration depth set by the parameter $\sigma_{\rm pert}$. We set $\sigma_{\rm pert}=\Rs/16$ in all of our simulations, as in M19. 
To comply with periodic boundary conditions, all wavelengths were harmonics of the box length, $k_{j}=2\pi n_{j}$, where $n_{j}$ is an integer, corresponding to a wavelength $\lambda_{j}=1/n_{j}$. In each simulation, we include all wavenumbers in the range $n_{j}=2-64$, corresponding to all available wavelengths in the range $\Rs/2 - 16\Rs$. 
Each perturbation mode is also assigned a symmetry mode, represented by the index $m_{j}$ in \equs{pertx} and \equm{perty}, and discussed in \se{khi}. As in M19, we only consider $m=0,1$. For each wavenumber $k_{j}$ we include both an $m=0$ mode and an $m=1$ mode. This results in a total of $N_{\rm pert}=2\times 63=126$ modes per simulation. 
Each mode is then given a random phase $\phi_{j} \in [0,2\pi)$. The stochastic variability from changing the 
random phases was extremely small, as shown in P18 and M19. The amplitude of each mode, $v_{0,j}$, was identical, with the rms amplitude normalised to $0.01\cs$.

%!!!!!!!!!!!!!!!!!!!!!!!!!!!!!!!!!!!!!!!!!!!
\subsubsection{Resolution and Refinement Scheme}
\label{sec:grid_res}
We used a statically refined grid with resolution decreasing away from the stream axis. The highest resolution region is ${\rm max}(|x|,|y|)<3\Rs$, with cell size $\Delta=2^{-10}$. For $\Rs=1/32$ this corresponds to 64 cells per stream diameter. The cell size increases by a factor of 2 every $3\Rs$ in the $x$ and $y$-directions, up to a maximal cell size of $2^{-6}$. The resolution is uniform along the $z$ direction, parallel to the stream axis. For uniform density cylinders, KHI surface modes are converged at this resolution (M19). 
We also ran two cases with a factor 2 higher resolution (128 cells per stream diameter) in order to test convergence of our results for self-gravitating streams. As described in \se{disruption} and \se{fragmentation}, we find that the majority of our results are indeed converged.

%!!!!!!!!!!!!!!!!!!!!!!!!!!!!!!!!!!!!!!!!!!!
\subsubsection{Simulations Without Self-Gravity}
\label{sec:no_grav}

\smallskip
In addition to the simulations of self-gravitating cylinders described above, we performed several simulations without self-gravity for comparison, hereafter our ``no-gravity" simulations. In the no-gravity simulations, the boundary conditions at $x=\pm 0.5$ and $y=\pm 0.5$ are simply zero gradients in all fluid variables, including the pressure. These were then initialised with the same density profiles as the corresponding self-gravitating streams, but with constant pressure throughout the simulation domain, since there is no gravitational field. We set the pressure to be the same as the pressure at the stream boundary in the corresponding self-gravity simulations, $P_{\rm no-gravity}(r)=P_{\rm self-gravity}(\Rs)$. This allows us to separate the effects of the density profile from those of self-gravity on the evolution of KHI.

%!!!!!!!!!!!!!!!!!!!!!!!!!!!!!!!!!!!!!!!!!!!
\subsection{Tracing the Two Fluids} 
\label{sec:tracer}
Following P18 and M19, we use a passive scalar field, $\psi(r,\varphi,z,t)$, to track the growth of the shear layer and the mixing of the two fluids. Initially, $\psi=1$ and $0$ at $r<\Rs$ and $r>\Rs$ respectively, and is then smoothed using \equs{ramp1}-\equm{ramp2}. During the simulation, $\psi$ is advected with the flow such that the density of stream-fluid in each cell is $\rhos=\psi\rho$, where $\rho$ is the total density in the cell. 

\smallskip
The volume-weighted average radial profile of the passive scalar is given by 
\be 
\label{eq:volume-averaged-colour}
\overline{\psi}(r,t) = \frac{\int_{-L/2}^{L/2}\int_{0}^{2\pi} \psi_{(r,\varphi ,z,t)} r~{\rm d\varphi \,dz}}{2\pi r L}.
\ee
{\no}The resulting profile is monotonic (neglecting small fluctuations on the grid scale) and can be used to define the edges of the shear layer around the stream interface, $r(\overline{\psi}=\epsilon)$ on the background side and $r(\overline{\psi}=1-\epsilon)$ on the stream side, where $\epsilon$ is an arbitrary threshold. We set $\epsilon=0.04$ following M19, though our results are not strongly dependent on this choice. The background-side thickness of the shear layer is then defined as 
\be 
\label{eq:hb_def}
\hb \equiv {\rm max_r}r(\overline{\psi}=\epsilon)-\Rs,
\ee
{\no}while the stream-side thickness is defined as 
\be 
\label{eq:hs_def}
\hs \equiv \Rs-{\rm min_r}r(\overline{\psi}=1-\epsilon).
\ee
{\no}While $\hb$ as defined in \equ{hb_def} is always well defined, at late times the perturbed region encompasses the entire stream and $\overline{\psi}(r=0)<1-\epsilon$. In this case, we define $\hs=\Rs$. The total width of the perturbed region is given by $h\equiv \hb+\hs$.

%%%%%%%%%%%%%%%%%%%%%%%%%%%%%%%%%%%%%%%%%%%%%%%%%%%%%%%
\section{Results} 
\label{sec:res} 

\smallskip
In this section we present the results of our numerical simulations. 
In \se{khivg}, we examine when the combined evolution of GI and KHI leads to the formation of long-lived clumps or to stream shredding, and compare to our theoretical predictions. In \se{disruption} and \se{fragmentation}, we discuss the late time evolution of the system in the cases when KHI and GI dominate, respectively.

%----------------------------------------------------------------
\subsection{KHI vs GI}
\label{sec:khivg}

\smallskip
As detailed in \se{combined}, we predict that 
a dense, self-gravitating filament shearing against a dilute background will either fragment into long-lived, bound clumps due to GI, or disrupt and mix into the background due to KHI, 
depending on the ratio of their respective timescales. The timescale for GI, $t_{\rm max}(\mu,\deltac)$, is well approximated by \equ{hunter} (see the Appendix~\se{surf_bod}). The timescales for KHI are $t_{\rm shear}(\Mb,\deltac)$ (\equnp{tau_shear}) or $t_{\rm dis}=(1+\delta^{1/2})t_{\rm shear}$ (\equnp{tau_diss}). For given values of $(\Mb,\deltac)$ there is a critical line mass ratio, $\mu_{\rm cr}$, such that for $\mu>\mucr$, $t_{\rm max}<t_{\rm shear}$ and GI will dominate. If $\mucr$ is small enough to be in the regime where GI is dominated by surface modes, then KHI will dominate for $\mu<\mucr$. However, if $\mucr$ is in the regime where GI is dominated by body modes, then the fate of the stream when $\mu<\mucr$ depends also on the ratio of $t_{\rm max}$ to $t_{\rm dis}$.

\begin{figure}
\includegraphics[trim={0cm 0.5cm 0cm 0cm},clip,width=0.49\textwidth]{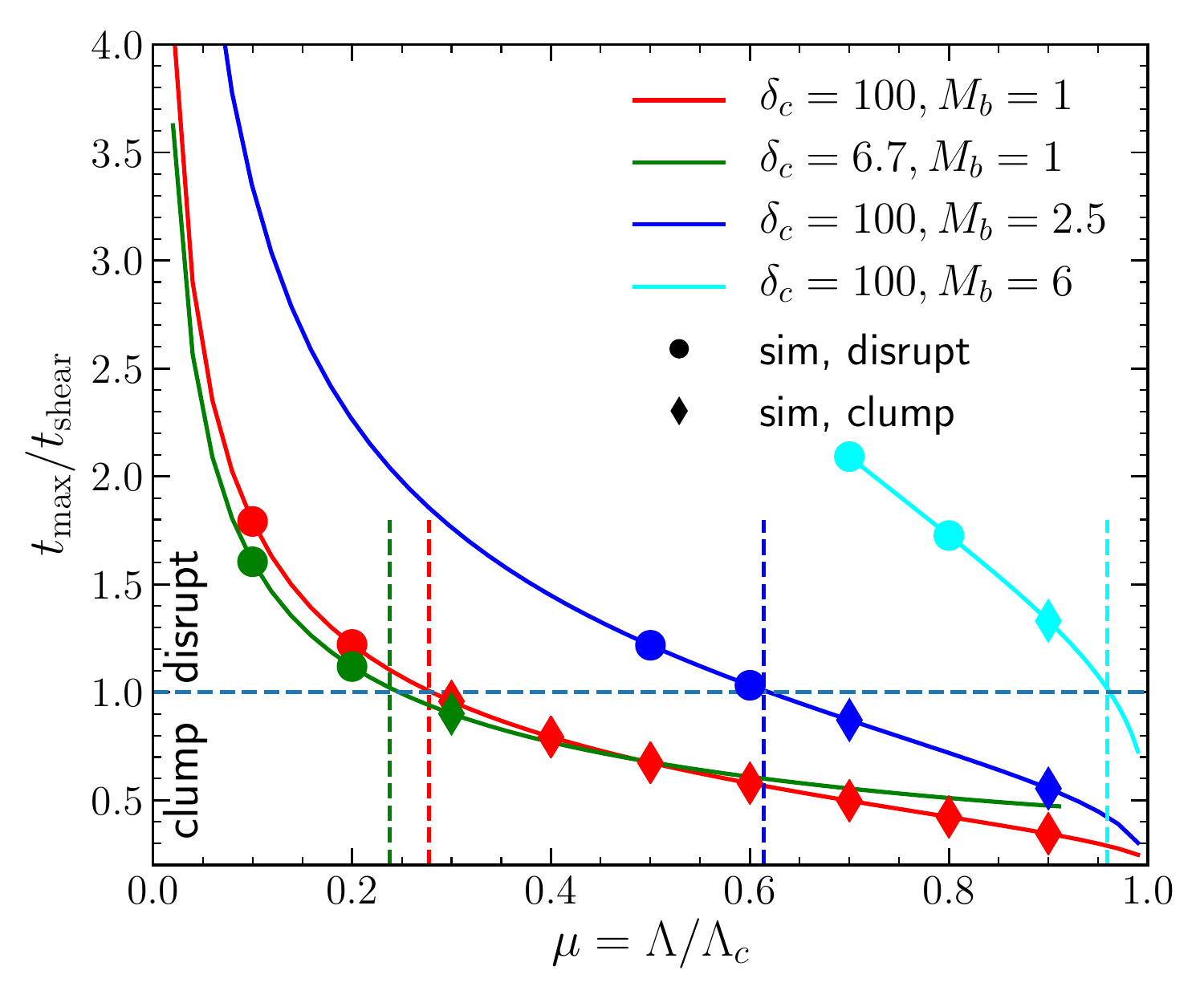}
\caption{Clump formation versus stream disruption according to our model, and in simulations. The solid lines show the ratio of the timescales for GI to form clumps, $t_{\rm max}$, and for KHI to destroy the contact discontinuity, $t_{\rm shear}$, as a function of the line-mass ratio, $\mu$. Different colours show different values of the Mach number and central density contrast, $\Mb$ and $\deltac$. Our model predicts that when this ratio is less than 1, marked by the horizontal dashed line, the stream should fragment and form clumps, while a ratio larger than one implies stream disruption by KHI. The transition occurs at a critical line-mass ratio, $\mucr\sim 0.28$, $0.36$, $0.62$, and $0.96$ for $(\Mb,\deltac)=(1.0,6.7)$, $(1.0,100)$, $(2.5,100)$, and $(6.0,100)$ respectively. The markers show simulation results, where circles indicate cases where the stream was disrupted by KHI and diamonds indicate cases where the stream fragmented to form clumps. Nearly all our simulations agree with our model, with circles lying above the dashed line and diamonds below it. The one exception is $(\Mb,\deltac,\mu)=(6.0,100,0.9)$, which is dominated by GI body modes rather than surface modes, and forms clumps despite $\mucr~0.96$.
}
\label{fig:tcompare}
\end{figure}

\smallskip
Solid curves in \fig{tcompare} show the ratio $t_{\rm max}/t_{\rm shear}$ as a function of $\mu$ for $(\Mb,\deltac)=(1.0,100)$, $(1.0,6.7)$, $(2.5,100)$, and $(6.0,100)$. The corresponding values of $\mucr$ are $\sim 0.36$, $0.28$, $0.62$, and $0.96$. Note the very weak dependence of $t_{\rm max}/t_{\rm shear}$ on $\deltac$ for $\Mb=1$, since $t_{\rm max}$ depends weakly on $\deltac$ for $\delta_c \gsim 4$, while $t_{\rm shear}$ depends weakly on $\deltac$ only through $\alpha(M_{\rm tot})$ (\equnp{alpha_fit}). The dependence of $t_{\rm max}/t_{\rm shear}$ on $\Mb$ is much stronger, since $t_{\rm shear}$ decreases roughly linearly with $\Mb$.

\begin{figure}
\includegraphics[trim={0cm 0.5cm 0cm 0cm},clip,width=0.49\textwidth]{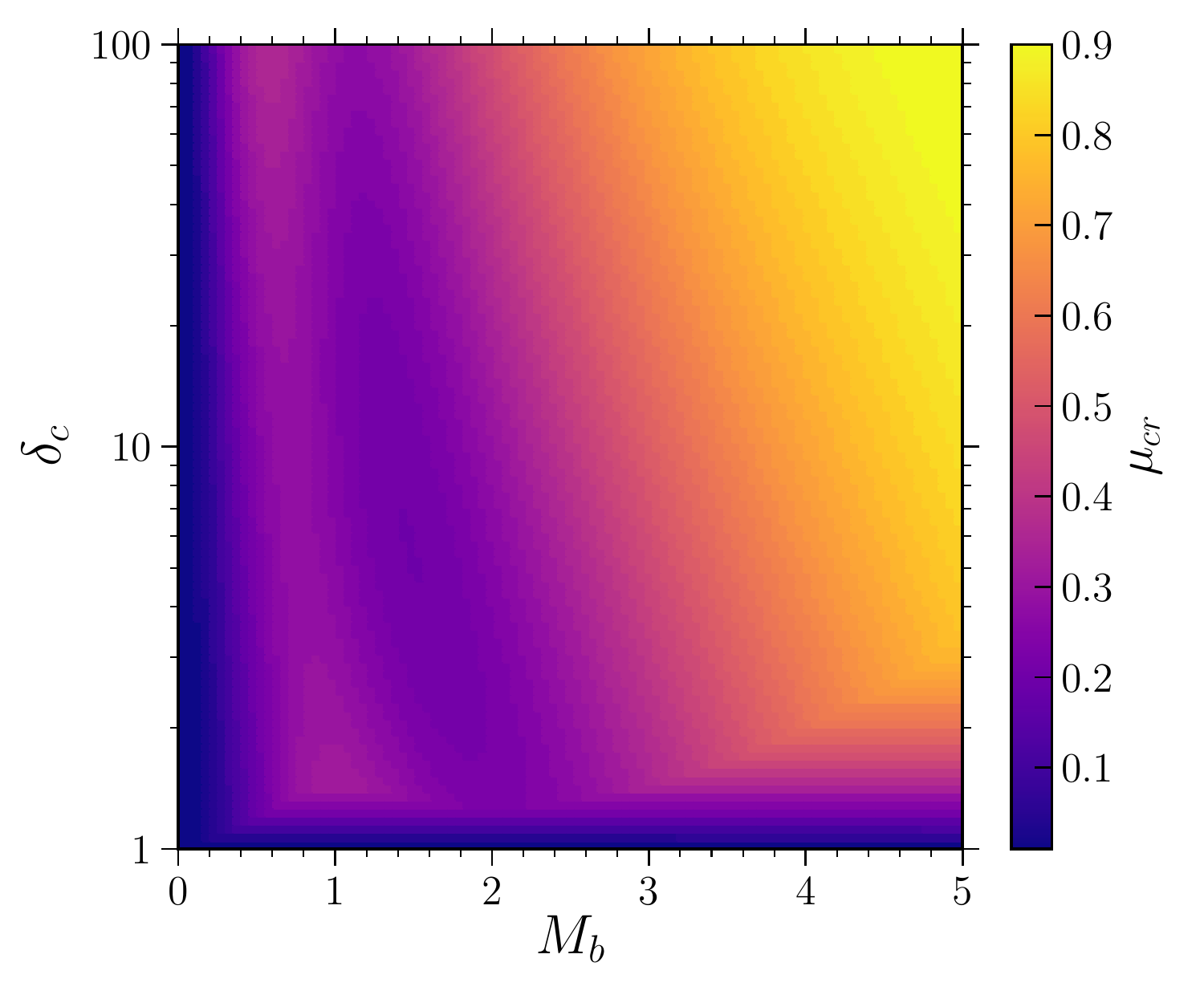}
\caption{Critical line-mass ratio, $\mucr$, for which $t_{\rm max}/t_{\rm shear}=1$, as a function of $\Mb$ and $\deltac$. For $\mu>\mu_{\rm cr}$, the stream will eventually fragment into clumps, while for $\mu<\mu_{\rm cr}$ KHI will disrupt the stream before fragmentation occurs. $\mucr$ tends to increase with $\Mb$, except for a narrow strip near $\Mb\sim 1.5$, and with $\deltac$, though the dependence on $\deltac$ is much weaker. For $\deltac\lsim 100$, $\mucr>0.5$ only for $\Mb\gsim 2.5$, suggesting that for large line-masses KHI can only overcome GI for very supersonic flows which are dominated by high-order azimuthal modes.
}
\label{fig:criticalmu}
\end{figure}

\smallskip
In \fig{criticalmu} we show $\mucr$ as a function of $\Mb$ and $\deltac$. The general trend is the same as inferred from \fig{tcompare}, namely $\mucr$ increases strongly with $\Mb$ and has only a slight tendency to increase with $\deltac$. The exception is a narrow strip near $\Mb\sim (1-2)$ where $\mucr$ decreases with $\Mb$. In this region, the increase of $t_{\rm shear}$ due to decreasing $\alpha$ is stronger than the decrease in $t_{\rm shear}$ due to increasing $V$, leading to a net increase in $t_{\rm shear}$ with $\Mb$ and thus a net decrease in $\mucr$. 
For density contrasts $\deltac\lsim 100$, $\mucr>0.5$ only for supersonic flows with $\Mb\gsim 2.5$. This implies that for massive streams, KHI can only overcome GI for highly supersonic flows (recall that the Mach number of the flow with respect to the sound speed in the stream is $\sim \delta^{1/2}\Mb$). In this regime, KHI is dominated by high-order azimuthal surface modes (see \se{khi}), which have a short eddy turnover time leading to rapid shear layer growth. 

\smallskip
Consider, for example, $\deltac\sim 30$. $\mucr$ increases from $\mucr<<1$ at $\Mb<<1$ towards $\mucr\gsim 0.3$ at $\Mb\sim 0.6$, then decreases to $\mucr\lsim 0.2$ at $\Mb\sim 1.2$, before strongly increasing at $\Mb>>1$. Thus, as $\Mb$ is increased from $\lsim 0.2$ to $\gsim 2$ for $\deltac\sim 30$ and $\mu\sim 0.25$, the stream fluctuates from being dominated by GI, to KHI, to GI, to KHI. The high $\Mb$ KHI regime is dominated by surface modes with high azimuthal wavenumber. While these modes are always unstable, at lower Mach numbers they tend to be sub-dominant compared to axisymmetric or helical modes, with $m=0,1$ (M19). 

\begin{table}
\centering
\begin{tabular}{c|c|c|c|c|c|c|c|c}
$\mu$ & $M_b$ & $\delta_c$ & $\delta$ & $t_{\rm max}$ & $t_{\rm shear}$ & 
$t_{\rm dis}$ & $t_{\rm sc}$ & $\lambda_{cr}$\\
\hline
0.1 & 1   & 100 & 92 &  2.39 & 1.34 & 14.15 & 0.64 & 5.95\\
0.2 & 1   & 100 & 84 &  2.46 & 2.01 & 20.43 & 0.90 & 5.95\\
0.3 & 1   & 100 & 76 &  2.54 & 2.65 & 25.65 & 1.11 & 5.95\\
0.4 & 1   & 100 & 67 &  2.63 & 3.32 & 30.51 & 1.28 & 5.95\\
0.5 & 1   & 100 & 58 &  2.76 & 4.09 & 35.38 & 1.44 & 5.98\\
0.6 & 1   & 100 & 49 &  2.93 & 5.06 & 40.62 & 1.58 & 5.98\\
0.7 & 1   & 100 & 40 &  3.18 & 6.39 & 46.77 & 1.73 & 6.02\\
0.8 & 1   & 100 & 30 &  3.60 & 8.52 & 55.01 & 1.88 & 6.09\\
0.9 & 1   & 100 & 18 &  4.56 & 13.20 & 69.68 & 2.06 & 6.23\\
\hline
0.1 & 1   & 6.7 & 6.2 & 3.21 & 2.00 & 6.97 & 0.64 & 7.02\\
0.2 & 1   & 6.7 & 5.6 & 3.41 & 3.04 & 10.26 & 0.90 & 7.16\\
0.3 & 1   & 6.7 & 5.1 & 3.66 & 4.07 & 13.22 & 1.11 & 7.36\\
\hline
0.5 & 2.5 & 100 & 58 &  2.76 & 2.27 & 19.57 & 1.44 & 5.98\\
0.6 & 2.5 & 100 & 49 &  2.93 & 2.84 & 22.77 & 1.58 & 5.98\\
0.7 & 2.5 & 100 & 40 &  3.18 & 3.65 & 26.71 & 1.73 & 6.02\\
0.9 & 2.5 & 100 & 18 & 4.56 & 8.22 & 43.40 & 2.06 & 6.23\\
\hline
0.7 & 6   & 100 & 40 &  3.18 & 1.52 & 11.13 & 1.73 & 6.02\\
0.8 & 6   & 100 & 30 &  3.60 & 2.09 & 13.47 & 1.88 & 6.09\\
0.9 & 6   & 100 & 18 &  4.56 & 3.43 & 18.09 & 2.06 & 6.23
\end{tabular}
\caption{ Parameters of simulations used for studying the evolution of streams undergoing both GI and KHI. The first three columns list the control parameters, namely the line-mass ratio $\mu$, Mach number $\Mb$, and the ratio of central density to background density at the stream boundary $\deltac$. The remaining six columns list derived parameters: the ratio of stream to background density on either side of the interface, $\delta$, the GI time scale, $t_{\rm max}$, the timescale for KHI to destroy the contact discontinuity, $t_{\rm shear}$, the timescale for KHI to destroy the entire stream, $t_{\rm dis}$, the stream sound crossing time, $t_{\rm sc}$, and the shortest unstable wavelength for GI, $\lambda_{\rm cr}$. All timescales are in units of the stream free-fall time, $t_{\rm ff}$, while $\lambda_{\rm cr}$ is in units of the stream radius, $\Rs$. For all cases, the fastest growing wavelength for GI is $\lambda_{\rm max}\lsim 2\lambda_{\rm cr}$. }
\label{tab:sim_clump}
\end{table}

\begin{figure}
\includegraphics[trim={0cm 0.4cm 0cm 0cm},clip,width=0.48\textwidth]{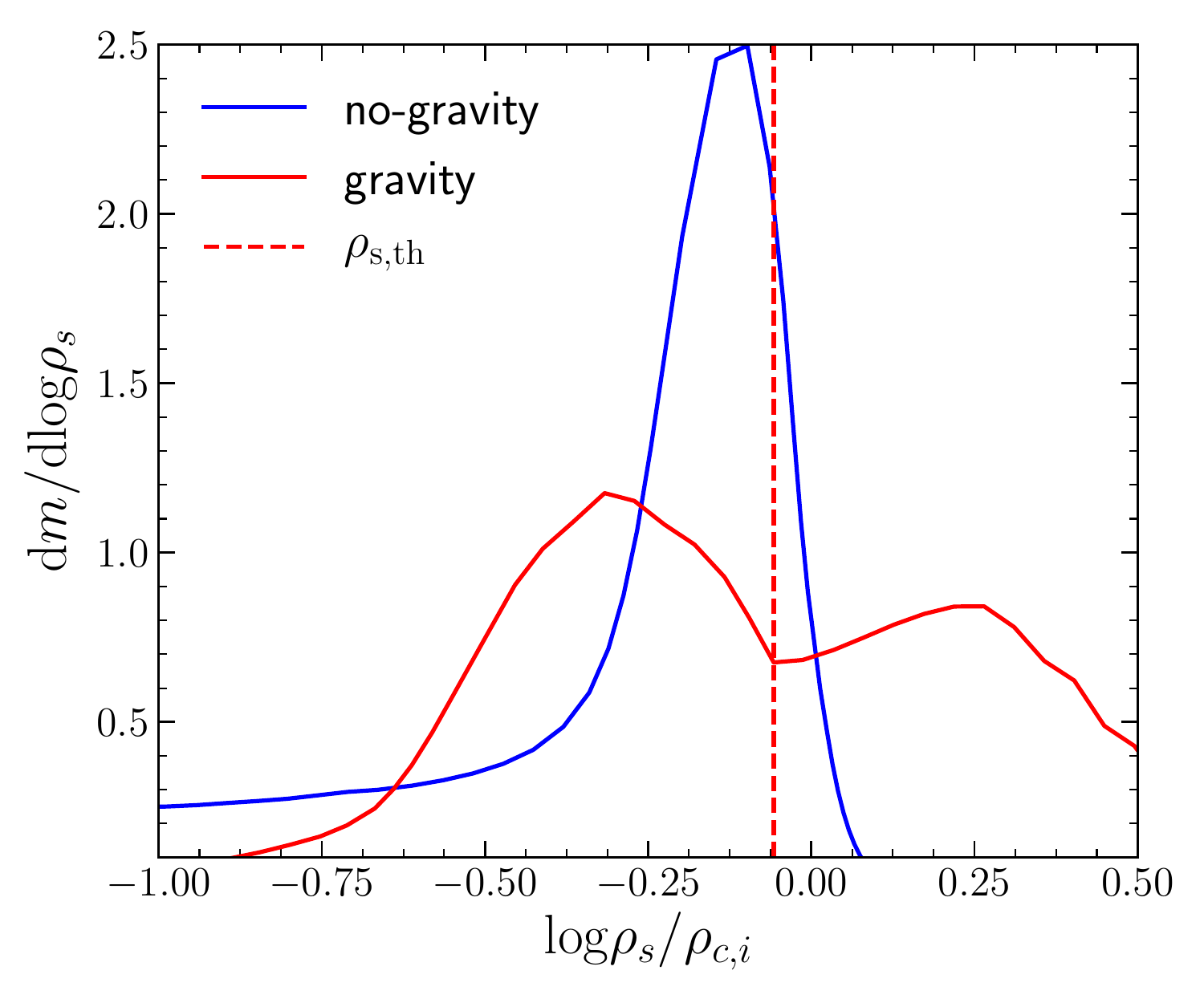}
\caption{Clump identification in the simulations. We show the PDFs of stream density, $\rho_{\rm s}=\psi\rho$, at $t=8t_{sc}$ for the no-gravity (blue) and gravity (red) simulations with $(\Mb,\deltac,\mu)=(1,100,0.9)$. While the no-gravity simulation exhibits a unimodal, roughly lognormal, PDF, the gravity simulation is bi-modal. Cells with densities higher than the break, $\rho_{\rm s,th}$ marked by the vertical dashed line, are associated with collapsed clumps.}
\label{fig:dpdf}
\end{figure}

\smallskip
To test our predictions, we performed a series of simulations with the same combinations of $(\Mb,\deltac)$ as shown in \fig{tcompare}, and different values of $\mu$. For $(\Mb,\deltac)=(1.0,100)$, we performed nine simulations spanning the line-mass range $\mu=0.1,\,0.2,\,...,\,0.9$. For the other combinations of $(\Mb,\deltac)$, we performed three to four simulations each, with $\mu$ spanning a small region around the predicted $\mucr$. The full list of simulations 
is presented in \tab{sim_clump}, along with several relevant parameters. 
%In addition to $\mu$, $\Mb$, and $\deltac$, we also list for each simulation the values of $\delta$, $t_{\rm max}$, $t_{\rm shear}$, and $t_{\rm dis}$, all calculated following \se{theory} and \se{sim}, as well as the stream sound crossing time, $\tsc$, defined as
The stream sound crossing time\footnote{The sound crossing times listed in \tab{sim_clump} refer to the self-gravity simulations only. In the no-gravity runs at $r<\Rs$, $\rho_{\rm no-gravity}(r)=\rho_{\rm gravity}(r)$ while $P_{\rm no-gravity}(r)=P_{\rm gravity}(\Rs)<P_{\rm gravity}(r)$. This results in a lower sound speed at each $r<\Rs$, and hence a longer sound crossing time.}, $\tsc$, is defined as
\be
\label{eq:sound_speed}
\tsc = 2\int_0^{R_s} 1/c_s(r) dr,
\ee
{\no}where $\cs(r)=(\gamma P(r)/\rho(r))^{1/2}$ is the sound speed at radius $r$.

\smallskip
The markers in \fig{tcompare} indicate for each of our simulations whether or not the stream has fragmented into long-lived collapsed clumps. To identify such clumps, we examine the PDF of stream fluid density, $\rhos=\psi\rho$. If the density distribution is a result of pure turbulence, we expect it to have a roughly lognormal shape. If, however, the highest density regions have collapsed due to gravity, we expect a break in the PDF at high densities  \citep[e.g.][]{VS08,Elmegreen11,Hopkins12,Federrath15}. An example is shown in \fig{dpdf}, where we show the density PDFs for the gravity and no-gravity simulations with $(\Mb,\deltac,\mu)=(1.0,100,0.9)$ at $t=8\tsc$. While the no-gravity simulation has a unimodal PDF which is roughly lognormal except at the lowest densities, the gravity simulation produces a bi-modal PDF, and we associate all cells with densities larger than the break density, $\rho_{\rm s,th}$, as being in clumps. As discussed in \se{fragmentation} below, these clumps are indeed long-lived. If a simulation never exhibits a similar break in the density PDF we determine that this simulation has not formed any clumps. In particular, isolated high density regions produced in no-gravity simulations at late times (see \figs{mpl1_maps} and \figss{mpl9_maps} below) are not clumps, but rather transient features associated with the high-density part of a turbulent PDF. 

\begin{figure*}
\includegraphics[trim={3.4cm 2.5cm 4.5cm 1.5cm},clip,width=0.99\textwidth]{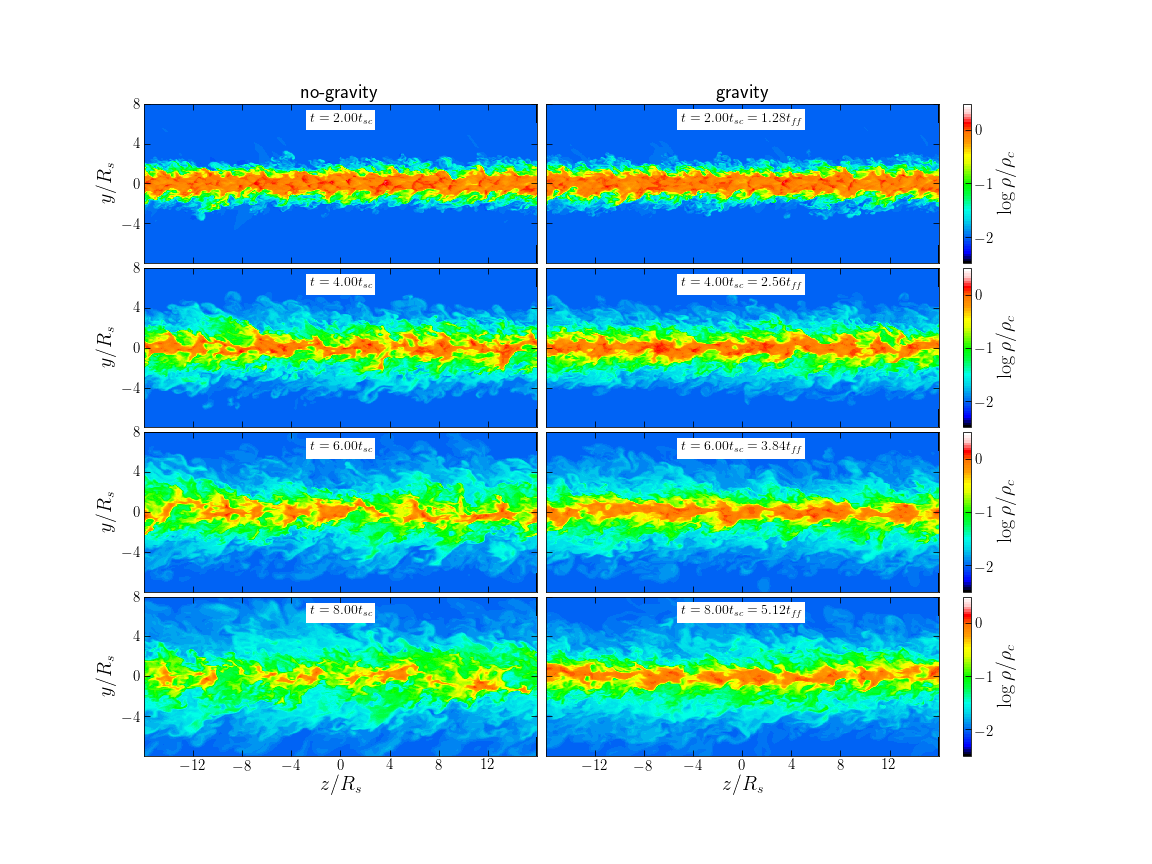}
\caption{
Evolution of streams with $\mu<\mucr$ undergoing KHI. Shown are snapshots of density normalised by the initial density along the stream axis, $\rhoc$, in a slice through the $yz$ plane showing an ``edge-on" view of the cylinder. The two columns show simulations with $(\Mb,\deltac,\mu)=(1.0,100,0.1)$ run without self-gravity (left) and with self-gravity (right). The snapshot times in units of the stream sound crossing time, $\tsc$, are listed in each panel. The evolution with and without gravity is very similar up until $t\sim 5\tsc$ and shows the formation of a turbulent shear layer penetrating into the stream and background and miximg the two fluids. At later times, the penetration of the shear layer into the background continues similarly, though self-gravity reduces the penetration into the stream, leaving more high density material near the stream axis.}
\label{fig:mpl1_maps}
\end{figure*}

\smallskip
All of our simulations with $\mu>\mucr(\Mb,\deltac)$ form gravitating clumps, as predicted by our model. Furthermore, for $\Mb=1,2.5$, when $\mucr\lsim 0.63$, streams in simulations with $\mu<\mucr(\Mb,\deltac)$ are disrupted by KHI and mixed into the background before forming bound clumps, as predicted by our model. In these cases, GI is dominated by surface modes, so the comparison of $t_{\rm max}$ and $t_{\rm shear}$ is justified. On the other hand, for $\Mb=6.0$, $\mucr=0.96$ is in the body mode regime for GI, and our simulation with $\mu=0.9$ fragments into bound clumps, as discussed in \se{combined}. However, in this same regime we find that streams with $\mu=0.8$ and $0.7$ are disrupted by KHI and do not form bound clumps. So the effect of GI body modes is to lower $\mucr$ from $\sim 0.96$ to $\sim 0.85$.
%On the other hand, such clumps are not produced in the same set of simulations with $\mu<\mucr$, where the stream is eventually disrupted by KHI and mixed into the background. However, for $(\Mb,\deltac)=(6.0,100)$, where $\mucr\sim 0.96$, the stream forms clump at $\mu=0.9<\mucr$ even though $t_{\rm max}>t_{\rm shear}$. Since $\mu=0.9$ is large enough for the stream to be unstable to GI body modes and $t_{\rm shear}<t_{\rm max}<t_{\rm dis}$, the relevant timescale to compare to gravitational instability is the time for shear layer to propagate into the stream to stop body modes as suggested in \se{combined}.

\smallskip
Overall, we conclude that our model adequately predicts the fate of streams under the combined effects of KHI and GI when GI surface modes dominate. When GI body modes dominate, the actual value of $\mucr$ is $\sim 10\%$ lower than our prediction, since the relevant timescale for KHI to prevent clump formation is no longer $t_{\rm shear}$.
%timescale needs to change to make consistent prediction. 
In the next two sections, we turn to studying the evolution of streams and clumps in the regime where each instability dominates.

%%%%%%%%%%%%%%%%%%%%%%%%%%%%%%%%%%%%%%%%%%%%%%%%%%%%%%%%%%%%
\subsection{Stream Disruption due to KHI}
\label{sec:disruption}

\smallskip
We here examine the evolution of streams with $\mu<\mucr(\Mb,\deltac)$, where KHI dominates over GI and prevents the formation of long-lived collapsed clumps. Specifically, we examine whether the self-gravity of the gas, while unable to completely overcome the KHI, affects its evolution in any way. 

\smallskip
\Fig{mpl1_maps} shows the evolution of streams with $(\Mb,\deltac,\mu)=(1.0,100,0.1)$, with and without self-gravity. 
At early times, $t\lsim 4\tsc$, the evolution in the two cases is extremely similar, and the shear layer seems to expand at roughly the same rate, mixing the two fluids and diluting the stream density. At later times, $t\gsim 6\tsc$, while the expansion of the shear layer into the background continues similarly in both simulations, the penetration into the stream has stalled in the gravity run. The self-gravity of the stream thus seems to partly shield its inner core from mixing and disruption. As we will show below, this is due to restoring buoyancy forces caused by the stream's gravitational field.

\begin{figure}
\includegraphics[trim={0cm 0.5cm 0cm 0cm},clip,width=0.48\textwidth]{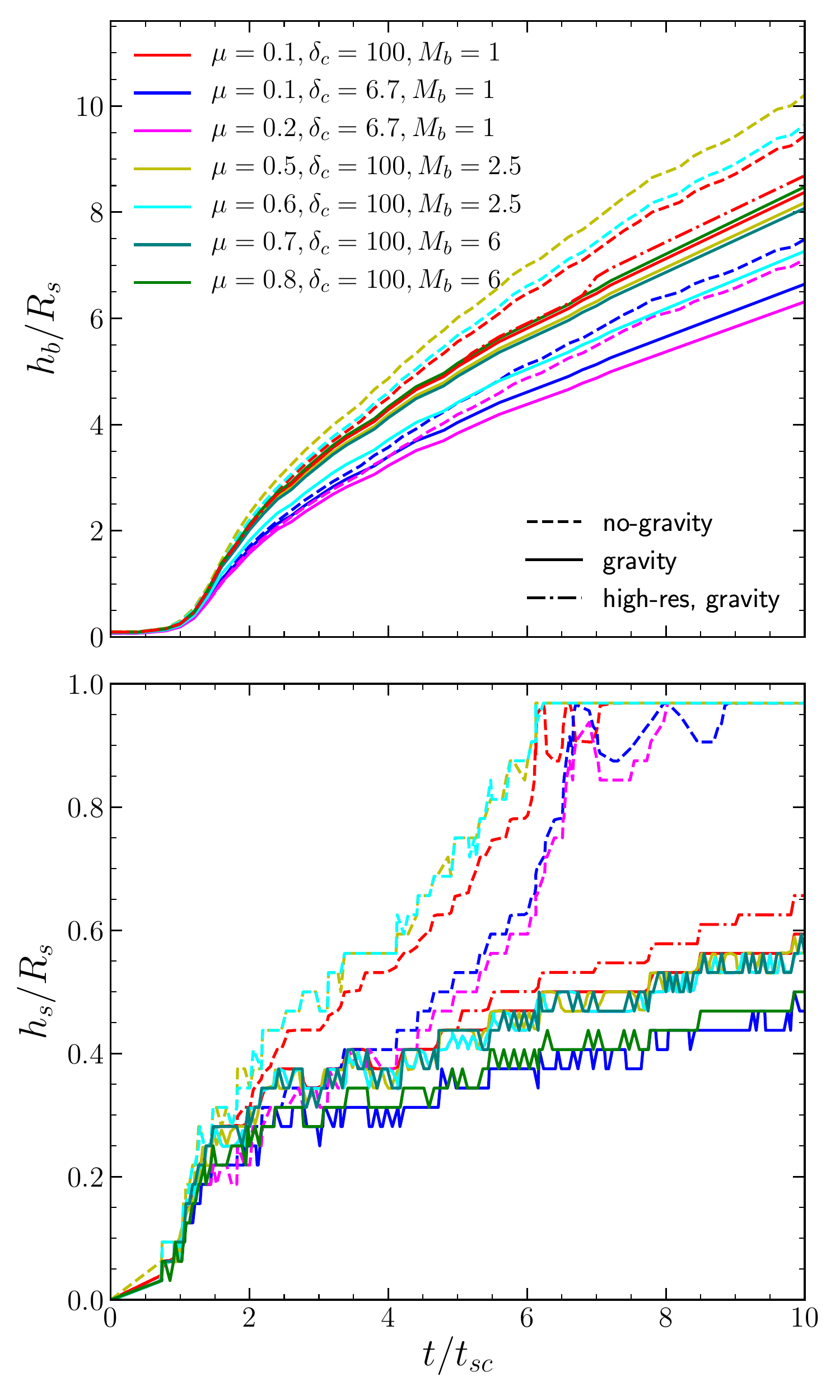}
\caption{ %\nmr{Please add the simulation with $\Mb=6.0,\mu=0.9$. Do this also in \figs{Ri} and \figss{momentum}.} 
Shear layer growth in simulations dominated by KHI, with $\mu<\mucr(\Mb,\deltac)$. We show the penetration depth of the shear layer into the background, $\hb$ (top), and into the stream, $\hs$ (bottom). These have been normalised by the stream radius, $\Rs$, while time on the x-axis has been normalised by the stream sound crossing time, $\tsc$. In each panel, solid lines show our fiducial simulations with self-gravity, while dashed lines show our no-gravity simulations. Different colours mark different combinations of $(\Mb,\deltac,\mu)$. The dot-dashed red line in each panel shows results from a simulation with $(\Mb,\deltac,\mu)=(1.0,100,0.1)$ and twice higher resolution. 
The penetration of the shear layer into the background proceeds similarly in simulations with and without gravity, while the penetration into the stream is qualitatively different with and without gravity. Without gravity, the shear layer consumes the entire stream at $t\sim t_{\rm dis}$ (\equnp{tau_diss}). However, with self-gravity $\hs\sim (0.3-0.4)\Rs$ at this time, regardless of $\mu$, likely caused by buoyancy stabilizing the inner stream.}
\label{fig:mixing}
\end{figure}

\begin{figure}
\includegraphics[trim={0cm 0.4cm 0cm 0cm},clip,width=0.48\textwidth]{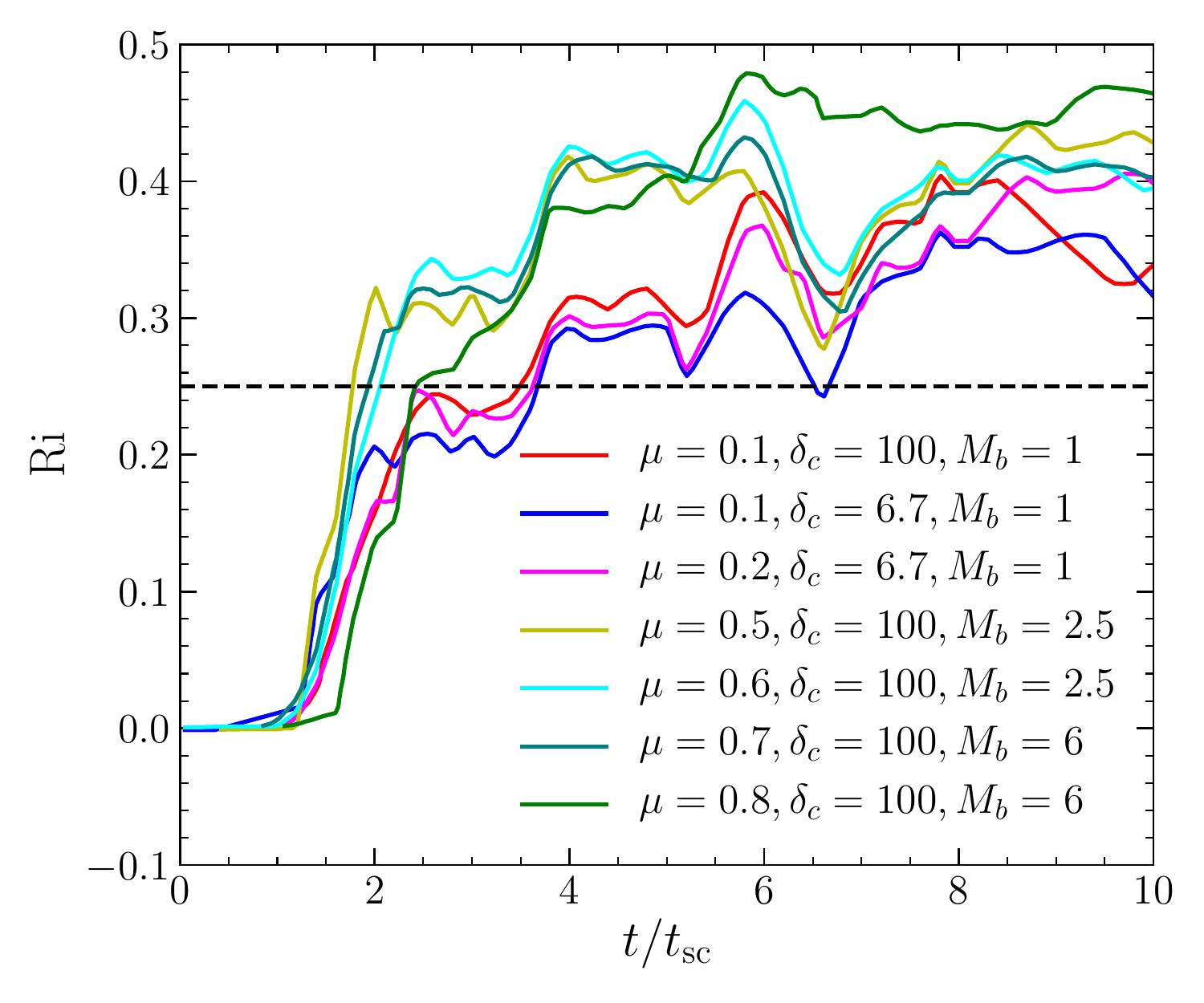}
\caption{Evolution of mass-weighed Richardson number ${\rm Ri}$ within the shear layer $[\Rs-\hs(t)]<r<\Rs$ over time. Value of ${\rm Ri}<1/4$ indicates that the buoyant force from gravity cannot stop mixing, resulting in quick growth of shear layer $h_s$ at early times. At late times when ${\rm Ri}>1/4$, the growth of shear layer slows down due to less mixing.}
\label{fig:Ri}
\end{figure}

\smallskip
We examine this more quantitatively in \fig{mixing}, where we compare the evolution of $\hb$ and $\hs$, the penetration of the shear layer into the background and stream respectively (\equnp{hs_growth}), in gravity and no-gravity simulations with $\mu<\mucr(\Mb,\deltac)$. Focusing on the top panel, we see that $\hb$ evolves similarly with and without self-gravity, and is consistent with the results of M19. During the first sound crossing time, $\hb$ remains roughly constant as the initial velocity perturbations trigger perturbations in the stream-background interface associated with growing eigenmodes of the system. Following this phase, $\hb$ grows approximately linearly following \equ{hs_growth} until it reaches $\hb\sim 2\Rs$. Up until this point, the gravity and no-gravity runs are nearly indistinguishable. Following this, the growth rate of $\hb$ is reduced by roughly half in both cases, as the developing turbulent cascade transfers power from the largest scales driving the expansion to smaller scales (M19). During this phase, the growth rate of $\hb$ is reduced in the gravity simulations, by $\sim 25\%$ for $\mu=(0.5-0.6)$ and $\sim 12\%$ for $\mu=(0.1-0.2)$. Overall, we find that the self-gravity of the stream has a relatively minor effect on the growth of $\hb$.

\smallskip
On the other hand, as inferred from \fig{mpl1_maps}, there is a qualitative difference in the evolution of $\hs$, shown in the bottom panel of \fig{mixing}. For the first $\sim 2\tsc$, until $\hs\sim 0.3\Rs$, the gravity and no-gravity runs evolve similarly. After this, the growth rate in the gravity runs is a factor $\sim 3$ smaller than in the no-gravity runs. In the latter, the shear layer reaches $\hs/\Rs=1$ and consumes the entire stream at $t\sim t_{\rm dis}$ (\equnp{tau_diss}), and the evolution is similar to that seen in M19 (see their figure B1). However, in the runs with self-gravity, $\hs\sim (0.3-0.4)\Rs$ at this time, and does not exceed $\sim 0.5\Rs$ at $t=10\tsc$. This is consistent with the visual impression in \fig{mpl1_maps}, where the density remains high in the interior of the self-gravitating stream even after the non-gravitating stream has been completely diluted. Although the growth rate of $\hs$ does not depend on $\mu$, there appears to be a tendency for more penetration for larger $\deltac$. 

\smallskip
We propose that the stalling of $\hs$ is due to restoring buoyancy forces in the stream interior. This can be seen by considering the Richardson number, ${\rm Ri}=[N_{\rm BV}/(\rmd u/\rmd r)]^2$, where $\rmd u/\rmd r$ is the gradient of longitudinal velocity inside the shear layer, and 
%the Froude number, $Fr=\sigma/(N_{\rm BV}l)$, where $\sigma$ is the turbulent velocity, $l$ is the characteristic length-scale of the turbulent eddies, and 
$N_{\rm BV}$ is the Brunt-Vais$\ddot{{\rm a}}$l$\ddot{{\rm a}}$ frequency,
\be 
\label{eq:NBV}
N_{\rm BV} = \left[\frac{g}{\gamma}\frac{\partial {\rm ln}K}{\partial r}\right]^{1/2},
\ee 
{\no}with $g(r)$ the magnitude of the gravitational field, $\gamma$ the adiabatic index, $K(r)=P(r)\rho^{-\gamma}(r)$ the entropy profile of the gas. Note that $K$ is piecewise constant in our initial conditions, with a non-zero gradient only at the stream-background interface. However, as the shear layer expands, mixing between the fluids creates a non-zero entropy gradient throughout the shear layer. Had our initial conditions been such that the initial stream was not isentropic, this may have increased $N_{\rm BV}$ and ${\rm Ri}$ in the stream interior.
%The growth of ${\rm Ri}$ is due to the growth of shearing layer, when the entropy contrast at the radius of the stream (see \equ{pressure_discontinuity}) slowly develops into the grwoing shearing layer. It is important to note that the entropy gradient is not limited to adiabatic streams and present in all streams with 2-fluids where pressure is continuous at the boundary for equilibrium, and the density contrast will introduce the entropy contrast.}

\smallskip
For a 2d plane-parallel system in a constant external gravitational field, it can be shown that a sufficient (but not necessary) criterion for buoyancy to stabilize the system against shearing induced mixing is that ${\rm Ri}>0.25$ \citep{Miles61,Howard61}. While our situation is more complex in that the geometry is cylindrical and the gravitational field is due to self-gravity rather than an external field\footnote{To our knowledge, no analogous criterion exists for the stability of self-gravitating flows or for cylindrical flows. Deriving such a criterion is beyond the scope of this paper.}, we may use this as a benchmark to asses the role of buoyancy in stabilizing the inner stream. In \fig{Ri} we show the mass weighed average of ${\rm Ri}$ within the shear layer, $[\Rs-\hs(t)]<r<\Rs$, as a function of time. In all simulations, ${\rm Ri}<<1$ at early times, and crosses ${\rm Ri}=0.25$ at $t\sim (2-3)\tsc$, corresponding to the sharp decline in the growth rate of $\hs$. Further growth of ${\rm Ri}$ is rather slow and it does not exceed ${\rm Ri}\sim (0.3-0.4)$. We find very similar behaviour when evaluating ${\rm Ri}$ locally at the inner boundary of the shear layer, $r=[\Rs-\hs(t)]$. This supports our assertion that buoyancy stabilizes the inner stream and slows the growth of $\hs$, significantly delaying stream disruption. 

\begin{figure}
\includegraphics[trim={0cm 0.5cm 0cm 0cm},clip,width=0.49\textwidth]{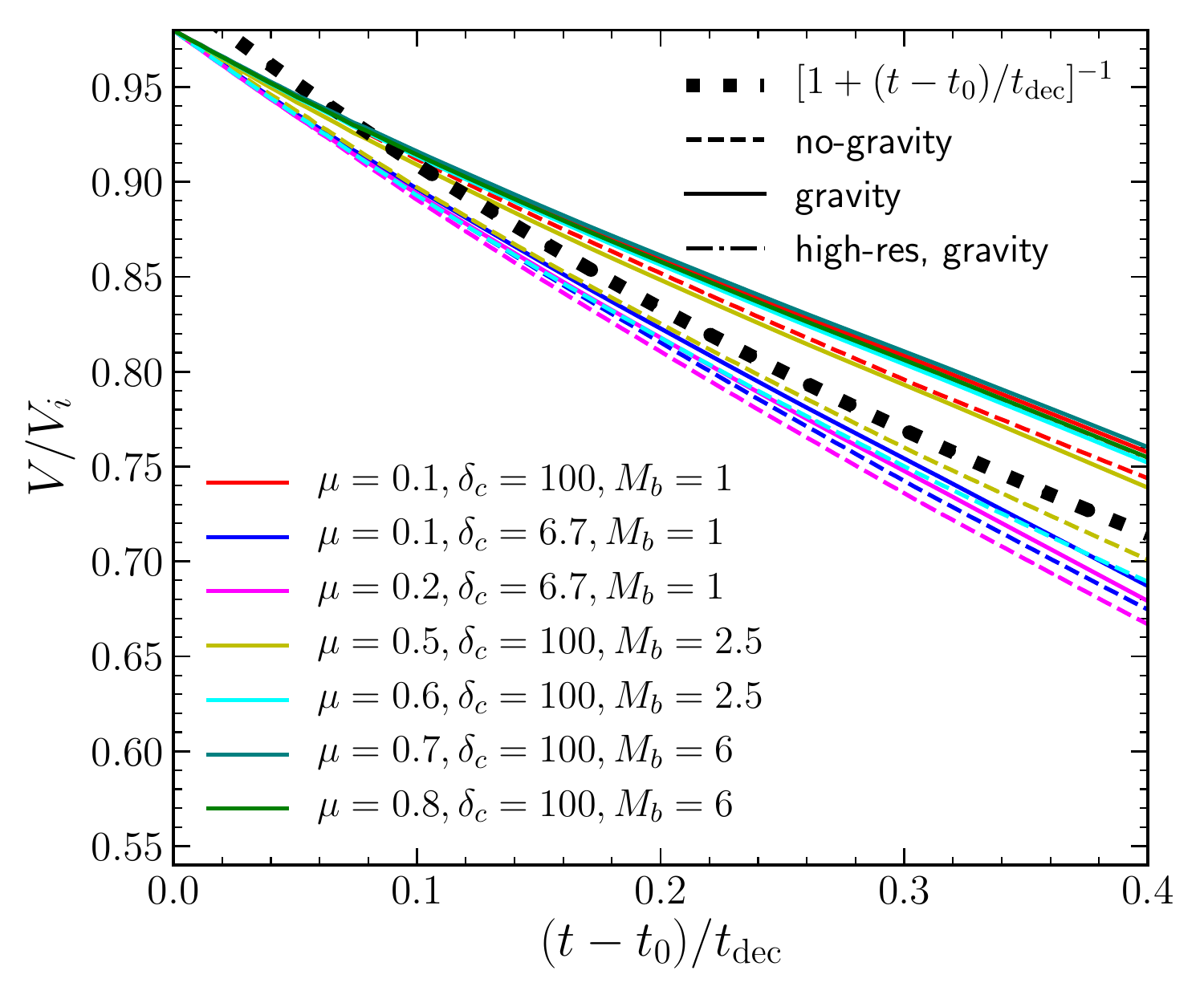}
\caption{Stream deceleration due to KHI. We show the centre of mass velocity of the stream fluid normalised by its initial value, as a function of time normalised by the predicted deceleration timescale, $t_{\rm dec}$ (\equnp{tau_dec}). The time axis has been set to zero at $t_0$, the time when the stream velocity is $98\%$ of its initial value. Solid (dashed) lines show simulations with (without) gravity, as in \fig{mixing}. The thick green dotted line shows the prediction for the deceleration rate due to KHI by M19 (\equnp{stream_deceleration}). The simulations with and without gravity behave similarly and closely follow the predicted deceleration rate. This is consistent with the similar behaviour of $\hb$, since the deceleration is primarily driven by entrainment of background material by the shear layer.
}
\label{fig:momentum}
\end{figure}

\smallskip
In \fig{momentum} we show the deceleration of streams in simulations with and without gravity. We show the centre of mass velocity of the stream fluid, i.e. weighted by the passive scalar $\psi$, normalised by its initial value, $V_{\rm i}$, as a function of time normalised by the predicted decelration timescale, $t_{\rm dec}$ (\equnp{tau_dec}). The time axis has been shifted to begin at $t_0$, the time when the stream velocity reaches $98\%$ of its initial value. In all cases, $t_0\sim \tsc$.
%This is done to focus on the deceleration itself rather than the ``incubation" period before it begins. 
The gravity and no-gravity simulations behave similarly, and both are well fit by the theoretical prediction (\equnp{stream_deceleration}). This was expected given the similarity between the evolution of $\hb$ in the gravity and no-gravity simulations (\fig{mixing}), since KHI-induced deceleration is primarily driven by entrainment of background material in the shear layer (P18,M19), not affected by buoyancy in the stream.

\smallskip
We~ran~the~$(\Mb,\deltac,\mu)=(1.0,100,0.1)$ simulation with a factor two higher spatial resolution throughout the simulation domain. The results of this simulation are shown in \figs{mixing} and \figss{momentum}. The evolution of $\hb$ and stream velocity, $V$, are nearly indistinguishable from our fiducial resolution. The penetration of the shear layer into the stream is slightly enhanced, with $\hs\sim 10\%$ larger in the high-resolution run at $t\sim 8\tsc$. This is still significantly less than the no-gravity simulation, supporting our conclusion that self-gravity significantly suppresses shear layer growth inside the stream.

%%%%%%%%%%%%%%%%%%%%%%%%%%%%%%%%%%%%%%%%%%%
\subsection{Stream Fragmentation due to GI}
\label{sec:fragmentation}

\begin{figure*}
\includegraphics[trim={4cm 2.5cm 5cm 2cm},clip,width=0.99\textwidth]{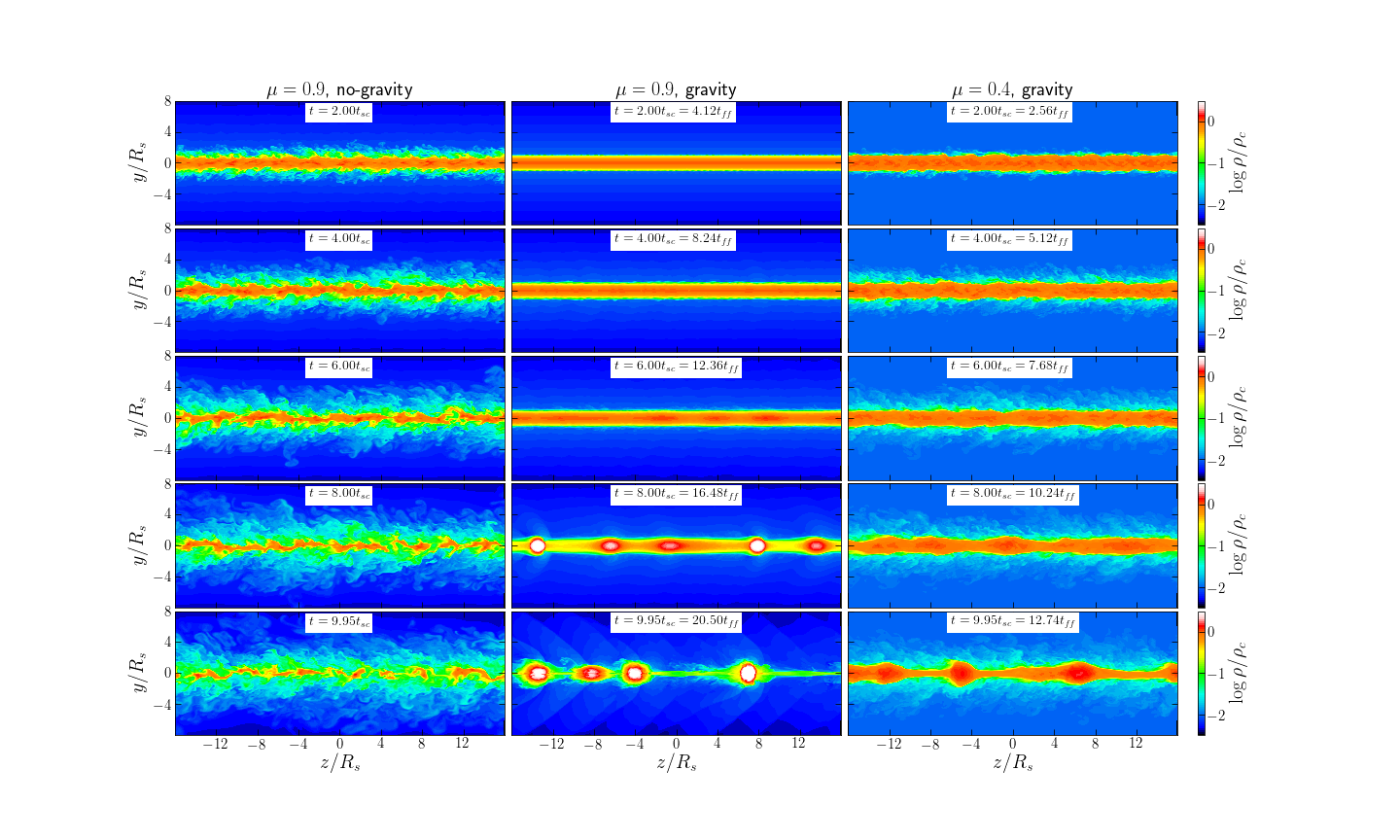}
\caption{
%\nmr{Based on footnote 5, replace the panels in the no-gravity run to correspond to snapshots at $t=2,4,6,8\tsc$ using the \textit{correct} $\tsc$ for the no-gravity simulation.}
Same as \fig{mpl1_maps}, but for simulations with $\mu>\mucr$. The three columns represent three different simulations, each with $(\Mb,\deltac)=(1.0,100)$. The left-hand column shows the no-gravity simulation with $\mu=0.9$,  
%a simulation with the initial density profile associated with $\mu=0.9$ but self-gravity turned off, while the centre and right-hand columns show simulations with self-gravity
while the centre and right-hand columns show the gravity simulations with $\mu=0.9$ and $0.4$ respectively. The snapshot times in units of the stream sound crossing time, $\tsc$, and free-fall time, $\tff$, are listed in each panel. At $t\sim 2\tsc$, a turbulent shear layer has developed in the non-gravitating simulation and the gravitating simulation with $\mu=0.4$, while the gravitating simulation with $\mu=0.9$ appears unperturbed. At later times, the shear layer consumes the non-gravitating stream as expected for KHI, while GI takes over in both simulations with gravity, resulting in dense clumps along the stream axis by $t\sim 10\tsc$. These clumps are separated by $\sim 6.5\Rs$, consistent with the shortest unstable mode predicted by H98 (see text).
}
\label{fig:mpl9_maps}
\end{figure*}

\smallskip
We here examine the evolution of streams undergoing GI in our simulations, and in particular the properties of clumps formed within them. Regardless of whether GI is dominated by surface or body modes in the linear regime, the end result is always expected to be the collapse of dense, long-lived clumps along the stream axis \citep[N87, H98,][]{Heigl16,Heigl18}. 

\smallskip
\Fig{mpl9_maps} shows the evolution of three simulations, each with $(\Mb,\deltac)=(1.0,100)$. The left-hand column shows the no-gravity simulation with $\mu=0.9$, while the centre and right-hand columns show the gravity simulations with $\mu=0.9$ and $0.4$, respectively. By $t=2\tsc$, the non-gravitating stream has developed a well defined shear layer which has penetrated into both the background and the stream, inducing a turbulent mixing zone and diluting the stream density. Meanwhile, the interior of the stream shows numerous density fluctuations caused by turbulence and shocks, with overdensities of up to $\gsim 1.5$ times the unperturbed density. %The fraction of unmixed fluid in the stream, with $\psi>0.96$, is \nmr{$\sim 80\%$}. 
At later times the shear layer continues to grow, reaching $\hs \sim 0.4\Rs$ at $t\sim 4\tsc$, when the fraction of unmixed fluid in the stream, with $\psi>0.96$, is $\sim 50\%$. This is very similar to the no-gravity simulation shown in the left-hand column of \fig{mpl1_maps}, and is consistent with the evolution of KHI in a constant density stream with $\delta=100$ (M19, figure B1), showing that the steep density profile associated with $\mu=0.9$ does not qualitatively alter the evolution. 

\smallskip
Comparing to the corresponding self-gravitating stream, we see that the initial KHI has been suppressed by the introduction of gravity. At $t=4\tsc$, %\sim 6.2\tff \sim 1.5t_{\rm max}$,
the stream appears relatively unperturbed, with no shear layer and only minor density perturbations. The fraction of unmixed fluid in the stream is $77\%$. By $t\sim 6\tsc$, % \sim 9.2\tff\sim 2.2 t_{\rm max}$, 
small density perturbations can be seen along the stream axis, with a wavelength of $\sim 6.5\Rs$, slightly larger than the shortest unstable wavelength for GI predicted by H98\footnote{While the fastest growing mode in this case is $\lambda_{\rm max} \sim 11\Rs$, corresponding to 3 clumps, the growth rate at $\sim 6.5\Rs$ is only $0.85$ times the growth rate at $\lambda_{\rm max}$, and the resulting power spectrum is roughly flat in the range $\lambda\sim (6.5-12)\Rs$.} (\tab{sim_clump}). These density peaks are associated with an axisymmetric distortion of the stream-background interface, despite the fact the the initial perturbations had an equal mix of axisymmetric ($m=0$) and helical ($m=1$) modes. As described in \se{sgi} and \se{khi}, GI is unstable only to $m=0$ modes, while the dominant KHI mode at late times has either a long-wavelength and $m=1$ or a short wavelength and $m>1$. This supports the fact that these density perturbations were not amplified by nonlinear KHI, but rather by GI.  By $t\sim 8\tsc$, these density perturbations have evolved into five dense clumps along the box length of $32\Rs$, two of which merge by $t\sim 10\tsc$. 
%\sim 10.2\tff\sim 2.9 t_{\rm max}$
%\sim 15.4\tff\sim 3.7 t_{\rm max}$. 
%We discuss the properties of these clumps below.

\smallskip
The evolution of the lower line-mass stream, with $\mu=0.4$, is different. Despite being in the regime where GI dominates over KHI (\fig{tcompare}), at early times the evolution appears dominated by KHI. By $t\sim 4\tsc$, %\sim 7.3\tff\sim 2.8 t_{\rm max}$, 
a shear layer has developed around the stream, turbulent density fluctuations are visible, and the fraction of unmixed fluid in the stream is $65\%$.  This is because the ratio $t_{\rm max}/t_{\rm shear}$ is larger and closer to 1, allowing KHI to develop further before GI takes over. However, by $t\sim 6\tsc$, % \sim 11.0\tff \sim 4.2 t_{\rm max}$, 
GI has begun to dominate, developing an axisymmetric pattern in the stream-background interface associated with density perturbations along the stream axis, characteristic of GI but not of nonlinear KHI. By $t\sim 8\tsc$, %\sim 14.6\tff \sim 5.6 t_{\rm max}$, 
five proto-clumps are visible along the stream axis, consistent with the predicted $\lambda_{\rm cr}$. Two of these clumps merge by $t\sim 10\tsc$, %\sim 18.3\tff \sim 7.0t_{\rm max}$, 
leaving four large clumps. Assymptotically, for both $\mu=0.4$ and $0.9$, the spacing between clumps is predicted to be $\lambda_{\rm max}\sim 11\Rs$, the fastest growing GI mode, corresponding to $3$ clumps across $32\Rs$. 

\smallskip
To study the properties of clumps in the simulations, we first select all cells with stream density greater than the break in the PDF of the corresponding snapshot, $\rho_{\rm s,th}$ (\fig{dpdf}). We then group together neighbouring cells above this threshold, removing groups containing fewer than 30 cells to avoid spurious density fluctuations. Varying $\rho_{\rm s,th}$ by 0.1 dex, or using $\rho$ rather than $\rhos$, does not change the number of identified clumps, changes the clump masses by $\lsim 20\%$, and the other clump properties discussed below by $\lsim 10\%$.

\begin{figure}
\includegraphics[trim={0.38cm 0.35cm 0.34cm 0cm},clip,width=0.49\textwidth]{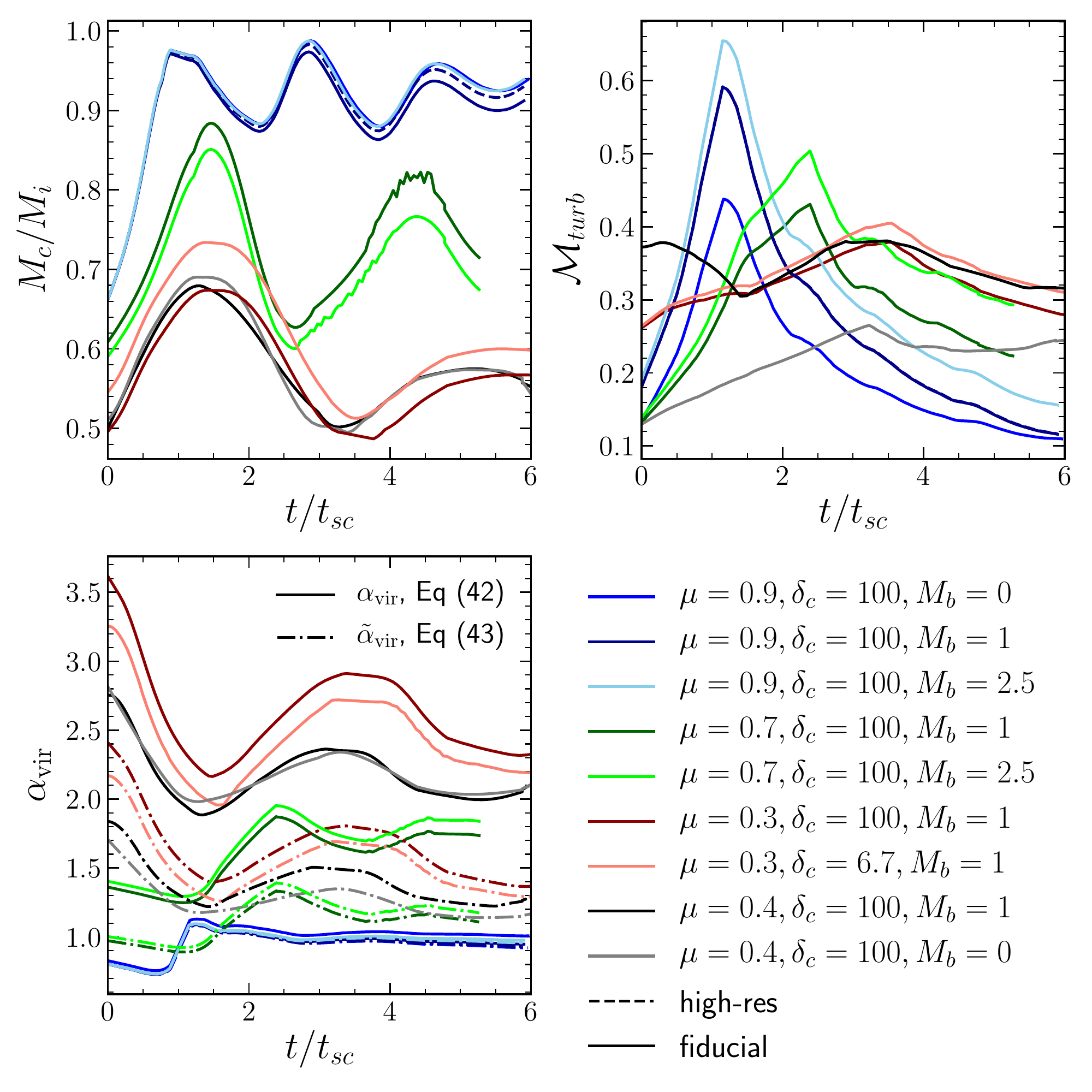}
\caption{%\nmr{A few notes on the new figure: (1) I assume the cyan line in the top right panel should be red? (2) Kindly add lines showing modified $\alpha$ for the bottom panel as well. (3) Please add the high resolution run to the $\mathcal{M}_{\rm turb}$ and $\alpha_{\rm vir}$ panels as well. (4) Do the red lines in the top row refer to $\mu=0.4$ or $\mu=0.3$? (5) The x axes label in the left-hand panels is missing. (6) You have added a simulation with $(\Mb,\deltac,\mu)=(2.5,100,0.9)$. This does not appear in \tab{sim_clump} or in \fig{tcompare}. Is this a typo in the figure legend? If not, we must add this simulation to both \tab{sim_clump} and \fig{tcompare} (7) I am not sure we really need two rows here. I think we can fit all six simulations in one row and it will still be clear. \dn{I agree that these plots are taking too much space for what they are worth. It would be better if we could combine them into one row.} Can use similar colours for each pair of simulations with similar $\mu$. Perhaps we can try it and see how it looks? (8) If you do use two rows, make sure the y axes are identical in the top and bottom rows. (9) If you do use two rows, make them tighter by only showing the x axes labels and tick numbers in the top row and reducing the white space in between the rows. } 
Evolution of clump properties, each shown as a function of time since clumps are first detected. Different coloured solid lines show different simulations as indicated in the legend. For clarity, we show results from only a few simulations bracketing the range of stream parameters examined, and thus the range of resulting clump properties. The dashed line in each panel shows the results of a simulation with $(\Mb,\deltac,\mu)=(1.0,100,0.9)$ and twice higher resolution than the fiducial value. \textit{Top-left panel:} clump mass normalised by the average initial stream mass per clump, $M_{\rm i}=M_{\rm stream}/N_{\rm clump}$. \textit{Top-right panel:} turbulent Mach number. \textit{Bottom-left panel:} clump virial parameter, with solid (dash-dotted) lines representing the virial parameter without (with) accounting for the external pressure (\equsnp{virial} and \equmnp{avir} respectively). Clumps forming in higher line-mass streams are more massive, have lower turbulent Mach numbers and lower virial parameters, though the dependence on $\Mb$ or $\deltac$ is extremely weak. For $\mu=0.9$, roughly $90\%$ of the initial stream mass winds up in clumps, which following collapse are in approximate virial equilibrium. For $\mu=0.3$, only $\lsim 50\%$ of the initial stream mass is in clumps, which are primarily confined by external pressure.
%Clumps forming in higher line-mass streams are more massive, have lower turbulent Mach numbers and lower virial parameters. This is primarily due to the larger degree of external pressure support for low line-mass streams present in the initial conditions, but also partly due to enhanced mixing and dilution in lower line-mass streams caused by more efficient KHI. For $\mu=0.9$, roughly $90\%$ of the initial stream mass winds up in clumps, which following collapse are in approximate virial equilibrium at the Jeans scale, supported primarily by thermal pressure. For $\mu=0.3$, only $\lsim 50\%$ of the initial stream mass is in clumps, the rest having mixed into the background due to KHI. The collapsed clumps are primarily confined by external pressure. In all cases, the turbulent pressure in the collapsed clumps is negligible, with turbulent Mach numbers $\sim (0.1-0.3)$ for $\mu=(0.9-0.3)$. At fixed $\mu$, the variation of clump properties with $\Mb$ and $\deltac$ is very small. \dn{Figure caption is a bit too long. Can we shorten it a bit (e.g., by refering some of the texts in the main text and minimizing overlaps)?}
}
\label{fig:evolution}
\end{figure}

\smallskip
\Fig{evolution} shows several properties of clumps identified in our simulations as a function of time, where $t=0$ is set to the first timestep where clumps have been identified. We show the clump mass, $M_{\rm c}$, the turbulent Mach number within the clumps, $\mathcal{M}_{\rm turb}=\sigma_{\rm turb}/\cs$, and the clump virial parameter, defined as 
\be
\label{eq:virial}
\alpha_{\rm vir} = \frac{5(\sigma_{\rm turb}^2+c_s^2)R}{3GM},
\end{equation}
{\no}where the factor $5/3$ comes from assuming a constant density profile inside the clump. If $\alpha_{\rm vir}\sim 1$, the clump is in virial equilibrium, while $\alpha_{\rm vir}<1$ implies the clump is collapsing and $\alpha_{\rm vir}>1$ implies it is unbound. For each property we display the average over all clumps identified in a given snapshot, typically four to five clumps.

\smallskip
Following the initial collapse when the clump mass grows significantly, it tends to saturate at a well defined value despite some oscillations. These oscillations, on the order of $\sim 10-20\%$, are due in part to our density threshold for clump cells, which is recalibrated at each snapshot. We have normalised the mass in \fig{evolution} by $M_i=M_{\rm stream}/N_{\rm clump}$, where $N_{\rm clump}$ is the number of clumps in the stream and  $M_{\rm stream}=\pi\Rs^2 L {\overline{\rhos}}$ is the initial stream mass with ${\overline{\rhos}}$ the mean density in the stream. $M_{\rm i}$ is thus the typical clump mass one would expect if the entire initial stream fragments into clumps. We find that $M_{\rm c}/M_{\rm i}$ increases with $\mu$, rising from $\sim (0.4-0.9)$ for $\mu=(0.3-0.9)$, independent of $\Mb$ or $\deltac$. 

\smallskip 
The spherical Jeans mass obtained using the average properties in the initial stream is $M_{\rm J}=(\pi^{5/2}/6){\overline{\cs}}^3 G^{-3/2} {\overline{\rhos}}^{-1/2}$. For $N_{\rm clumps}=4$ and $L=32\Rs$ we obtain $M_{\rm i}/M_{\rm J}\sim 0.14(\tsc/\tff)^3$, with $\tsc\simeq 2\Rs/{\overline{\cs}}$ and $\tff=(4G{\overline{\rhos}})^{-1/2}$. This corresponds to $M_{\rm i}/M_{\rm J}\sim (0.2-1.2)$ for $\mu=(0.3-0.9)$ (\tab{sim_clump}), yielding clump masses $M_{\rm c}\sim (0.1-1)M_{\rm J}$. 
For small $\mu$, when the density profile in the initial stream is roughly constant, the Bonnor-Ebert mass (\equnp{BE_mass}) is $M_{\rm BE}\sim 0.5M_{\rm J}$. 
%For $\mu=0.3$, ${\overline{\rhos}}\sim 0.79\rhoc\sim 1.04\rho(\Rs^-)$. Inserting this into the expression for the Bonnor-Ebert mass (\equnp{BE_mass}), together with $P_{\rm ext}=\rho(\Rs^-)\cs^2/\gamma$, yields $M_{\rm BE}\sim 0.5M_{\rm J}>M_{\rm c}$. 
In general, $M_{\rm BE}>M_{\rm c}$ for $\mu<1$.

\smallskip
The turbulent Mach number increases by a factor of $\sim 3$ as $\mu$ is decreased from 0.9 to 0.3. However, in all cases $\mathcal{M}_{\rm turb}$ is $\lsim 0.3$ asymptotically, and does not exceed $\sim 0.6$ during the initial collapse of the clump. Turbulent support is thus negligible compared to thermal pressure. The clump virial parameter increases from $\alpha_{\rm vir}\sim 1$ for $\mu=0.9$, consistent with $M_{\rm c}/M_{\rm J}\sim 1$ in this case, to $\alpha_{\rm vir}\sim 2.3$ for $\mu=0.3$. 

\smallskip
The additional support for clumps in simulations with lower values of $\mu$ comes from the external pressure, which also played a larger role in confining the initial stream. This can be seen by considering the full virial parameter including the surface pressure term \citep[e.g.][Chapter 6]{Krumholz15}. We approximate this as 
\be
\label{eq:avir}
{\tilde{\alpha}}_{\rm vir} = \frac{5(\sigma_{\rm turb}^2+c_s^2-\gamma P_{\rm ext}/\rho)R}{3GM}.
\ee
{\no}This is shown by dot-dashed lines in the rightmost panel of \fig{evolution}. For $\mu=0.9$, the external pressure is negligible and the two virial parameters are nearly identical. However, for $\mu=0.3$, ${\tilde{\alpha}}_{\rm vir}\sim 1.4$, indicating that the clumps in this case are primarily confined by external pressure. While this is still larger than 1, \equ{avir} is only an approximation, assuming a spherical clump with constant density and uniform external pressure. Properly accounting for the density profile within the clump tends to reduce the virial parameter compared to \equs{virial}-\equm{avir} \citep[e.g.][]{M17}. Given this, a value of ${\tilde{\alpha}}_{\rm vir}\sim 1.4$ is indicative of the clumps being in approximate virial equilibrium due to a combination of gravitational and pressure confinement. 

\smallskip
Contrary to the strong dependence of clump properties on $\mu$, their dependence on $(\Mb,\deltac)$ at fixed $\mu$ is extremely weak. $M_{\rm c}$ and $\alpha_{\rm vir}$ vary by only a few percent as $\deltac$ varies from $6.7-100$ or $\Mb$ from $1-2.5$. Furthermore, clumps formed in simulations of pure GI, with $(\Mb,\deltac)=(0,100)$ (see the Appendix~\se{surf_bod}), have masses only $\sim 10\%$ larger than those in simulations with $\Mb=1$ for both $\mu=0.9$ and $0.4$. We conclude that once GI dominates over KHI and leads to clump formation, KHI has little effect on the resulting clump properties even if $\mu$ only slightly exceeds $\mucr$. 
%This is expected, since smaller values of $\mu$ imply that KHI will be more prominent in the early stages of stream evolution (\fig{mpl9_maps}). This leads to more mixing of the stream and the background before fragmentation and clump formation occur, diluting the total stream mass which can form the clumps. For streams with $(\Mb,\deltac)=(1.0,100)$, we find that $M_{\rm c}/M_{\rm i}$ ranges from $\sim (0.4-0.9)$ for $\mu=(0.3-0.9)$. 

%The clump properties are also measured in simulations of $M_b=2.5,\delta_c=100,\mu=0.9$, and found that the resulting clump properties do not vary much ($<9\%$) from simulations of $M_b=1$ implying that the collapse from GI do not depend on Mach number. 

\smallskip
To check convergence, we repeated the $(\Mb,\deltac,\mu)=(1.0,100,0.9)$ simulation with a factor two higher spatial resolution, and show the results in \fig{evolution}. No significant change was found in the number of clumps, their formation time, or their properties. The clump mass increases by $\lsim 4\%$, while $\mathcal{M}_{\rm turb}$ and $\alpha_{\rm vir}$ are unchanged. We conclude that our fiducial resolution is sufficient to resolve the stream fragmentation and resulting clumps.

\smallskip
In summary, clumps forming in higher line-mass streams are more massive, have lower turbulent Mach numbers and lower virial parameters. This is primarily due to the larger degree of external pressure support for low line-mass streams present in the initial conditions, with a small contribution from enhanced mixing and dilution in lower line-mass streams caused by more efficient KHI. At fixed $\mu$, the variation of clump properties with $\Mb$ and $\deltac$ is very small. For $\mu=0.9$, roughly $90\%$ of the initial stream mass winds up in clumps, which following collapse are in approximate virial equilibrium at the thermal Jeans scale. For $\mu=0.3$, only $\lsim 50\%$ of the initial stream mass is in clumps, the rest having mixed into the background due to KHI. The collapsed clumps have $M_{\rm c}\sim (0.1-0.2)M_{\rm J}$, and are confined by external pressure. In all cases, the turbulent pressure in the collapsed clumps is negligible, with turbulent Mach numbers $\sim (0.1-0.3)$ for $\mu=(0.9-0.3)$.

%%%%%%%%%%%%%%%%%%%%%%%%%%%%%%%%%%%%%%%%%%%%%%%%%%%%%%%%%%%%
\section{discussion}
\label{sec:disc}

%!!!!!!!!!!!!!!!!!!!!!!!!!!!!!!!!!!!!!!!!!!!!!!!!!!!!!!!!!!!!
\subsection{Astrophysical Applications}
\label{sec:astr}

%Our results indicate that the previous studies of KHI and self-gravity in filamentary streams are very much applicable in each regime where the perturbation is dominated by each instability. \dn{I suggest to start by highlighting the most original and interesting results/predictions from this paper, instead of emphasizing how our work confirms the applicability of previous works.} When the line-mass is small, the streams will be disrupted by KHI as the mixing region grows and decelerates as the momentum is transferred outward. The growth rate of the mixing layer and the deceleration rate is decreased by the presence of self-gravity and the effect is more apparent as the line-mass $\mu$ increases. On the other hand, when the line-mass is large, the streams collapse due to GI. The properties of resulting clumps, however, will differ depending on the line-mass. As the line-mass gets smaller, the turbulence inside the clumps is higher, the clump is less bounded and more pressure-confined, and more masses are transported out of stream, due to some KHI activity at the start.

\smallskip
Our results on the combined evolution of KHI and GI in self-gravitating filaments have several astrophysical implications. In this section, we highlight potential applications for studies of star-forming filaments in the ISM and for cold streams feeding massive galaxies at high redshift. %Though we caution that without the inclusion of additional physical processes outlined in (\se{phys}), the application is likely to be qualitative in nature.

%---------------------------------------------------------------
\subsubsection{High-z Intergalactic Streams}

\smallskip
Massive galaxies with baryonic masses $\gsim 10^{11}\msun$ at $z\sim (1-4)$ reside in halos with virial masses $M_{\rm vir}\gsim 10^{12}\msun$. The CGM of these galaxies is thought to contain hot gas with $T\gsim 10^6\K$ in approximate hydrostatic equilibrium. However, the star-formation rates measured in these galaxies of $\gsim 100\sy$ is significantly larger than expected from the cooling of the hot CGM, and their prevalence exceeds that expected from mergers \citep{Dekel09}. As outlined in \se{intro}, such galaxies are fed by cold, $T\sim 10^4\K$ gas streams from the cosmic web, which efficiently penetrate the hot halo all the way to the central galaxy \citep{Keres05,db06,Ocvirk08,Dekel09,CDB,FG11,vdv11}. The shearing against the hot CGM makes these streams susceptible to KHI. This has motivated several detailed studies of KHI in such systems, with $\delta\sim (30-100)$ and $\Mb\sim (0.5-2)$ (M16; P18; M19). As cosmological simulations lack the spatial resolution to properly resolve KHI in the streams, these studies have been idealized, accounting thus far only for non-radiative hydrodynamics without gravity. 

\smallskip
These studies find that sufficiently narrow streams, with $\Rs/\Rv\lsim (0.005-0.05)$ where $\Rv$ is the halo virial radius, will disrupt in the CGM before reaching the central galaxy. The threshold value of $\Rs$ depends on $(\Mb,\delta)$. However, our results suggest that in a certain regime of parameter space, self-gravity may stabilize streams and halt their disruption. Even if the line mass is very low compared to the critical value, $\mu\sim 0.1$, we find that buoyancy can prevent the shear layer from penetrating the inner stream (\figs{mpl1_maps} and \figss{mixing}). For $\delta=100$, we find that the penetration rate of the shear layer into the stream is reduced by a factor of $\gsim 3$ when self-gravity is included (\fig{mixing}). This implies that the previous estimates of the upper limit on the radius of streams that can disrupt in the CGM should be reduced by a similar factor, namely $\Rs/\Rv\lsim (0.0015-0.015)$. Very narrow streams may thus survive the journey to the central galaxy, though they are likely to reach it somewhat wider and more diluted than they began. 

\smallskip
M19 also found that typical streams can significantly decelerate in the CGM, dissipating $\sim (10-50)\%$ of their bulk kinetic energy before the central galaxy. If this energy is subsequently radiated away, it can significantly contribute to the Ly$\alpha$ emission observed in the CGM of massive high-$z$ galaxies. Our results show that the self-gravity of the gas is unlikely to alter this conclusion, because the deceleration rates and the entrainment of background mass are unaffected (\figs{mixing} and \figss{momentum}). 

\smallskip
Other studies have suggested that at higher redshift, $z\gsim 5$, the streams feeding massive galaxies may be gravitationally unstable, with $\mu\sim 1$ \citep{M18a}. These authors speculated that such streams could gravitationally fragment while still in the halo, and that this could lead to the formation of metal-poor globular clusters and stars directly in the halos of high-$z$ galaxies. While this study did not account for KHI in the streams, our results suggest that this is unlikely to affect their conclusions, since for $\mu\sim 1$ and $\Mb\sim 1$, GI is unaffected by KHI (\figs{mpl9_maps} and \figss{evolution}). We note that for cosmic web filaments far from haloes, only GI operates as no shear is expected.

%---------------------------------------------------------------
\subsubsection{ISM Filaments}
\label{sec:disc_1_2}

\smallskip
As outlined in \se{intro}, numerous filametary structures are observed in the ISM, in particular in star-forming regions such as giant molecular clouds. While much attention has been payed to the gravitational stability and fragmentation of such filaments, these studies do not consider KHI induced by shearing motions between the filament and its surroundings. This is despite the fact that strong shearing motions and even signatures of KHI have been detected in molecular clouds and around filaments \citep[e.g.][]{Rodriguez92,Berne10,Berne12}. Numerical simulations of molecular clouds in the central molecular zone have also revealed strong shearing motions which generate turbulence and reduce the SFR by a factor of $\sim 7$ compared to nearby clouds \citep{Federrath16}. It is thus important to consider how KHI might affect the fragmentation of ISM filaments. 

\smallskip
We note that in this case, the shearing motion is thought to be due to a background ``wind" flowing across a roughly static filament, rather than a stream flowing through a static background. However, due to Galilean invariance, these two scenarios should behave identically.

\smallskip
The regions surrounding ISM filaments are often extremely turbulent, with turbulent Mach numbers of order 10 or higher, and the filaments themselves are often supervirial, with $\mu\gsim 1$. This is obviously very different from our initial conditions of a smooth filament in hydrostatic equilibrium (see \se{phys}). However, subvirial filaments with $\mu<1$ have been observed (\citealp{Henshaw16}; \citealp{Hacar18}, and references therein; \citealp{Orkisz19}). In some cases, these low line-mass filaments host pre-stellar cores which are at least partly supported by external pressure \citep{kirk17,Seo18}, consistent with GI in filaments with $\mucr<\mu<1$ (\fig{evolution}). If $\mu$ is known, this will constrain $\mucr$, which in turn can be used to place constraints on the properties of the confining medium, and in particular on $\Mb$, the velocity of the shear flow between the filament and the background (\fig{criticalmu}). 

%\subsubsection{Tidally Disrupted Streams}
\subsubsection{Tidal Disruption Events}
\label{sec:disc_1_3}
Stars that wander too close to a supermassive black hole, such as found in the centres of most massive galaxies, can be disrupted by the strong tidal forces exerted by the black hole \citep{Rees88}. Following the disruption, the stellar debris often evolves into a gas stream which partly accretes onto the black hole producing a luminous flare. Following their formation, the tidal shear of the black hole renders these streams gravitationally stable, so bound clumps are unlikely to form along the stream. Furthermore, the streams can be treated as approximately in hydrostatic equilibrium in the cylindrically-radial direction \citep{Coughlin2015,Coughlin2016,Coughlin2016b}. Recently, it has been argued that interactions between the debris stream and the ambient tenuous gas near the galactic centre can render such streams unstable to KHI, with nominal disruption times shorter than the infall timescale of the stream onto the black hole \citep{Bonnerot2016}. If true, this would significantly reduce the expected luminosity of the accretion flare. However, as we have shown, even in weakly self-gravitating streams, total stream disruption is significantly delayed due to buoyancy within the stream. 
%The parameters of these streams are explored in \citet{Bonnerot2016} and can be estimated to be around $M_b\approx 1, \mu=0.1-0.5, \delta<10$. 
This would mean that KHI in the streams below $\mucr$ will be stopped by buoyancy, and the decrease in the flare-luminosity predicted by \citet{Bonnerot2016} may be overestimated. Such a scenario can be tested with dedicated simulations.
%Tidally disrupted streams can form when stars falling into the galactic center gets tidally destroyed by the gravity of the black hole, resulting in a cylindrical gas stream \citep{Rees88}. These streams have been observed in various observations and has been extensively studied for gravitational instabilities. The gravitational instabilities here, however, are mostly dominated by gravity of the black hole  \citep{Coughlin2015,Coughlin2016,Coughlin2016b}. The resulting stream and its fallback into the black hole and associated energy release has been used as an indirect detection of black holes at the center of the galaxies. However, latest studies have found that these streams may experience KHI disruption which can destroy the streams and reduce the energy of accretion of stream onto black hole and luminosity detectable \citep{Bonnerot2016}. However, our results show that KH instabilities can be stopped by buoyancy before completely destroying the stream. This implies that the reduction of energy in previous study may be overestimated, and all streams will be preserved except that it will have lost about half of its mass while doing so.}

%!!!!!!!!!!!!!!!!!!!!!!!!!!!!!!!!!!!!!!!!!!!!!!!!!!!!!!!!!!!!
\subsection{Caveats and Additional Physical Effects}
\label{sec:phys}

\smallskip
While our analysis has focused on elucidating the interplay between KHI and GI in filaments, applications of our results to astrophysical scenarios require careful consideration of additional physical processes that have not yet been taken into account.
These include the assumed isentropic initial conditions and lack of radiative cooling, the assumption of line mass ratios $\mu<1$ and hydrostatic equilibrium in the initial conditions, the lack of magnetic fields, and (in the case of cold streams feeding massive galaxies at high redshift) the lack of a dark matter component to the gravitational potential. In this section, we speculate as to the possible effects of these processes, all of which will be explored in future work.

% Cooling
\smallskip
Radiative cooling is clearly very important for both ISM filaments and intergalactic gas streams. Both of these are expected to have cooling times much shorter than their sound crossing times, which is why they are often modeled as isothermal. Radiative cooling can either enhance or suppress KHI in the linear regime, depending on the slope of the cooling function and on the ratio of the cooling time in each fluid to the sound crossing time \citep{Massaglia92,Bodo93,Vietri97,Hardee97,Xu00}. However, when these ratios are either much larger or much smaller than unity, the linear growth rates are similar to the adiabatic case at longitudinal wavelengths $\lambda\gsim \Rs$ (Mandelker et al., in prep.). Even in this case, cooling can substantially alter the nonlinear evolution of KHI \citep{Vietri97,Stone97,Xu00,Micono00}, though the net effect again depends on details of the cooling function and the stream parameters. Some authors have found that cooling leads to more violent disruption of the stream \citep{Stone97,Xu00}, while others have found that it prevents stream disruption by limiting the penetration of the shear layer into the stream \citep{Vietri97,Micono00}. 
%Stream deceleration, however, is found to be similar to the adiabatic case \citep{Micono00}. 
If shear layer growth is suppressed and the contact discontinuity maintained, then $t_{\rm shear}$ will increase and $\mucr$ will decrease (\fig{tcompare}). Thus, the regime where GI dominates over KHI will expand. Furthermore, it is also found that KHI in a cooling medium leads to much larger density fluctuations, and to the formation of dense knots and filaments inside the stream. These are likely to further enhance GI and filament fragmentation. Cooling is also likely to allow the clumps to collapse to higher densities and reach lower temperatures, thus decreasing their Jeans mass and leading to further fragmentation and collapse.

% magnetic fields
\smallskip
Magnetic fields are likely to be dynamically important in ISM filaments. This can have a stabilizing effect on GI, especially when $\mu<1$ (e.g. N87, H98), and also on KHI, where magnetic fields parallel to the flow have been found to stabilise high-$m$ modes and suppress shear layer growth \citep{Ferrari81,Birkinshaw90}. It is therefore unclear what the net effect will be in terms of the competition between these two processes, and this will likely depend sensitively on the properties of the field. For intergalactic gas streams at high redshift, magnetic fields are likely dynamically unimportant \citep[e.g.][]{Bagchi02}. Nevertheless, they may significantly weaken thermal conductivity and viscosity, which will influence the width of the shear layer (M19) and thus affect the instability. All these effects should be accounted for simultaneously in future work.

% dark matter
\smallskip
When considering intergalactic gas streams, we must also account for the contribution of the host dark matter filament to the gravitational potential. To our knowledge, the gravitational stability of a gas stream embedded in a dark matter filament has not been studied. The dark matter may stabilise the stream by making it more buoyant, or it may destabilise the stream by increasing the inward radial gravitational force, thus requiring non-thermal turbulent motions to support the stream against radial collapse. This may also suppress KHI by further limiting shear layer growth and stream disruption (see \figs{mpl1_maps}-\figss{mixing}). The central dark matter halo into which the streams are flowing will also affect their evolution. The central potential focuses the stream into a conical shape with its radius decreasing towards the halo centre, $\Rs\propto r$. \citep{Dekel09,Voort12}. This decreases the KHI timescales, which are proportional to $\Rs$ (\equsnp{tau_diss}-\equmnp{tau_shear}). However, this focusing also increases the stream density, with $\rho\propto \Rs^{-2}\propto r^{-2}$, resulting in a decrease of the free-fall time, $\tff\propto \rho^{-1/2}\propto r$. Since $t_{\rm max}\propto \tff$, the ratio $t_{\rm max}/t_{\rm shear}$ is unlikely to vary significantly throughout the halo, as is the critical line-mass ratio, $\mucr$. However, this must be studied in more detail, as must the effect of gravitational acceleration towards the halo centre on the evolution of KHI and GI in intergalactic cold streams.

% hydrostatic equilibrium, turbulence
\smallskip
Throughout our analysis, we assumed that filaments began in hydrostatic equilibrium, and without any internal non-thermal support such as turbulence or vorticity. This is unlikely to be the case for either ISM filaments or intergalactic streams. Theoretical studies of GI in ISM filaments growing self-consistently via radial accretion have shown that turbulence builds up inside the stream with Mach numbers of order unity and contributes to its support \citep{Heitsch13,Clarke16,Clarke17,Heigl18b}. Despite this, the filament was found to fragment when its line mass reached the critical value for hydrostatic equilibrium, namely at $\mu\gsim 1$, in a similar manner to the $\mu<1$ filaments considered here, leading to the formation of Jeans-scale clumps \citep{Clarke16,Clarke17}. It is unclear how these results will change in the presence of KHI. Likewise, it has been suggested that accretion onto cosmic gas streams from the intergalactic medium creates specific profiles \citep{FG84,Birnboim16}, induces roughly sonic turbulence \citep{M18a} and vorticity \citep{Codis12,Codis15,Laigle15}, and grows streams to $\mu>1$ \citep{M18a}. Such non-equilibrium effects must be considered in order to describe stream evolution.

%%%%%%%%%%%%%%%%%%%%%%%%%%%%%%%%%%%%%%%%%%%%%%%%%%%%%%
\section{Summary and Conclusions}
\label{sec:conc}

\smallskip
Self-gravitating gaseous filaments are ubiquitous in astrophysics, from sub-pc filaments within the interstellar medium, to Mpc scale streams feeding galaxies along the cosmic web. As such, they may be subject to gravitational instability (GI), which leads to stream fragmentation and to the formation of long-lived, collapsed clumps along the stream axis. In many cases, such filaments are also susceptible to Kelvin-Helmholtz Instability (KHI) due to a shear flow against a confining background medium, which acts to mix the filament with the background fluid via a turbulent shear layer. Motivated by this, we have performed the first ever study of the evolution of a self-gravitating filament or stream undergoing KHI, using simple analytic models and hydrodynamic simulations. Such a system is characterised by three dimensionless parameters: the Mach number of the stream with respect to the sound speed in the (static) background, $\Mb$, the ratio of the central density in the stream to the background density outside the stream, $\deltac$, and the ratio of the mass-per-unit-length (line-mass) of the stream to the maximal line-mass for which initial hydrostatic equilibrium is possible, $\mu$. The current analysis is restricted to filaments with $\mu<1$ initially in hydrostatic equilibrium. 
Our main results can be summarised as follows:

\begin{enumerate}

\smallskip
\item The competition between GI and KHI is governed by the ratio of the timescale for linear growth of the fastest growing GI mode, $t_{\rm max}$, and the relevant nonlinear KHI timescale. When GI is dominated by surface modes, this is the time for the KHI-induced shear layer to expand to a size comparable to the stream radius and destroy the initial contact discontinuity, $t_{\rm shear}$. If $t_{\rm max}/t_{\rm shear}<1$, GI causes the stream to fragment into long-lived clumps and suppresses mixing with the background medium. Likewise, if $t_{\rm max}/t_{\rm shear}>1$, KHI mixes the stream with the background medium, dilutes its density and suppresses clump formation (\fig{tcompare}). Regardless, the stream is always unstable. When GI is dominated by body modes, clumps may form even when $t_{\rm max}$ is slightly longer than $t_{\rm shear}$, since the contact discontinuity no longer plays a role in GI.

\smallskip
\item The timescale criterion can be rephrased as a criterion on the line-mass ratio $\mu$. If this is smaller than a critical value which depends on the Mach number and density contrast, $\mucr(\Mb,\deltac)$, then KHI will win and mix the stream and background. However, if $\mu>\mucr(\Mb,\deltac)$, the stream will fragment into long-lived, bound clumps. $\mucr$ increases strongly with $\Mb$, and has a weak tendency to increase with $\deltac$ (\fig{criticalmu}). For $\Mb\lsim 2.5$ we have $\mucr\lsim 0.5$. At larger Mach numbers, when KHI is dominated by high-order azimuthal surface modes, $\mucr\lsim 0.9$. In practice, values of $\mucr>0.9$ are not relevant, as GI body modes will cause clump formation even if $\mu\lsim \mucr$.

\smallskip
\item When $\mu<\mucr$, the evolution of KHI outside the stream boundary is similar to the case of a non-gravitating uniform density stream, studied in detail by \citet{M18b}. Self-gravity slows the expansion of the shear layer into the background by less than $20\%$ for large $\mu$, and significantly less than that for smaller $\mu$ (\fig{mixing}). Consequently, the stream deceleration due to entrainment of background mass in the shear layer is also unaffected, and follows the analytical prediction (\fig{momentum}). 

\smallskip
\item However, gravity does qualitatively affect the penetration of the shear layer into the stream. At $t\gsim (2-3)\tsc$, the penetration rate of the shear layer into the stream is slowed by a factor of $\gsim 3$ compared to the no-gravity case (\fig{mixing}). This is due to restoring buoyancy forces in the stream interior, corresponding to values of the Richardson number, ${\rm Ri}>0.25$ (\fig{Ri}). This significantly slows the final disruption of the stream by KHI, as a dense central core remains partly shielded against mixing (\fig{mpl1_maps}).

\smallskip
\item 
The clumps that form by GI when $\mu>\mucr$ are largely unaffected by KHI. They are typically less massive than the Jeans mass, and supported partially by external pressure. However, as $\mu\rightarrow 1$ the clumps approach the Jeans mass and the external pressure support becomes negligible. In all cases, the internal turbulent motions are subsonic and turbulent pressure support is negligible, though the turbulent Mach number increases towards lower $\mu$ (\fig{evolution}). KHI seems to have a minor effect on the clump properties, which are largely insensitive to the Mach number of the flow, even in the static limit, $\Mb=0$.

\smallskip
\item Our finding that self-gravity may shield the inner core of filaments from disruption by KHI, implies that recent studies of KHI in gas streams feeding massive galaxies at high-$z$ may have overestimated the disruption of these streams in the CGM. However, the dissipation and deceleration rates should not be affected. Additionally, our finding that GI induced fragmentation only occurs when $\mu>\mucr(\Mb,\deltac)$ can be used to place constraints on the properties and kinematics of the confining medium surrounding low mass filaments in the ISM. However, in order to properly address these phenomena, additional physics such as radiative cooling, magnetic fields, external gravitational potential, and non-thermal turbulent motions, will have to be added to our models.
\end{enumerate}

%%%%%%%%%%%%%%%%%%%%%%%%%%%%%%%%%%%%%%%%%%%%%%%%%%%%%%%%
\section*{Acknowledgments} 
%We thank Frank van den Bosch, Andreas Burkert, Diederik Kruijssen, and Xun Shi for insightful conversations, and for their thoughts on earlier versions of this manuscript. We thank Shuo Kong for helpful discussions. 
We thank Romain Teyssier for many helpful suggestions while running the simulations. We thank Frank van den Bosch, Frederic Bournaud, Andreas Burkert, Drummond Fielding, Shuo Kong, Diederik Kruijssen, and Xun Shi for insightful discussions. NM acknowledges support from the Klaus Tschira Foundation through the HITS Yale Program in Astrophysics (HYPA). The simulations were performed on the Omega and Grace HPC clusters at Yale. This work is supported in part by the facilities and staff of the Yale Center for Research Computing. AD was partly supported by the grants BSF 2014-273, GIF I-1341-303.7/2016 and NSF AST-1405962.

%%%%%%%%%%%%%%%%%%%%%%%%%%%%%%%%%%%%%%%%%%%%%%%%%%%%%%%%
\bibliography{biblio}

\begin{thebibliography}{}
\makeatletter
\relax
\def\mn@urlcharsother{\let\do\@makeother \do\$\do\&\do\#\do\^\do\_\do\%\do\~}
\def\mn@doi{\begingroup\mn@urlcharsother \@ifnextchar [ {\mn@doi@}
  {\mn@doi@[]}}
\def\mn@doi@[#1]#2{\def\@tempa{#1}\ifx\@tempa\@empty \href
  {http://dx.doi.org/#2} {doi:#2}\else \href {http://dx.doi.org/#2} {#1}\fi
  \endgroup}
\def\mn@eprint#1#2{\mn@eprint@#1:#2::\@nil}
\def\mn@eprint@arXiv#1{\href {http://arxiv.org/abs/#1} {{\tt arXiv:#1}}}
\def\mn@eprint@dblp#1{\href {http://dblp.uni-trier.de/rec/bibtex/#1.xml}
  {dblp:#1}}
\def\mn@eprint@#1:#2:#3:#4\@nil{\def\@tempa {#1}\def\@tempb {#2}\def\@tempc
  {#3}\ifx \@tempc \@empty \let \@tempc \@tempb \let \@tempb \@tempa \fi \ifx
  \@tempb \@empty \def\@tempb {arXiv}\fi \@ifundefined
  {mn@eprint@\@tempb}{\@tempb:\@tempc}{\expandafter \expandafter \csname
  mn@eprint@\@tempb\endcsname \expandafter{\@tempc}}}

\bibitem[\protect\citeauthoryear{{Andr{\'e}} et~al.,}{{Andr{\'e}}
  et~al.}{2010}]{Andre10}
{Andr{\'e}} P.,  et~al., 2010, \mn@doi [\aap] {10.1051/0004-6361/201014666},
  \href {http://adsabs.harvard.edu/abs/2010A%26A...518L.102A} {518, L102}

\bibitem[\protect\citeauthoryear{{Andr{\'e}}, {Di Francesco}, {Ward-Thompson},
  {Inutsuka}, {Pudritz}  \& {Pineda}}{{Andr{\'e}} et~al.}{2014}]{Andre14}
{Andr{\'e}} P.,  {Di Francesco} J.,  {Ward-Thompson} D.,  {Inutsuka} S.-I.,
  {Pudritz} R.~E.,   {Pineda} J.~E.,  2014, \mn@doi [Protostars and Planets VI]
  {10.2458/azu_uapress_9780816531240-ch002}, \href
  {http://adsabs.harvard.edu/abs/2014prpl.conf...27A} {pp 27--51}

\bibitem[\protect\citeauthoryear{{Arrigoni Battaia}, {Prochaska}, {Hennawi},
  {Obreja}, {Buck}, {Cantalupo}, {Dutton}  \& {Macci{\`o}}}{{Arrigoni Battaia}
  et~al.}{2018}]{Arrigoni18}
{Arrigoni Battaia} F.,  {Prochaska} J.~X.,  {Hennawi} J.~F.,  {Obreja} A.,
  {Buck} T.,  {Cantalupo} S.,  {Dutton} A.~A.,   {Macci{\`o}} A.~V.,  2018,
  \mn@doi [\mnras] {10.1093/mnras/stx2465}, \href
  {http://adsabs.harvard.edu/abs/2018MNRAS.473.3907A} {473, 3907}

\bibitem[\protect\citeauthoryear{{Arzoumanian} et~al.,}{{Arzoumanian}
  et~al.}{2011}]{Arzoumanian11}
{Arzoumanian} D.,  et~al., 2011, \mn@doi [\aap] {10.1051/0004-6361/201116596},
  \href {http://adsabs.harvard.edu/abs/2011A%26A...529L...6A} {529, L6}

\bibitem[\protect\citeauthoryear{{Bagchi}, {Ensslin}, {Miniati}, {Stalin},
  {Singh}, {Raychaudhury}  \& {Humeshkar}}{{Bagchi} et~al.}{2002}]{Bagchi02}
{Bagchi} J.,  {Ensslin} T.~A.,  {Miniati} F.,  {Stalin} C.~S.,  {Singh} M.,
  {Raychaudhury} S.,   {Humeshkar} N.~B.,  2002, \mn@doi [New Astronomy]
  {10.1016/S1384-1076(02)00137-9}, \href
  {http://adsabs.harvard.edu/abs/2002NewA....7..249B} {7, 249}

\bibitem[\protect\citeauthoryear{{Banerjee}, {V{\'a}zquez-Semadeni},
  {Hennebelle}  \& {Klessen}}{{Banerjee} et~al.}{2009}]{Banerjee09}
{Banerjee} R.,  {V{\'a}zquez-Semadeni} E.,  {Hennebelle} P.,   {Klessen} R.~S.,
   2009, \mn@doi [\mnras] {10.1111/j.1365-2966.2009.15115.x}, \href
  {http://adsabs.harvard.edu/abs/2009MNRAS.398.1082B} {398, 1082}

\bibitem[\protect\citeauthoryear{{Bassett} \& {Woodward}}{{Bassett} \&
  {Woodward}}{1995}]{Bassett95}
{Bassett} G.~M.,  {Woodward} P.~R.,  1995, \mn@doi [\apj] {10.1086/175385},
  \href {http://adsabs.harvard.edu/abs/1995ApJ...441..582B} {441, 582}

\bibitem[\protect\citeauthoryear{{Bern{\'e}} \& {Matsumoto}}{{Bern{\'e}} \&
  {Matsumoto}}{2012}]{Berne12}
{Bern{\'e}} O.,  {Matsumoto} Y.,  2012, \mn@doi [\apjl]
  {10.1088/2041-8205/761/1/L4}, \href
  {http://adsabs.harvard.edu/abs/2012ApJ...761L...4B} {761, L4}

\bibitem[\protect\citeauthoryear{{Bern{\'e}}, {Marcelino}  \&
  {Cernicharo}}{{Bern{\'e}} et~al.}{2010}]{Berne10}
{Bern{\'e}} O.,  {Marcelino} N.,   {Cernicharo} J.,  2010, \mn@doi [\nat]
  {10.1038/nature09289}, \href
  {http://adsabs.harvard.edu/abs/2010Natur.466..947B} {466, 947}

\bibitem[\protect\citeauthoryear{{Birkinshaw}}{{Birkinshaw}}{1984}]{Birkinshaw84}
{Birkinshaw} M.,  1984, \mn@doi [\mnras] {10.1093/mnras/208.4.887}, \href
  {http://adsabs.harvard.edu/abs/1984MNRAS.208..887B} {208, 887}

\bibitem[\protect\citeauthoryear{{Birkinshaw}}{{Birkinshaw}}{1990}]{Birkinshaw90}
{Birkinshaw} M.,  1990, {The Stability of Jets}

\bibitem[\protect\citeauthoryear{{Birnboim}, {Padnos}  \& {Zinger}}{{Birnboim}
  et~al.}{2016}]{Birnboim16}
{Birnboim} Y.,  {Padnos} D.,   {Zinger} E.,  2016, \mn@doi [\apjl]
  {10.3847/2041-8205/832/1/L4}, \href
  {http://adsabs.harvard.edu/abs/2016ApJ...832L...4B} {832, L4}

\bibitem[\protect\citeauthoryear{{Bodo}, {Massaglia}, {Rossi}, {Trussoni}  \&
  {Ferrari}}{{Bodo} et~al.}{1993}]{Bodo93}
{Bodo} G.,  {Massaglia} S.,  {Rossi} P.,  {Trussoni} E.,   {Ferrari} A.,  1993,
  \mn@doi [Physics of Fluids] {10.1063/1.858863}, \href
  {http://adsabs.harvard.edu/abs/1993PhFl....5..405B} {5, 405}

\bibitem[\protect\citeauthoryear{{Bodo}, {Rossi}, {Massaglia}, {Ferrari},
  {Malagoli}  \& {Rosner}}{{Bodo} et~al.}{1998}]{Bodo98}
{Bodo} G.,  {Rossi} P.,  {Massaglia} S.,  {Ferrari} A.,  {Malagoli} A.,
  {Rosner} R.,  1998, \aap, \href
  {http://adsabs.harvard.edu/abs/1998A%26A...333.1117B} {333, 1117}

\bibitem[\protect\citeauthoryear{{Bogey}, {Marsden}  \& {Bailly}}{{Bogey}
  et~al.}{2011}]{Bogey11}
{Bogey} C.,  {Marsden} O.,   {Bailly} C.,  2011, \mn@doi [Physics of Fluids]
  {10.1063/1.3555634}, \href
  {http://adsabs.harvard.edu/abs/2011PhFl...23c5104B} {23, 035104}

\bibitem[\protect\citeauthoryear{{Bond}, {Kofman}  \& {Pogosyan}}{{Bond}
  et~al.}{1996}]{Bond96}
{Bond} J.~R.,  {Kofman} L.,   {Pogosyan} D.,  1996, \mn@doi [\nat]
  {10.1038/380603a0}, \href {http://adsabs.harvard.edu/abs/1996Natur.380..603B}
  {380, 603}

\bibitem[\protect\citeauthoryear{{Bonnerot}, {Rossi}  \& {Lodato}}{{Bonnerot}
  et~al.}{2016}]{Bonnerot2016}
{Bonnerot} C.,  {Rossi} E.~M.,   {Lodato} G.,  2016, \mn@doi [\mnras]
  {10.1093/mnras/stw486}, \href
  {http://adsabs.harvard.edu/abs/2016MNRAS.458.3324B} {458, 3324}

\bibitem[\protect\citeauthoryear{{Bonnor}}{{Bonnor}}{1956}]{Bonnor56}
{Bonnor} W.~B.,  1956, \mn@doi [\mnras] {10.1093/mnras/116.3.351}, \href
  {http://adsabs.harvard.edu/abs/1956MNRAS.116..351B} {116, 351}

\bibitem[\protect\citeauthoryear{{Borisova} et~al.,}{{Borisova}
  et~al.}{2016}]{Borisova16}
{Borisova} E.,  et~al., 2016, \mn@doi [\apj] {10.3847/0004-637X/831/1/39},
  \href {http://adsabs.harvard.edu/abs/2016ApJ...831...39B} {831, 39}

\bibitem[\protect\citeauthoryear{{Bouch{\'e}}, {Murphy}, {Kacprzak},
  {P{\'e}roux}, {Contini}, {Martin}  \& {Dessauges-Zavadsky}}{{Bouch{\'e}}
  et~al.}{2013}]{Bouche13}
{Bouch{\'e}} N.,  {Murphy} M.~T.,  {Kacprzak} G.~G.,  {P{\'e}roux} C.,
  {Contini} T.,  {Martin} C.~L.,   {Dessauges-Zavadsky} M.,  2013, \mn@doi
  [Science] {10.1126/science.1234209}, \href
  {http://adsabs.harvard.edu/abs/2013Sci...341...50B} {341, 50}

\bibitem[\protect\citeauthoryear{{Bouch{\'e}} et~al.,}{{Bouch{\'e}}
  et~al.}{2016}]{Bouche16}
{Bouch{\'e}} N.,  et~al., 2016, \mn@doi [\apj] {10.3847/0004-637X/820/2/121},
  \href {http://adsabs.harvard.edu/abs/2016ApJ...820..121B} {820, 121}

\bibitem[\protect\citeauthoryear{{Cantalupo}, {Arrigoni-Battaia}, {Prochaska},
  {Hennawi}  \& {Madau}}{{Cantalupo} et~al.}{2014}]{Cantalupo14}
{Cantalupo} S.,  {Arrigoni-Battaia} F.,  {Prochaska} J.~X.,  {Hennawi} J.~F.,
  {Madau} P.,  2014, \mn@doi [\nat] {10.1038/nature12898}, \href
  {http://adsabs.harvard.edu/abs/2014Natur.506...63C} {506, 63}

\bibitem[\protect\citeauthoryear{{Ceverino}, {Dekel}  \& {Bournaud}}{{Ceverino}
  et~al.}{2010}]{CDB}
{Ceverino} D.,  {Dekel} A.,   {Bournaud} F.,  2010, \mn@doi [\mnras]
  {10.1111/j.1365-2966.2010.16433.x}, \href
  {http://adsabs.harvard.edu/abs/2010MNRAS.404.2151C} {404, 2151}

\bibitem[\protect\citeauthoryear{{Chandrasekhar} \& {Fermi}}{{Chandrasekhar} \&
  {Fermi}}{1953}]{Chandrasekhar53}
{Chandrasekhar} S.,  {Fermi} E.,  1953, \mn@doi [\apj] {10.1086/145732}, \href
  {http://adsabs.harvard.edu/abs/1953ApJ...118..116C} {118, 116}

\bibitem[\protect\citeauthoryear{{Clarke}, {Whitworth}  \& {Hubber}}{{Clarke}
  et~al.}{2016}]{Clarke16}
{Clarke} S.~D.,  {Whitworth} A.~P.,   {Hubber} D.~A.,  2016, \mn@doi [\mnras]
  {10.1093/mnras/stw407}, \href
  {http://adsabs.harvard.edu/abs/2016MNRAS.458..319C} {458, 319}

\bibitem[\protect\citeauthoryear{{Clarke}, {Whitworth}, {Duarte-Cabral}  \&
  {Hubber}}{{Clarke} et~al.}{2017}]{Clarke17}
{Clarke} S.~D.,  {Whitworth} A.~P.,  {Duarte-Cabral} A.,   {Hubber} D.~A.,
  2017, \mn@doi [\mnras] {10.1093/mnras/stx637}, \href
  {http://adsabs.harvard.edu/abs/2017MNRAS.468.2489C} {468, 2489}

\bibitem[\protect\citeauthoryear{{Codis}, {Pichon}, {Devriendt}, {Slyz},
  {Pogosyan}, {Dubois}  \& {Sousbie}}{{Codis} et~al.}{2012}]{Codis12}
{Codis} S.,  {Pichon} C.,  {Devriendt} J.,  {Slyz} A.,  {Pogosyan} D.,
  {Dubois} Y.,   {Sousbie} T.,  2012, \mn@doi [\mnras]
  {10.1111/j.1365-2966.2012.21636.x}, \href
  {http://adsabs.harvard.edu/abs/2012MNRAS.427.3320C} {427, 3320}

\bibitem[\protect\citeauthoryear{{Codis}, {Pichon}  \& {Pogosyan}}{{Codis}
  et~al.}{2015}]{Codis15}
{Codis} S.,  {Pichon} C.,   {Pogosyan} D.,  2015, \mn@doi [\mnras]
  {10.1093/mnras/stv1570}, \href
  {http://adsabs.harvard.edu/abs/2015MNRAS.452.3369C} {452, 3369}

\bibitem[\protect\citeauthoryear{{Colless} et~al.,}{{Colless}
  et~al.}{2003}]{Colless03}
{Colless} M.,  et~al., 2003, VizieR Online Data Catalog, \href
  {http://adsabs.harvard.edu/abs/2003yCat.7226....0C} {7226}

\bibitem[\protect\citeauthoryear{{Coughlin} \& {Nixon}}{{Coughlin} \&
  {Nixon}}{2015}]{Coughlin2015}
{Coughlin} E.~R.,  {Nixon} C.,  2015, \mn@doi [\apjl]
  {10.1088/2041-8205/808/1/L11}, \href
  {http://adsabs.harvard.edu/abs/2015ApJ...808L..11C} {808, L11}

\bibitem[\protect\citeauthoryear{{Coughlin}, {Nixon}, {Begelman}, {Armitage}
  \& {Price}}{{Coughlin} et~al.}{2016a}]{Coughlin2016b}
{Coughlin} E.~R.,  {Nixon} C.,  {Begelman} M.~C.,  {Armitage} P.~J.,   {Price}
  D.~J.,  2016a, \mn@doi [\mnras] {10.1093/mnras/stv2511}, \href
  {http://adsabs.harvard.edu/abs/2016MNRAS.455.3612C} {455, 3612}

\bibitem[\protect\citeauthoryear{{Coughlin}, {Nixon}, {Begelman}  \&
  {Armitage}}{{Coughlin} et~al.}{2016b}]{Coughlin2016}
{Coughlin} E.~R.,  {Nixon} C.,  {Begelman} M.~C.,   {Armitage} P.~J.,  2016b,
  \mn@doi [\mnras] {10.1093/mnras/stw770}, \href
  {http://adsabs.harvard.edu/abs/2016MNRAS.459.3089C} {459, 3089}

\bibitem[\protect\citeauthoryear{{Danovich}, {Dekel}, {Hahn}  \&
  {Teyssier}}{{Danovich} et~al.}{2012}]{Danovich12}
{Danovich} M.,  {Dekel} A.,  {Hahn} O.,   {Teyssier} R.,  2012, \mn@doi
  [\mnras] {10.1111/j.1365-2966.2012.20751.x}, \href
  {http://adsabs.harvard.edu/abs/2012MNRAS.422.1732D} {422, 1732}

\bibitem[\protect\citeauthoryear{{Dekel} \& {Birnboim}}{{Dekel} \&
  {Birnboim}}{2006}]{db06}
{Dekel} A.,  {Birnboim} Y.,  2006, \mn@doi [\mnras]
  {10.1111/j.1365-2966.2006.10145.x}, \href
  {http://adsabs.harvard.edu/abs/2006MNRAS.368....2D} {368, 2}

\bibitem[\protect\citeauthoryear{{Dekel} et~al.,}{{Dekel}
  et~al.}{2009a}]{Dekel09}
{Dekel} A.,  et~al., 2009a, \mn@doi [\nat] {10.1038/nature07648}, \href
  {http://adsabs.harvard.edu/abs/2009Natur.457..451D} {457, 451}

\bibitem[\protect\citeauthoryear{{Dekel}, {Sari}  \& {Ceverino}}{{Dekel}
  et~al.}{2009b}]{DSC}
{Dekel} A.,  {Sari} R.,   {Ceverino} D.,  2009b, \mn@doi [\apj]
  {10.1088/0004-637X/703/1/785}, \href
  {http://adsabs.harvard.edu/abs/2009ApJ...703..785D} {703, 785}

\bibitem[\protect\citeauthoryear{{Dimotakis}}{{Dimotakis}}{1991}]{Dimotakis}
{Dimotakis} P.~E.,  1991, Turbulent free shear layer mixing and combustion.
  Tech. rep.

\bibitem[\protect\citeauthoryear{{Ebert}}{{Ebert}}{1955}]{Ebert55}
{Ebert} R.,  1955, \zap, \href
  {http://adsabs.harvard.edu/abs/1955ZA.....37..217E} {37, 217}

\bibitem[\protect\citeauthoryear{{Elmegreen}}{{Elmegreen}}{2011}]{Elmegreen11}
{Elmegreen} B.~G.,  2011, \mn@doi [\apj] {10.1088/0004-637X/731/1/61}, \href
  {http://adsabs.harvard.edu/abs/2011ApJ...731...61E} {731, 61}

\bibitem[\protect\citeauthoryear{{Faucher-Gigu{\`e}re}, {Kere{\v s}}  \&
  {Ma}}{{Faucher-Gigu{\`e}re} et~al.}{2011}]{FG11}
{Faucher-Gigu{\`e}re} C.-A.,  {Kere{\v s}} D.,   {Ma} C.-P.,  2011, \mn@doi
  [\mnras] {10.1111/j.1365-2966.2011.19457.x}, \href
  {http://adsabs.harvard.edu/abs/2011MNRAS.417.2982F} {417, 2982}

\bibitem[\protect\citeauthoryear{{Federrath} \& {Banerjee}}{{Federrath} \&
  {Banerjee}}{2015}]{Federrath15}
{Federrath} C.,  {Banerjee} S.,  2015, \mn@doi [\mnras] {10.1093/mnras/stv180},
  \href {http://adsabs.harvard.edu/abs/2015MNRAS.448.3297F} {448, 3297}

\bibitem[\protect\citeauthoryear{{Federrath} et~al.,}{{Federrath}
  et~al.}{2016}]{Federrath16}
{Federrath} C.,  et~al., 2016, \mn@doi [\apj] {10.3847/0004-637X/832/2/143},
  \href {http://adsabs.harvard.edu/abs/2016ApJ...832..143F} {832, 143}

\bibitem[\protect\citeauthoryear{{Ferrari}, {Trussoni}  \&
  {Zaninetti}}{{Ferrari} et~al.}{1981}]{Ferrari81}
{Ferrari} A.,  {Trussoni} E.,   {Zaninetti} L.,  1981, \mn@doi [\mnras]
  {10.1093/mnras/196.4.1051}, \href
  {http://adsabs.harvard.edu/abs/1981MNRAS.196.1051F} {196, 1051}

\bibitem[\protect\citeauthoryear{{Fillmore} \& {Goldreich}}{{Fillmore} \&
  {Goldreich}}{1984}]{FG84}
{Fillmore} J.~A.,  {Goldreich} P.,  1984, \mn@doi [\apj] {10.1086/162070},
  \href {http://adsabs.harvard.edu/abs/1984ApJ...281....1F} {281, 1}

\bibitem[\protect\citeauthoryear{{Fischera} \& {Martin}}{{Fischera} \&
  {Martin}}{2012}]{Fischera12}
{Fischera} J.,  {Martin} P.~G.,  2012, \mn@doi [\aap]
  {10.1051/0004-6361/201218961}, \href
  {http://adsabs.harvard.edu/abs/2012A%26A...542A..77F} {542, A77}

\bibitem[\protect\citeauthoryear{{Freundlich}, {Jog}  \& {Combes}}{{Freundlich}
  et~al.}{2014}]{Freundlich14}
{Freundlich} J.,  {Jog} C.~J.,   {Combes} F.,  2014, \mn@doi [\aap]
  {10.1051/0004-6361/201323325}, \href
  {http://adsabs.harvard.edu/abs/2014A%26A...564A...7F} {564, A7}

\bibitem[\protect\citeauthoryear{{Fridman} \& {Poliachenko}}{{Fridman} \&
  {Poliachenko}}{1984}]{Fridman84}
{Fridman} A.~M.,  {Poliachenko} V.~L.,  1984, {Physics of gravitating systems.
  II - Nonlinear collective processes: Nonlinear waves, solitons, collisionless
  shocks, turbulence. Astrophysical applications}

\bibitem[\protect\citeauthoryear{{Fumagalli} et~al.,}{{Fumagalli}
  et~al.}{2017}]{Fumagalli17}
{Fumagalli} M.,  et~al., 2017, \mn@doi [\mnras] {10.1093/mnras/stx1896}, \href
  {http://adsabs.harvard.edu/abs/2017MNRAS.471.3686F} {471, 3686}

\bibitem[\protect\citeauthoryear{{Genel}, {Dekel}  \& {Cacciato}}{{Genel}
  et~al.}{2012}]{Genel12}
{Genel} S.,  {Dekel} A.,   {Cacciato} M.,  2012, \mn@doi [\mnras]
  {10.1111/j.1365-2966.2012.21652.x}, \href
  {http://adsabs.harvard.edu/abs/2012MNRAS.tmp.3466G} {p.~3466}

\bibitem[\protect\citeauthoryear{{Ginolfi} et~al.,}{{Ginolfi}
  et~al.}{2017}]{Ginolfi17}
{Ginolfi} M.,  et~al., 2017, \mn@doi [\mnras] {10.1093/mnras/stx712}, \href
  {http://adsabs.harvard.edu/abs/2017MNRAS.468.3468G} {468, 3468}

\bibitem[\protect\citeauthoryear{{G{\'o}mez} \&
  {V{\'a}zquez-Semadeni}}{{G{\'o}mez} \&
  {V{\'a}zquez-Semadeni}}{2014}]{Gomez14}
{G{\'o}mez} G.~C.,  {V{\'a}zquez-Semadeni} E.,  2014, \mn@doi [\apj]
  {10.1088/0004-637X/791/2/124}, \href
  {http://adsabs.harvard.edu/abs/2014ApJ...791..124G} {791, 124}

\bibitem[\protect\citeauthoryear{{Hacar}, {Tafalla}, {Forbrich}, {Alves},
  {Meingast}, {Grossschedl}  \& {Teixeira}}{{Hacar} et~al.}{2018}]{Hacar18}
{Hacar} A.,  {Tafalla} M.,  {Forbrich} J.,  {Alves} J.,  {Meingast} S.,
  {Grossschedl} J.,   {Teixeira} P.~S.,  2018, \mn@doi [\aap]
  {10.1051/0004-6361/201731894}, \href
  {http://adsabs.harvard.edu/abs/2018A%26A...610A..77H} {610, A77}

\bibitem[\protect\citeauthoryear{{Hansen}, {Aizenman}  \& {Ross}}{{Hansen}
  et~al.}{1976}]{hansen76}
{Hansen} C.~J.,  {Aizenman} M.~L.,   {Ross} R.~L.,  1976, \mn@doi [\apj]
  {10.1086/154542}, \href {http://adsabs.harvard.edu/abs/1976ApJ...207..736H}
  {207, 736}

\bibitem[\protect\citeauthoryear{{Hardee} \& {Stone}}{{Hardee} \&
  {Stone}}{1997}]{Hardee97}
{Hardee} P.~E.,  {Stone} J.~M.,  1997, \mn@doi [\apj] {10.1086/304208}, \href
  {http://adsabs.harvard.edu/abs/1997ApJ...483..121H} {483, 121}

\bibitem[\protect\citeauthoryear{{Hardee}, {Clarke}  \& {Howell}}{{Hardee}
  et~al.}{1995}]{Hardee95}
{Hardee} P.~E.,  {Clarke} D.~A.,   {Howell} D.~A.,  1995, \mn@doi [\apj]
  {10.1086/175389}, \href {http://adsabs.harvard.edu/abs/1995ApJ...441..644H}
  {441, 644}

\bibitem[\protect\citeauthoryear{{Harford} \& {Hamilton}}{{Harford} \&
  {Hamilton}}{2011}]{Harford11}
{Harford} A.~G.,  {Hamilton} A.~J.~S.,  2011, \mn@doi [\mnras]
  {10.1111/j.1365-2966.2011.19220.x}, \href
  {http://adsabs.harvard.edu/abs/2011MNRAS.416.2678H} {416, 2678}

\bibitem[\protect\citeauthoryear{{Harford}, {Hamilton}  \& {Gnedin}}{{Harford}
  et~al.}{2008}]{Harford08}
{Harford} A.~G.,  {Hamilton} A.~J.~S.,   {Gnedin} N.~Y.,  2008, \mn@doi
  [\mnras] {10.1111/j.1365-2966.2008.13608.x}, \href
  {http://adsabs.harvard.edu/abs/2008MNRAS.389..880H} {389, 880}

\bibitem[\protect\citeauthoryear{{Heigl}, {Burkert}  \& {Hacar}}{{Heigl}
  et~al.}{2016}]{Heigl16}
{Heigl} S.,  {Burkert} A.,   {Hacar} A.,  2016, \mn@doi [\mnras]
  {10.1093/mnras/stw2271}, \href
  {http://adsabs.harvard.edu/abs/2016MNRAS.463.4301H} {463, 4301}

\bibitem[\protect\citeauthoryear{{Heigl}, {Burkert}  \& {Gritschneder}}{{Heigl}
  et~al.}{2018a}]{Heigl18b}
{Heigl} S.,  {Burkert} A.,   {Gritschneder} M.,  2018a, \mn@doi [\mnras]
  {10.1093/mnras/stx3145}, \href
  {http://adsabs.harvard.edu/abs/2018MNRAS.474.4881H} {474, 4881}

\bibitem[\protect\citeauthoryear{{Heigl}, {Gritschneder}  \& {Burkert}}{{Heigl}
  et~al.}{2018b}]{Heigl18}
{Heigl} S.,  {Gritschneder} M.,   {Burkert} A.,  2018b, \mn@doi [\mnras]
  {10.1093/mnrasl/sly146}, \href
  {http://adsabs.harvard.edu/abs/2018MNRAS.481L...1H} {481, L1}

\bibitem[\protect\citeauthoryear{{Heitsch}}{{Heitsch}}{2013}]{Heitsch13}
{Heitsch} F.,  2013, \mn@doi [\apj] {10.1088/0004-637X/769/2/115}, \href
  {http://adsabs.harvard.edu/abs/2013ApJ...769..115H} {769, 115}

\bibitem[\protect\citeauthoryear{{Hennebelle} \& {Andr{\'e}}}{{Hennebelle} \&
  {Andr{\'e}}}{2013}]{Hennebelle13}
{Hennebelle} P.,  {Andr{\'e}} P.,  2013, \mn@doi [\aap]
  {10.1051/0004-6361/201321761}, \href
  {http://adsabs.harvard.edu/abs/2013A%26A...560A..68H} {560, A68}

\bibitem[\protect\citeauthoryear{{Henshaw}, {Longmore}  \&
  {Kruijssen}}{{Henshaw} et~al.}{2016}]{Henshaw16}
{Henshaw} J.~D.,  {Longmore} S.~N.,   {Kruijssen} J.~M.~D.,  2016, \mn@doi
  [\mnras] {10.1093/mnrasl/slw168}, \href
  {http://adsabs.harvard.edu/abs/2016MNRAS.463L.122H} {463, L122}

\bibitem[\protect\citeauthoryear{{Hily-Blant} \& {Falgarone}}{{Hily-Blant} \&
  {Falgarone}}{2009}]{HilyBlant09}
{Hily-Blant} P.,  {Falgarone} E.,  2009, \mn@doi [\aap]
  {10.1051/0004-6361/200912296}, \href
  {http://adsabs.harvard.edu/abs/2009A%26A...500L..29H} {500, L29}

\bibitem[\protect\citeauthoryear{{Hopkins}, {Kere{\v s}}, {Murray}, {Quataert}
  \& {Hernquist}}{{Hopkins} et~al.}{2012}]{Hopkins12}
{Hopkins} P.~F.,  {Kere{\v s}} D.,  {Murray} N.,  {Quataert} E.,   {Hernquist}
  L.,  2012, \mn@doi [\mnras] {10.1111/j.1365-2966.2012.21981.x}, \href
  {http://adsabs.harvard.edu/abs/2012MNRAS.427..968H} {427, 968}

\bibitem[\protect\citeauthoryear{Howard}{Howard}{1961}]{Howard61}
Howard L.~N.,  1961, \mn@doi [Journal of Fluid Mechanics]
  {10.1017/S0022112061000317}, 10, 509–512

\bibitem[\protect\citeauthoryear{{Huchra} et~al.,}{{Huchra}
  et~al.}{2005}]{Huchra05}
{Huchra} J.,  et~al., 2005, in {Fairall} A.~P.,  {Woudt} P.~A.,  eds,
  Astronomical Society of the Pacific Conference Series Vol. 329, Nearby
  Large-Scale Structures and the Zone of Avoidance. p. Fairall

\bibitem[\protect\citeauthoryear{{Hunter}, {Whitaker}  \& {Lovelace}}{{Hunter}
  et~al.}{1997}]{Hunter97}
{Hunter} Jr. J.~H.,  {Whitaker} R.~W.,   {Lovelace} R.~V.~E.,  1997, \mn@doi
  [\apj] {10.1086/304154}, \href
  {http://adsabs.harvard.edu/abs/1997ApJ...482..852H} {482, 852}

\bibitem[\protect\citeauthoryear{{Hunter}, {Whitaker}  \& {Lovelace}}{{Hunter}
  et~al.}{1998}]{Hunter98}
{Hunter} Jr. J.~H.,  {Whitaker} R.~W.,   {Lovelace} R.~V.~E.,  1998, \mn@doi
  [\apj] {10.1086/306428}, \href
  {http://adsabs.harvard.edu/abs/1998ApJ...508..680H} {508, 680}

\bibitem[\protect\citeauthoryear{{Inoue} \& {Yoshida}}{{Inoue} \&
  {Yoshida}}{2018}]{Inoue18}
{Inoue} S.,  {Yoshida} N.,  2018, \mn@doi [\mnras] {10.1093/mnras/stx2978},
  \href {http://adsabs.harvard.edu/abs/2018MNRAS.474.3466I} {474, 3466}

\bibitem[\protect\citeauthoryear{{Inutsuka} \& {Miyama}}{{Inutsuka} \&
  {Miyama}}{1992}]{Inutsuka92}
{Inutsuka} S.-I.,  {Miyama} S.~M.,  1992, \mn@doi [\apj] {10.1086/171162},
  \href {http://adsabs.harvard.edu/abs/1992ApJ...388..392I} {388, 392}

\bibitem[\protect\citeauthoryear{{Jackson}, {Finn}, {Chambers}, {Rathborne}  \&
  {Simon}}{{Jackson} et~al.}{2010}]{Jackson10}
{Jackson} J.~M.,  {Finn} S.~C.,  {Chambers} E.~T.,  {Rathborne} J.~M.,
  {Simon} R.,  2010, \mn@doi [\apjl] {10.1088/2041-8205/719/2/L185}, \href
  {http://adsabs.harvard.edu/abs/2010ApJ...719L.185J} {719, L185}

\bibitem[\protect\citeauthoryear{{Kere{\v s}}, {Katz}, {Weinberg}  \&
  {Dav{\'e}}}{{Kere{\v s}} et~al.}{2005}]{Keres05}
{Kere{\v s}} D.,  {Katz} N.,  {Weinberg} D.~H.,   {Dav{\'e}} R.,  2005, \mn@doi
  [\mnras] {10.1111/j.1365-2966.2005.09451.x}, \href
  {http://adsabs.harvard.edu/abs/2005MNRAS.363....2K} {363, 2}

\bibitem[\protect\citeauthoryear{{Kirk} et~al.,}{{Kirk} et~al.}{2013}]{Kirk13}
{Kirk} J.~M.,  et~al., 2013, \mn@doi [\mnras] {10.1093/mnras/stt561}, \href
  {http://adsabs.harvard.edu/abs/2013MNRAS.432.1424K} {432, 1424}

\bibitem[\protect\citeauthoryear{{Kirk} et~al.,}{{Kirk} et~al.}{2017}]{kirk17}
{Kirk} H.,  et~al., 2017, \mn@doi [\apj] {10.3847/1538-4357/aa8631}, \href
  {http://adsabs.harvard.edu/abs/2017ApJ...846..144K} {846, 144}

\bibitem[\protect\citeauthoryear{{Kruijssen} et~al.,}{{Kruijssen}
  et~al.}{2019}]{Kruijssen19}
{Kruijssen} J.~M.~D.,  et~al., 2019, \mn@doi [\mnras] {10.1093/mnras/stz381},
  \href {http://adsabs.harvard.edu/abs/2019MNRAS.484.5734K} {484, 5734}

\bibitem[\protect\citeauthoryear{{Krumholz}}{{Krumholz}}{2015}]{Krumholz15}
{Krumholz} M.~R.,  2015, arXiv e-prints, \href
  {http://adsabs.harvard.edu/abs/2015arXiv151103457K} {}

\bibitem[\protect\citeauthoryear{{Laigle} et~al.,}{{Laigle}
  et~al.}{2015}]{Laigle15}
{Laigle} C.,  et~al., 2015, \mn@doi [\mnras] {10.1093/mnras/stu2289}, \href
  {http://adsabs.harvard.edu/abs/2015MNRAS.446.2744L} {446, 2744}

\bibitem[\protect\citeauthoryear{{Larson}}{{Larson}}{1985}]{Larson85}
{Larson} R.~B.,  1985, \mn@doi [\mnras] {10.1093/mnras/214.3.379}, \href
  {http://adsabs.harvard.edu/abs/1985MNRAS.214..379L} {214, 379}

\bibitem[\protect\citeauthoryear{{Leclercq} et~al.,}{{Leclercq}
  et~al.}{2017}]{Leclercq17}
{Leclercq} F.,  et~al., 2017, \mn@doi [\aap] {10.1051/0004-6361/201731480},
  \href {http://adsabs.harvard.edu/abs/2017A%26A...608A...8L} {608, A8}

\bibitem[\protect\citeauthoryear{{Mandelker}, {Padnos}, {Dekel}, {Birnboim},
  {Burkert}, {Krumholz}  \& {Steinberg}}{{Mandelker} et~al.}{2016}]{M16}
{Mandelker} N.,  {Padnos} D.,  {Dekel} A.,  {Birnboim} Y.,  {Burkert} A.,
  {Krumholz} M.~R.,   {Steinberg} E.,  2016, \mn@doi [\mnras]
  {10.1093/mnras/stw2267}, \href
  {http://adsabs.harvard.edu/abs/2016MNRAS.463.3921M} {463, 3921}

\bibitem[\protect\citeauthoryear{{Mandelker}, {Dekel}, {Ceverino}, {DeGraf},
  {Guo}  \& {Primack}}{{Mandelker} et~al.}{2017}]{M17}
{Mandelker} N.,  {Dekel} A.,  {Ceverino} D.,  {DeGraf} C.,  {Guo} Y.,
  {Primack} J.,  2017, \mn@doi [\mnras] {10.1093/mnras/stw2358}, \href
  {http://adsabs.harvard.edu/abs/2017MNRAS.464..635M} {464, 635}

\bibitem[\protect\citeauthoryear{{Mandelker}, {van Dokkum}, {Brodie}, {van den
  Bosch}  \& {Ceverino}}{{Mandelker} et~al.}{2018}]{M18a}
{Mandelker} N.,  {van Dokkum} P.~G.,  {Brodie} J.~P.,  {van den Bosch} F.~C.,
  {Ceverino} D.,  2018, \mn@doi [\apj] {10.3847/1538-4357/aaca98}, \href
  {http://adsabs.harvard.edu/abs/2018ApJ...861..148M} {861, 148}

\bibitem[\protect\citeauthoryear{{Mandelker}, {Nagai}, {Aung}, {Dekel},
  {Padnos}  \& {Birnboim}}{{Mandelker} et~al.}{2019}]{M18b}
{Mandelker} N.,  {Nagai} D.,  {Aung} H.,  {Dekel} A.,  {Padnos} D.,
  {Birnboim} Y.,  2019, \mn@doi [\mnras] {10.1093/mnras/stz012}, \href
  {http://adsabs.harvard.edu/abs/2019MNRAS.484.1100M} {484, 1100}

\bibitem[\protect\citeauthoryear{{Martin}, {Chang}, {Matuszewski}, {Morrissey},
  {Rahman}, {Moore}  \& {Steidel}}{{Martin} et~al.}{2014a}]{Martin14a}
{Martin} D.~C.,  {Chang} D.,  {Matuszewski} M.,  {Morrissey} P.,  {Rahman} S.,
  {Moore} A.,   {Steidel} C.~C.,  2014a, \mn@doi [\apj]
  {10.1088/0004-637X/786/2/106}, \href
  {http://adsabs.harvard.edu/abs/2014ApJ...786..106M} {786, 106}

\bibitem[\protect\citeauthoryear{{Martin}, {Chang}, {Matuszewski}, {Morrissey},
  {Rahman}, {Moore}, {Steidel}  \& {Matsuda}}{{Martin}
  et~al.}{2014b}]{Martin14b}
{Martin} D.~C.,  {Chang} D.,  {Matuszewski} M.,  {Morrissey} P.,  {Rahman} S.,
  {Moore} A.,  {Steidel} C.~C.,   {Matsuda} Y.,  2014b, \mn@doi [\apj]
  {10.1088/0004-637X/786/2/107}, \href
  {http://adsabs.harvard.edu/abs/2014ApJ...786..107M} {786, 107}

\bibitem[\protect\citeauthoryear{{Massaglia}, {Trussoni}, {Bodo}, {Rossi}  \&
  {Ferrari}}{{Massaglia} et~al.}{1992}]{Massaglia92}
{Massaglia} S.,  {Trussoni} E.,  {Bodo} G.,  {Rossi} P.,   {Ferrari} A.,  1992,
  \aap, \href {http://adsabs.harvard.edu/abs/1992A%26A...260..243M} {260, 243}

\bibitem[\protect\citeauthoryear{{Micono}, {Bodo}, {Massaglia}, {Rossi},
  {Ferrari}  \& {Rosner}}{{Micono} et~al.}{2000}]{Micono00}
{Micono} M.,  {Bodo} G.,  {Massaglia} S.,  {Rossi} P.,  {Ferrari} A.,
  {Rosner} R.,  2000, \aap, \href
  {http://adsabs.harvard.edu/abs/2000A%26A...360..795M} {360, 795}

\bibitem[\protect\citeauthoryear{{Mikha{\v i}lovski{\v i}} \&
  {Fridman}}{{Mikha{\v i}lovski{\v i}} \& {Fridman}}{1972}]{Mikhailovskii72}
{Mikha{\v i}lovski{\v i}} A.~B.,  {Fridman} A.~M.,  1972, Soviet Journal of
  Experimental and Theoretical Physics, \href
  {http://adsabs.harvard.edu/abs/1972JETP...34..243M} {34, 243}

\bibitem[\protect\citeauthoryear{Miles}{Miles}{1961}]{Miles61}
Miles J.~W.,  1961, \mn@doi [Journal of Fluid Mechanics]
  {10.1017/S0022112061000305}, 10, 496–508

\bibitem[\protect\citeauthoryear{{Moeckel} \& {Burkert}}{{Moeckel} \&
  {Burkert}}{2015}]{Moeckel15}
{Moeckel} N.,  {Burkert} A.,  2015, \mn@doi [\apj]
  {10.1088/0004-637X/807/1/67}, \href
  {http://adsabs.harvard.edu/abs/2015ApJ...807...67M} {807, 67}

\bibitem[\protect\citeauthoryear{{Molinari} et~al.,}{{Molinari}
  et~al.}{2010}]{Molinari10}
{Molinari} S.,  et~al., 2010, \mn@doi [\aap] {10.1051/0004-6361/201014659},
  \href {http://adsabs.harvard.edu/abs/2010A%26A...518L.100M} {518, L100}

\bibitem[\protect\citeauthoryear{{Murray}, {White}, {Blondin}  \&
  {Lin}}{{Murray} et~al.}{1993}]{Murray93}
{Murray} S.~D.,  {White} S.~D.~M.,  {Blondin} J.~M.,   {Lin} D.~N.~C.,  1993,
  \mn@doi [\apj] {10.1086/172540}, \href
  {http://adsabs.harvard.edu/abs/1993ApJ...407..588M} {407, 588}

\bibitem[\protect\citeauthoryear{{Nagasawa}}{{Nagasawa}}{1987}]{Nagasawa}
{Nagasawa} M.,  1987, \mn@doi [Progress of Theoretical Physics]
  {10.1143/PTP.77.635}, \href
  {http://adsabs.harvard.edu/abs/1987PThPh..77..635N} {77, 635}

\bibitem[\protect\citeauthoryear{{Nelson}, {Vogelsberger}, {Genel}, {Sijacki},
  {Kere{\v s}}, {Springel}  \& {Hernquist}}{{Nelson} et~al.}{2013}]{Nelson13}
{Nelson} D.,  {Vogelsberger} M.,  {Genel} S.,  {Sijacki} D.,  {Kere{\v s}} D.,
  {Springel} V.,   {Hernquist} L.,  2013, \mn@doi [\mnras]
  {10.1093/mnras/sts595}, \href
  {http://adsabs.harvard.edu/abs/2013MNRAS.429.3353N} {429, 3353}

\bibitem[\protect\citeauthoryear{{Nelson}, {Genel}, {Pillepich},
  {Vogelsberger}, {Springel}  \& {Hernquist}}{{Nelson} et~al.}{2016}]{Nelson16}
{Nelson} D.,  {Genel} S.,  {Pillepich} A.,  {Vogelsberger} M.,  {Springel} V.,
   {Hernquist} L.,  2016, \mn@doi [\mnras] {10.1093/mnras/stw1191}, \href
  {http://adsabs.harvard.edu/abs/2016MNRAS.460.2881N} {460, 2881}

\bibitem[\protect\citeauthoryear{{Ocvirk}, {Pichon}  \& {Teyssier}}{{Ocvirk}
  et~al.}{2008}]{Ocvirk08}
{Ocvirk} P.,  {Pichon} C.,   {Teyssier} R.,  2008, \mn@doi [\mnras]
  {10.1111/j.1365-2966.2008.13763.x}, \href
  {http://adsabs.harvard.edu/abs/2008MNRAS.390.1326O} {390, 1326}

\bibitem[\protect\citeauthoryear{{Orkisz} et~al.,}{{Orkisz}
  et~al.}{2019}]{Orkisz19}
{Orkisz} J.~H.,  et~al., 2019, arXiv e-prints, \href
  {http://adsabs.harvard.edu/abs/2019arXiv190202077O} {}

\bibitem[\protect\citeauthoryear{{Ostriker}}{{Ostriker}}{1964a}]{Ostriker64}
{Ostriker} J.,  1964a, \mn@doi [\apj] {10.1086/148005}, \href
  {http://adsabs.harvard.edu/abs/1964ApJ...140.1056O} {140, 1056}

\bibitem[\protect\citeauthoryear{{Ostriker}}{{Ostriker}}{1964b}]{Ostriker64b}
{Ostriker} J.,  1964b, \mn@doi [\apj] {10.1086/148057}, \href
  {http://adsabs.harvard.edu/abs/1964ApJ...140.1529O} {140, 1529}

\bibitem[\protect\citeauthoryear{{Padnos}, {Mandelker}, {Birnboim}, {Dekel},
  {Krumholz}  \& {Steinberg}}{{Padnos} et~al.}{2018}]{P18}
{Padnos} D.,  {Mandelker} N.,  {Birnboim} Y.,  {Dekel} A.,  {Krumholz} M.~R.,
  {Steinberg} E.,  2018, \mn@doi [\mnras] {10.1093/mnras/sty789}, \href
  {http://adsabs.harvard.edu/abs/2018MNRAS.477.3293P} {477, 3293}

\bibitem[\protect\citeauthoryear{{Padoan}, {Juvela}, {Goodman}  \&
  {Nordlund}}{{Padoan} et~al.}{2001}]{Padoan01}
{Padoan} P.,  {Juvela} M.,  {Goodman} A.~A.,   {Nordlund} {\AA}.,  2001,
  \mn@doi [\apj] {10.1086/320636}, \href
  {http://adsabs.harvard.edu/abs/2001ApJ...553..227P} {553, 227}

\bibitem[\protect\citeauthoryear{{Palmeirim} et~al.,}{{Palmeirim}
  et~al.}{2013}]{Palmeirim13}
{Palmeirim} P.,  et~al., 2013, \mn@doi [\aap] {10.1051/0004-6361/201220500},
  \href {http://adsabs.harvard.edu/abs/2013A%26A...550A..38P} {550, A38}

\bibitem[\protect\citeauthoryear{{Payne} \& {Cohn}}{{Payne} \&
  {Cohn}}{1985}]{Payne_Cohn85}
{Payne} D.~G.,  {Cohn} H.,  1985, \mn@doi [\apj] {10.1086/163104}, \href
  {http://adsabs.harvard.edu/abs/1985ApJ...291..655P} {291, 655}

\bibitem[\protect\citeauthoryear{{Prochaska}, {Lau}  \& {Hennawi}}{{Prochaska}
  et~al.}{2014}]{Prochaska14}
{Prochaska} J.~X.,  {Lau} M.~W.,   {Hennawi} J.~F.,  2014, \mn@doi [\apj]
  {10.1088/0004-637X/796/2/140}, \href
  {http://adsabs.harvard.edu/abs/2014ApJ...796..140P} {796, 140}

\bibitem[\protect\citeauthoryear{{Rees}}{{Rees}}{1988}]{Rees88}
{Rees} M.~J.,  1988, \mn@doi [\nat] {10.1038/333523a0}, \href
  {http://adsabs.harvard.edu/abs/1988Natur.333..523R} {333, 523}

\bibitem[\protect\citeauthoryear{{Robertson}, {Kravtsov}, {Gnedin}, {Abel}  \&
  {Rudd}}{{Robertson} et~al.}{2010}]{Robertson10}
{Robertson} B.~E.,  {Kravtsov} A.~V.,  {Gnedin} N.~Y.,  {Abel} T.,   {Rudd}
  D.~H.,  2010, \mn@doi [\mnras] {10.1111/j.1365-2966.2009.15823.x}, \href
  {http://adsabs.harvard.edu/abs/2010MNRAS.401.2463R} {401, 2463}

\bibitem[\protect\citeauthoryear{{Rodriguez-Franco}, {Martin-Pintado},
  {Gomez-Gonzalez}  \& {Planesas}}{{Rodriguez-Franco}
  et~al.}{1992}]{Rodriguez92}
{Rodriguez-Franco} A.,  {Martin-Pintado} J.,  {Gomez-Gonzalez} J.,   {Planesas}
  P.,  1992, \aap, \href {http://adsabs.harvard.edu/abs/1992A%26A...264..592R}
  {264, 592}

\bibitem[\protect\citeauthoryear{{Schneider} \& {Elmegreen}}{{Schneider} \&
  {Elmegreen}}{1979}]{Schneider79}
{Schneider} S.,  {Elmegreen} B.~G.,  1979, \mn@doi [\apjs] {10.1086/190609},
  \href {http://adsabs.harvard.edu/abs/1979ApJS...41...87S} {41, 87}

\bibitem[\protect\citeauthoryear{{Seo} et~al.,}{{Seo} et~al.}{2018}]{Seo18}
{Seo} Y.~M.,  et~al., 2018, arXiv e-prints, \href
  {https://ui.adsabs.harvard.edu/\#abs/2018arXiv181206121S} {p.
  arXiv:1812.06121}

\bibitem[\protect\citeauthoryear{{Smith}, {Glover}, {Klessen}  \&
  {Fuller}}{{Smith} et~al.}{2016}]{Smith16}
{Smith} R.~J.,  {Glover} S.~C.~O.,  {Klessen} R.~S.,   {Fuller} G.~A.,  2016,
  \mn@doi [\mnras] {10.1093/mnras/stv2559}, \href
  {http://adsabs.harvard.edu/abs/2016MNRAS.455.3640S} {455, 3640}

\bibitem[\protect\citeauthoryear{{Springel} et~al.,}{{Springel}
  et~al.}{2005}]{Springel05}
{Springel} V.,  et~al., 2005, \mn@doi [\nat] {10.1038/nature03597}, \href
  {http://adsabs.harvard.edu/abs/2005Natur.435..629S} {435, 629}

\bibitem[\protect\citeauthoryear{{Stone}, {Xu}  \& {Hardee}}{{Stone}
  et~al.}{1997}]{Stone97}
{Stone} J.~M.,  {Xu} J.,   {Hardee} P.,  1997, \mn@doi [\apj] {10.1086/304209},
  \href {http://adsabs.harvard.edu/abs/1997ApJ...483..136S} {483, 136}

\bibitem[\protect\citeauthoryear{{Tegmark} et~al.,}{{Tegmark}
  et~al.}{2004}]{Tegmark04}
{Tegmark} M.,  et~al., 2004, \mn@doi [\apj] {10.1086/382125}, \href
  {http://adsabs.harvard.edu/abs/2004ApJ...606..702T} {606, 702}

\bibitem[\protect\citeauthoryear{{Teyssier}}{{Teyssier}}{2002}]{Teyssier02}
{Teyssier} R.,  2002, \mn@doi [\aap] {10.1051/0004-6361:20011817}, \href
  {http://adsabs.harvard.edu/abs/2002A%26A...385..337T} {385, 337}

\bibitem[\protect\citeauthoryear{{Toro}, {Spruce}  \& {Speares}}{{Toro}
  et~al.}{1994}]{Toro94}
{Toro} E.~F.,  {Spruce} M.,   {Speares} W.,  1994, \mn@doi [Shock Waves]
  {10.1007/BF01414629}, 4, 25

\bibitem[\protect\citeauthoryear{{V{\'a}zquez-Semadeni}, {Gonz{\'a}lez},
  {Ballesteros-Paredes}, {Gazol}  \& {Kim}}{{V{\'a}zquez-Semadeni}
  et~al.}{2008}]{VS08}
{V{\'a}zquez-Semadeni} E.,  {Gonz{\'a}lez} R.~F.,  {Ballesteros-Paredes} J.,
  {Gazol} A.,   {Kim} J.,  2008, \mn@doi [\mnras]
  {10.1111/j.1365-2966.2008.13778.x}, \href
  {http://adsabs.harvard.edu/abs/2008MNRAS.390..769V} {390, 769}

\bibitem[\protect\citeauthoryear{{Vietri}, {Ferrara}  \& {Miniati}}{{Vietri}
  et~al.}{1997}]{Vietri97}
{Vietri} M.,  {Ferrara} A.,   {Miniati} F.,  1997, \apj, \href
  {http://adsabs.harvard.edu/abs/1997ApJ...483..262V} {483, 262}

\bibitem[\protect\citeauthoryear{{Xu}, {Hardee}  \& {Stone}}{{Xu}
  et~al.}{2000}]{Xu00}
{Xu} J.,  {Hardee} P.~E.,   {Stone} J.~M.,  2000, \mn@doi [\apj]
  {10.1086/317094}, \href {http://adsabs.harvard.edu/abs/2000ApJ...543..161X}
  {543, 161}

\bibitem[\protect\citeauthoryear{{Zel'dovich}}{{Zel'dovich}}{1970}]{Zeldovich70}
{Zel'dovich} Y.~B.,  1970, Astronomy and Astrophysics, 5, 84

\bibitem[\protect\citeauthoryear{{Zinger}, {Dekel}, {Birnboim}, {Kravtsov}  \&
  {Nagai}}{{Zinger} et~al.}{2016}]{Zinger16}
{Zinger} E.,  {Dekel} A.,  {Birnboim} Y.,  {Kravtsov} A.,   {Nagai} D.,  2016,
  \mn@doi [\mnras] {10.1093/mnras/stw1283}, \href
  {http://adsabs.harvard.edu/abs/2016MNRAS.461..412Z} {461, 412}

\bibitem[\protect\citeauthoryear{{van Leer}}{{van Leer}}{1977}]{vanLeer77}
{van Leer} B.,  1977, \mn@doi [Journal of Computational Physics]
  {10.1016/0021-9991(77)90094-8}, 23, 263

\bibitem[\protect\citeauthoryear{{van de Voort} \& {Schaye}}{{van de Voort} \&
  {Schaye}}{2012}]{Voort12}
{van de Voort} F.,  {Schaye} J.,  2012, \mn@doi [\mnras]
  {10.1111/j.1365-2966.2012.20949.x}, \href
  {http://adsabs.harvard.edu/abs/2012MNRAS.423.2991V} {423, 2991}

\bibitem[\protect\citeauthoryear{{van de Voort}, {Schaye}, {Booth}, {Haas}  \&
  {Dalla Vecchia}}{{van de Voort} et~al.}{2011}]{vdv11}
{van de Voort} F.,  {Schaye} J.,  {Booth} C.~M.,  {Haas} M.~R.,   {Dalla
  Vecchia} C.,  2011, \mn@doi [\mnras] {10.1111/j.1365-2966.2011.18565.x},
  \href {http://adsabs.harvard.edu/abs/2011MNRAS.414.2458V} {414, 2458}

\makeatother
\end{thebibliography}
\bibliographystyle{mnras}

%%%%%%%%%%%%%%%%%%%%%%%%%%%%%%%%%%%%%%%%%%%%%%%%%%%%%%%%
\appendix

\section{GI Growth Rates and Surface Vs Body Modes}
\label{sec:surf_bod}

\smallskip
For an incompressible pressure-confined cylinder, the linear growth-time of the fastest growing GI mode is given by \equ{hunter}, based on H98. In our case, however, the stream is highly compressible and its density can be far from constant (\fig{profiles}), so it is unclear whether \equ{hunter} remains valid. On the other hand, for isothermal cylinders confined by a zero density background, the ratio $t_{\rm max}/t_{\rm ff}$ does not vary much with line-mass (N87). If the same is true for a non-isothermal cylinder confined by an arbitrary density background, then \equ{hunter} may also apply to our case. To test this, we performed simulations with self-gravity but without shear flow, $M_b=0$, and with a single perturbation wavelength, $\lambda=1/4=8\Rs$. We examined two different values of the line-mass, $\mu=0.4$ and $0.9$, with $\deltac=100$. Based on the analysis of N87, the former is expected to be unstable to surface modes, while the latter to body modes. 

\begin{figure}
\includegraphics[width=0.49\textwidth]{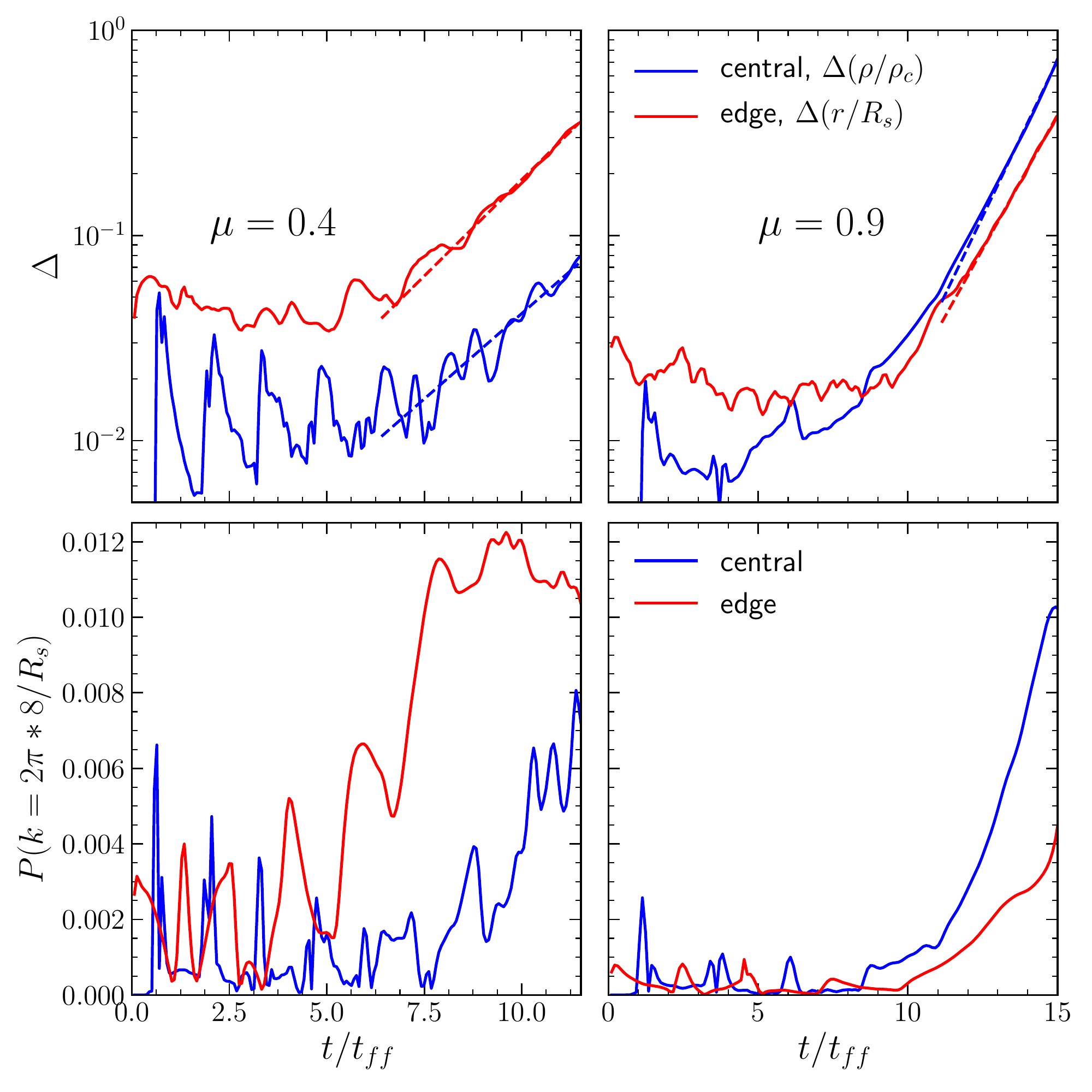}
\caption{Growth of a single wavelength perturbation, $k=2\pi\times 8/R_s$, due to GI in simulations with line-mass $\mu=0.4$ (left) and $0.9$ (right). {\it Top panels:} the perturbation amplitude $\Delta$, measured as the rms density fluctuations along the stream axis (blue lines, \equ{edge_pert}) or as the deformation of the stream-background interface (red lines, \equ{cen_pert}). Both definitions yield similar evolution. After a perturbation sound crossing time, the perturbations grow exponentially, with best fit exponential growth rates, shown by dashed lines, within $\lsim 10\%$ of those predicted by H98. {\it Bottom panels:} the power of the density perturbation at the perturbed wavelength, $\lambda=8R_s$, measured along the stream axis (blue lines) and near the stream edge (red lines). For $\mu=0.4~(0.9)$ perturbations near the edge (centre) contain more power and grow faster. This is consistent with $\mu=0.4~(0.9)$ being a surface (body) mode. }
\label{fig:growth_time}
\end{figure}

\smallskip
\fig{growth_time} shows the perturbation amplitude as a function of time, measured in each simulation at both the surface of the stream and along its axis. The former is defined by the deformation of the stream-background interface, 
\be
\label{eq:edge_pert}
\Delta \left(r/R_s\right) = ({\rm max}-{\rm min})\left(r/R_s\right), 
%\Delta \left(\frac{r}{R_s}\right) = ({\rm max}-{\rm min})\left(\frac{r}{R_s}\right), 
\ee 
{\no}where $r$ specifies the radius at which the tracer variable $\psi=0.5$. For the perturbation along the stream axis we use the density
\be
\label{eq:cen_pert}
\Delta \left(\rho/\rho_c\right) = {\rm rms}\left[\rho(t,r=0)/\rho_c-1\right].
%\Delta \left(\frac{\rho}{\rho_c}\right) = {\rm rms}\left(\frac{\rho(t,r=0)}{\rho_c}-1\right).
\ee
%\smallskip
As can be seen in the top panels of \fig{growth_time}, these two measurements of the perturbation amplitude yield similar growth rates. At $t\lsim 7t_{\rm ff}$, the perturbation amplitude is roughly constant. This is approximately four times the stream sound crossing time, $\tsc \propto 2\Rs/\cs$ (see \equnp{sound_speed} and \tab{sim_clump}), which is the sound crossing time of the perturbation wavelength, $\lambda = 8\Rs$. This is the coherence time of the perturbation, during which the initial velocity perturbation is converted into a growing eigenmode of the system (see M16 and M19 for a discussion of a similar phenomenon in KHI). At later times the perturbation amplitude is well fit by  $\Delta \propto \exp{(\omega t)}$. The best fit growth rates are $\omega_{\rm fit}\approx 0.96\omega_{\rm Hunter}$ for $\mu=0.4$, and $\omega_{\rm fit}\approx 0.90\omega_{\rm Hunter}$ for $\mu=0.9$, 
where the profile deviates further from the constant density assumption in H98. We conclude that the growth time for GI can be approximated by the H98 dispersion relation even for large values of the line-mass, $\mu$. 

The bottom panels compare the 1-D power spectra of density perturbations along the stream axis, at $r=0$, and near its edge, at $r\lsim \Rs$. We show here the power measured at the wavelength $8R_s$, but note that the results are nearly identical when showing the total power, as power at all other scales is small. For $\mu=0.4$, the power near the stream edge is larger, while the opposite is true for $\mu=0.9$, consistent with these two simulations corresponding to surface and body modes, respectively.

\end{document}